\documentclass[11pt]{article}
\usepackage{verbatim,amsmath,amssymb}
\usepackage{epsfig,float,color}
\usepackage[font=small,labelfont=bf]{caption}
\usepackage{geometry}
\usepackage{setspace}
\usepackage{natbib}
\usepackage{amsthm,mathrsfs,amsfonts,dsfont} 
\usepackage{bm}
\usepackage{hyperref}
\usepackage[labelfont=bf]{caption}

\definecolor{darkblue}{rgb}{0,0,1}
\definecolor{col1}{rgb}{1,0,1}
\definecolor{col2}{rgb}{0,0.5,0}
\definecolor{col3}{rgb}{0.5,0,1}
\definecolor{col4}{rgb}{0.1,.75,0}
\hypersetup{pdftex=true, colorlinks=true, breaklinks=true, linkcolor=darkblue, menucolor=darkblue, pagecolor=darkblue, citecolor=darkblue, urlcolor=darkblue}

\newtheoremstyle{rem}%name
{6pt}%Space above
{6pt}%Space below
{\small}%Body font
{}%Indent amount
{\bf}%Theorem head font
{:}%Punctuation after theorem head
{.5em}%Space after theorem head
{}%Theorem head spec(can be left empty, meaning ‘normal’)

\theoremstyle{rem}
\newtheorem{remark}{Remark}[section]

\newcommand{\bitm}{\begin{itemize}}
\newcommand{\eitm}{\end{itemize}}
\newcommand{\bnumr}{\begin{enumerate}}
\newcommand{\enumr}{\end{enumerate}}

\newcommand{\mrT}{\mathrm{T}}

\newcommand {\eqb}[1]{\begin{equation}\begin{array}{#1}}
\newcommand {\eqe}{\end{array}\end{equation}}

\newcommand {\esb}[1]{\begin{equation*}\begin{array}{#1}}
\newcommand {\ese}{\end{array}\end{equation*}}
\newcommand {\ds}{\displaystyle}

\newcommand {\pa}[2]{\frac{\partial{#1}}{\partial{#2}}}
\newcommand {\pad}[2]{\frac{\mathrm{d}{#1}}{\mathrm{d}{#2}}}

\newcommand {\back}{\! \! \!}
\newcommand {\is}{\back &=& \back}
\newcommand {\dis}{\back &:=& \back}

\newcommand {\plus}{\back &+& \back}
\newcommand {\mi}{\back &-& \back}

\newcommand {\norm}[1]{\|#1\|}

\newcommand {\grad}{\mathrm{grad}\,}
\newcommand {\divz}{\mathrm{div}\,}

\newcommand {\dif}{\mathrm{d}}

\newcommand {\II}{{I\kern-.3em I}}
\newcommand {\III}{{I\kern-.3em I\kern-.3em I}}

\newcommand {\mra}{\mathrm{a}}
\newcommand {\mrb}{\mathrm{b}}
\newcommand {\mrc}{\mathrm{c}}
\newcommand {\mrd}{\mathrm{d}}
\newcommand {\mre}{\mathrm{e}}

\newcommand {\mri}{\mathrm{i}}

\newcommand {\mrk}{\mathrm{k}}

\newcommand {\mrm}{\mathrm{m}}
\newcommand {\mrn}{\mathrm{n}}

\newcommand {\mrp}{\mathrm{p}}

\newcommand {\mrr}{\mathrm{r}}
\newcommand {\mrs}{\mathrm{s}}
\newcommand {\mrt}{\mathrm{t}}
\newcommand {\mru}{\mathrm{u}}

\newcommand {\ba}{\boldsymbol{a}}
\newcommand {\bb}{\boldsymbol{b}}

\newcommand {\bg}{\boldsymbol{g}}

\newcommand {\bi}{\boldsymbol{i}}

\newcommand {\bn}{\boldsymbol{n}}

\newcommand {\bq}{\boldsymbol{q}}

\newcommand {\bt}{\boldsymbol{t}}

\newcommand {\bv}{\boldsymbol{v}}

\newcommand {\bx}{\boldsymbol{x}}

\newcommand {\bmu}{\boldsymbol{\mu}}

\newcommand {\bet}{\mbox{\boldmath$\eta$}}

\newcommand {\bD}{\boldsymbol{D}}
\newcommand {\bE}{\boldsymbol{E}}
\newcommand {\bF}{\boldsymbol{F}}

\newcommand {\bS}{\boldsymbol{S}}

\newcommand {\bX}{\boldsymbol{X}}

\newcommand {\bveps}{\mbox{\boldmath$\epsilon$}}

\newcommand {\bsig}{\mbox{\boldmath$\sigma$}}

\newcommand {\bone}{\mathbf{1}}

\newcommand {\bbE}{\mathbb{E}}

\newcommand {\bbK}{\mathbb{K}}

\newcommand {\bbR}{\mathbb{R}}
\newcommand {\bbS}{\mathbb{S}}

\newcommand {\bbU}{\mathbb{U}}

\newcommand {\IR}{{\rm\kern.24em
   \vrule width.02em height1.53ex depth-.05ex
   \kern-.3em R}}
\newcommand {\ic}{{\rm\kern.20em
   \vrule width.02em height1.0ex depth-.05ex
   \kern-.22em c}}
\newcommand {\ia}{{\rm\kern.20em
   \vrule width.02em height1.05ex depth-.0ex
   \kern-.25em a}}
\newcommand {\IC}{{\rm\kern.24em
   \vrule width.02em height1.4ex depth-.05ex
   \kern-.26em C}}
\newcommand {\ID}{{\rm\kern.34em
   \vrule width.02em height1.5ex depth-.05ex
   \kern-.36em D}}
\newcommand {\IS}{{\rm\kern.24em
   \vrule width.02em height1.6ex depth.05ex
   \kern-.26em S}}
\newcommand {\IT}{{\rm\kern.50em
   \vrule width.02em height1.55ex depth-.05ex
   \kern-.52em T}}

\newcommand {\IE}{{\rm\kern.24em
   \vrule width.02em height1.55ex depth-.05ex
   \kern-.33em E}}
\newcommand {\IEa}{{\rm\kern.24em
   \vrule width.02em height1.55ex depth-.05ex
   \kern-.33em E}^{1}_{ijkl}}
\newcommand {\IEb}{{\rm\kern.24em
   \vrule width.02em height1.55ex depth-.05ex
   \kern-.33em E}^{2}_{ijkl}}

\newcommand {\sB}{\mathcal{B}}

\newcommand {\sD}{\mathcal{D}}

\newcommand {\sL}{\mathcal{L}}

\newcommand {\sP}{\mathcal{P}}

\newcommand {\sS}{\mathcal{S}}

\newcommand {\Ass}[2]{\kern 0.9ex \vrule width0.45em height0.2ex depth0ex \kern -2.1ex \bigwedge_{#1}^{#2}}
\newcommand {\ASS}[2]{\kern 1.45ex \vrule width0.5em height0.2ex depth0ex \kern -2.65ex \bigwedge_{#1}^{#2}}

\newcommand{\unde}[1]{\mathsf{#1}}	
\newcommand{\undet}{\bm{\unde{t}}}
\newcommand{\undetc}{\bm{\unde{t}}_\unde{c}}

\newcommand{\undeEc}{\bm{\unde{E}}_\unde{c}}
\newcommand{\Jc}{J_\mrc}
\newcommand{\nc}{n_\mrc}
\newcommand{\Nc}{N_\mrc}
\newcommand{\Lg}{\mathring\bg_\mrt}
\newcommand{\Lge}{\mathring\bg_\mre}
\newcommand{\Lgi}{\mathring\bg_\mri}

\newcommand{\bexk}{\bar\bb_k}
\newcommand{\rexk}{\bar r_k}

\begin{document}

\begin{center}
\Large{\bf{A chemo-mechano-thermodynamical contact theory \\ for adhesion, friction and (de)bonding reactions}}\\

\end{center}

\renewcommand{\thefootnote}{\fnsymbol{footnote}}

\begin{center}
\large{Roger A.~Sauer$^{\mra,\mrb,\mrc,}$\footnote[1]{corresponding author, email: sauer@aices.rwth-aachen.de}, Thang X.~Duong$^\mra$ and Kranthi K.~Mandadapu$^{\mrd,\mre}$
}\\
\vspace{4mm}

\small{\textit{
$^\mra$Aachen Institute for Advanced Study in Computational Engineering Science (AICES), \\ 
RWTH Aachen University, Templergraben 55, 52056 Aachen, Germany \\[1.1mm]
$^\mrb$Department of Mechanical Engineering, Indian Institute of Technology Kanpur, UP 208016, India \\[1.1mm]
$^\mrc$Faculty of Civil and Environmental Engineering, Gda\'{n}sk University of Technology, ul.~Narutowicza 11/12, 80-233 Gda\'{n}sk, Poland \\[1.1mm]
$^\mrd$Department of Chemical and Biomolecular Engineering, University of California at Berkeley, \\
110 Gilman Hall, Berkeley, CA 94720-1460, USA \\[1.1mm]
$^\mre$Chemical Sciences Division, Lawrence Berkeley National Laboratory, CA 94720, USA}}

\end{center}

\vspace{-4mm}

\renewcommand{\thefootnote}{\arabic{footnote}}

\begin{center}

\small{Published\footnote{This pdf is the personal version of an article whose journal version is available at \href{http://dx.doi.org/10.1177/10812865211032723}{https:/\!/journals.sagepub.com}} 
in \textit{Mathematics and Mechanics of Solids}, \href{http://dx.doi.org/10.1177/10812865211032723}{DOI: 10.1177/10812865211032723} \\
Submitted on 26 February 2021; Revised on 25 June 2021; Accepted on 27 June 2021} 

\end{center}

\vspace{-3mm}

%\doublespacing

\rule{\linewidth}{.15mm}
{\bf Abstract}

This work presents a self-contained continuum formulation for coupled chemical, mechanical and thermal contact interactions.
The formulation is very general and hence admits arbitrary geometry, deformation and material behavior.
All model equations are derived rigorously from the balance laws of mass, momentum, energy and entropy in the framework of irreversible thermodynamics,
thus exposing all the coupling present in the field equations and constitutive relations.
In the process, the conjugated kinematic and kinetic variables for mechanical, thermal and chemical contact are identified, and the analogies between mechanical, thermal and chemical contact are highlighted.
Particular focus is placed on the thermodynamics of chemical bonding distinguishing between exothermic and endothermic contact reactions.
Distinction is also made between long-range, non-touching surface interactions and short-range, touching contact.
For all constitutive relations examples are proposed and discussed comprehensively with particular focus on their coupling. 
Finally, three analytical test cases are presented that illustrate the thermo-chemo-mechanical contact coupling and are useful for verifying computational models.
While the main novelty is the extension of existing contact formulations to chemical contact, the presented formulation also sheds new light on thermo-mechanical contact, since it is consistently derived from basic principles using only few assumptions.


\vspace{-5mm}
\rule{\linewidth}{.15mm}

\vspace{6mm}

{\Large\bf List of main field variables and their sets}\\[-7mm]

\begin{tabbing}
$\bone$ \quad~ \= identity tensor in $\bbR^3$  \\
$\ba_\alpha$ \> $:=\partial\bx/\partial\xi^\alpha$, $\alpha\in\{1,2\}$; covariant tangent vectors at surface point $\bx$; unit:~[1] or [m] \\
$\ba^\alpha$ \> contravariant tangent vectors at $\bx$; $\alpha\in\{1,2\}$; unit:~[1] or [m$^{-1}$] depending on $\xi^\alpha$  \\
$\ba^\mrp_\alpha$ \> and $\ba_\mrp^\alpha$; tangent vectors at $\bx_\mrp\in\sS_1$  \\
$\bexk$ \> prescribed, mass-specific body force at $\bx_k\in\sB_k$; unit:~[N/kg] \\ 
$\sB_k$ \> set of points contained in body $k$ \\ 
$\partial\sB_k$ \> set of points on the surface of body $k$ \\
$\partial_q\sB_k$ \> $\subset\partial\sB_k$; boundary of body $k$ where heat flux $\bar q_k$ is prescribed \\
$\partial_t\sB_k$ \> $\subset\partial\sB_k$; boundary of body $k$ where traction $\bar\bt_k$ is prescribed \\
$\bD_k$ \> rate of deformation tensor at $\bx_k\in\sB_k$; unit: [1/s] \\
$\bD^\mri_k$ \> inelastic part of $\bD_k$ \\
$\dif a_k$ \> area element on the current surface $\sS_k$; unit:~[m$^2$] \\
$\dif v_k$ \> volume element in the current body $\sB_k$; unit:~[m$^3$] \\
$\bE_k$ \> Green-Lagrange strain tensor at $\bx_k\in\sB_k$; unit-free \\
$\bE^\mre_k$ \> elastic part of $\bE_k$ \\
$\bE^\mri_k$ \> inelastic part of $\bE_k$ \\
$\bbE$ \> total energy of the two-body system; unit:~[J] \\
$e_\mrc$ \> bond-specific contact enthalpy on surface $\sS_\mrc$ (per bonding site); unit: [J] \\
$\phi$ \> non-dimensional bonding state at $\bx_\mrc\in\sS_\mrc$ \\
$\psi_k$ \> mass-specific Helmholtz free bulk energy at $\bx_k\in\sB_k$; unit [J/kg] \\
$\unde{\Psi}_{\!k}$ \> $=\rho_{0k}\,\psi_k$; Helmholtz free bulk energy density per undeformed volume of $\sB_k$; unit:~[J/m$^3$] \\
$\psi_{12}$ \> Helmholtz free interaction energy between $\bx_1\in\sS_1$ and $\bx_2\in\sS_2$; unit:~[J] \\
$\psi_\mrc$ \> bond-specific Helmholtz free contact energy on surface $\sS_\mrc$ (per bonding site); unit:~[J] \\
$\unde{\Psi_{\!c}}$ \> $= \Nc\,\psi_\mrc$; Helmholtz free contact energy density per undeformed area; unit:~[J/m$^2$] \\
$\bF_k$ \> deformation gradient at $\bx_k\in\sB_k$; unit-free \\
$\bg$ \> $:=\bx_2-\bx_1$; gap vector between surface points $\bx_1\in\sS_1$ and $\bx_2\in\sS_2$; unit:~[m] \\
$g_\mrn$ \> $:=\bg\cdot\bn$; normal gap \\
$\bg_\mrt$ \> $:=\bg-g_\mrn\bn = (\ba_\alpha\otimes\ba^\alpha)\bg$; tangential gap vector \\
$\Lg$ \> $=\dot\xi^\alpha_\mrp\,\ba^\mrp_\alpha$; Lie derivative of $\bg_\mrt$ at $\bx_1=\bx_\mrp$; unit:~[m/s] \\
$\bg_\mre$ \> reversible (elastic) part of $\bg$ associated with sticking contact \\
$\bg_\mri$ \> irreversible (inelastic) part of $\bg$ associated with sliding contact \\
$\eta_k^\mre$ \> mass-specific external bulk entropy production rate at $\bx_k\in\sB_k$; unit:~[J/(kg\,K\,s)] \\
$\eta_k^\mri$ \> mass-specific internal bulk entropy production rate at $\bx_k\in\sB_k$; unit:~[J/(kg\,K\,s)] \\
$\eta_\mrc^\mri$ \> area-specific internal contact entropy production rate at $\bx_\mrc\in\sS_\mrc$; unit:~[J/(K\,m$^2$\,s)] \\
$h$ \> heat transfer coefficient between body $\sB_1$ and body $\sB_2$; unit:~[J/(K\,m$^2$\,s)] \\
$h_k$ \> heat transfer coefficient between body $\sB_k$ and an interfacial medium; unit:~[J/(K\,m$^2$\,s)] \\
$J_k$ \> change of volume at $\bx_k\in\sB_k$; unit-free \\
$J_{\mrs k}$ \> change of area at $\bx_k\in\sS_k$; unit-free \\
$\Jc$ \> reference value for the area change; either $\Jc=J_{\mrs 1}$ or $\Jc=J_{\mrs 2}$ \\
$k$ \> $\in\{1,2\}$; body index \\
$\bbK$ \> kinetic energy of the two-body system; unit:~[J] \\
$\mu_\mrc$ \> chemical contact potential per bonding site at $\bx_\mrc\in\sS_\mrc$; unit:~[J] \\
$M_\mrc$ \> $= \nc\,\mu_\mrc$; chemical contact potential per current area at $\bx_\mrc\in\sS_\mrc$; unit:~[J/m$^2$] \\
$\unde{M_c}$ \> $= \Nc\,\mu_\mrc$; chemical contact potential per reference area at $\bx_\mrc\in\sS_\mrc$; unit:~[J/m$^2$] \\
$n_k$ \> current density of bonding sites at $\bx_k\in\sS_k$ and time $t > 0$; unit:~[1/m$^2$] \\
$N_k$ \> $=J_{\mrs k}\,n_k$; initial density of bonding sites at $\bx_k\in\sS_k$; unit:~[1/m$^2$] \\
$\nc$ \> reference value for the current bonding site density; either $\nc=n_1$ or $\nc=n_2$ \\
$\Nc$ \> $=J_\mrc\,n_\mrc$; reference value for the initial bonding site density; either $\Nc=N_1$ or $\Nc=N_2$ \\
$n^\mrb_k$ \> current density of bonded bonding sites at $\bx_k\in\sS_k$ and time $t > 0$; unit:~[1/m$^2$] \\
$n^{\mru\mrb}_k$ \> current density of unbonded bonding sites at $\bx_k\in\sS_k$ and time $t > 0$; unit:~[1/m$^2$] \\
$n_\mrb$  \> $:=n^\mrb_1$ in case $n^\mrb_2=n^\mrb_1$ \\
$\bn_k$ \> outward unit normal vector at $\bx_k\in\sS_k$ \\
$\bn_\mrp$ \> $:=\bn_1(\xi^\alpha_\mrp,t)$; outward unit normal vector at $\bx_\mrp\in\sS_1$ \\
$\sP_k$ \> subset of $\sB_k$ or $\sS_k$ \\
$p_\mrc$ \> contact pressure, i.e.~normal part of $\bt_\mrc$; unit:~[N/m$^2$] \\
$q_k$ \> heat influx per current area; unit:~[J/(m$^2$\,s)] \\
$\bq_k$ \> heat flux vector per current area; unit:~[J/(m$^2$\,s)] \\
$\bar q_k$ \> prescribed heat influx per current area at $\bx_k\in\sS_k$; unit:~[J/(m$^2$\,s)] \\
$q^\mrc_k$ \> contact heat influx per current area at $\bx_k\in\sS_\mrc$; unit:~[J/(m$^2$\,s)] \\
$q^\mrc_\mrm$ \> $:=(q^\mrc_1+q^\mrc_2)/2$; mean contact heat influx into $\sB_1$ and $\sB_2$ \\
$q^\mrc_\mrt$ \> $:=(q^\mrc_1-q^\mrc_2)/2$; transfer heat flux from body $\sB_2$ to body $\sB_1$\\
$\tilde q_k$ \> entropy influx per current area; unit:~J/(K\,m$^2$\,s)] \\
$\tilde\bq_k$ \> entropy flux vector per current area; unit:~[J/(K\,m$^2$\,s)] \\
$\bar{\tilde q}_k$ \> prescribed entropy influx per current area at $\bx_k\in\sS_k$; unit:~[J/(K\,m$^2$\,s)] \\
$\tilde q^\mrc_k$ \> contact entropy influx per current area at $\bx_\mrc\in\sS_\mrc$; unit:~[J/(K\,m$^2$\,s)] \\
$\rho_k$ \> current mass density at $\bx_k\in\sB_k$ and time $t > 0$; unit:~[kg/m$^3$] \\
$\rho_{0k}$ \> $=J_k\,\rho_k$; initial mass density at $\bx_k\in\sB_k$; unit:~[kg/m$^3$] \\
$\rexk$ \> prescribed, mass-specific heat source at $\bx_k\in\sB_k$; unit:~[J/(kg\,s)] \\
$R_\mrc$ \> bonding reaction rate per current area; unit:~[1/(m$^2$s)] \\
$\bsig_{\!k}$ \> Cauchy stress tensor at $\bx_k\in\sB_k$; unit:~[N/m$^2$] \\
$\bS_k$ \> second Piola-Kirchhoff stress tensor at $\bx_k\in\sS_k$; unit:~[N/m$^2$] \\
$s_k$ \> mass-specific bulk entropy at $\bx_k\in\sB_k$; unit:~[J/(kg\,K)]  \\
$\unde{S}_k$ \> $=\rho_{0k}\,s_k$; bulk entropy density per undeformed volume of $\sB_k$; unit:~[J/(K\,m$^3)$] \\
$s_\mrc$ \> bond-specific internal contact entropy on surface $\sS_\mrc$ (per bonding site); unit:~[J/K]  \\
$\unde{S_c}$ \> $= \Nc\,s_\mrc$; internal contact entropy density per undeformed area; unit:~[J/(K\,m$^2)$] \\
$\bbS$ \> total entropy of the two-body system; unit:~[J/K] \\
$\sS_k$ \> $\subset\partial\sB_k$; set of points defining the contact surface of body $k$; $\sS_1\,\hat{=}$ `master', $\sS_2\,\hat{=}$ `slave' \\
$\sS_\mrc$ \> set of points defining the shared contact surface; in case $\sS_1\approx\sS_2=:\sS_\mrc$ \\
$t$ \> time; unit:~[s] \\
$\bt_k$ \> surface traction per current area at $\bx_k\in\sS_k$; unit:~[N/m$^2$] \\
$\bar\bt_k$ \> prescribed surface traction per current area at $\bx_k\in\sS_k$; unit:~[N/m$^2$] \\
$\bar\undet_k$ \> prescribed surface traction per reference area at $\bx_k\in\sS_k$; unit:~[N/m$^2$] \\
$\bt_k^\mrc$ \> contact traction per current area at $\bx_\mrc\in\sS_\mrc$; unit:~[N/m$^2$] \\
$\undet_k^\unde{c}$ \> $=J_{\mrs k}\,\bt_k^\mrc$; contact traction per reference area at $\bx_\mrc\in\sS_\mrc$; unit:~[N/m$^2$] \\
$\bt_\mrc$  \> $:=\bt^\mrc_1$ in case $\bt^\mrc_2=-\bt^\mrc_1$; contact traction on the master surface $\sS_1$ \\
$\undetc$ \> $:=\Jc\,\bt_\mrc$; either $\undetc=\undet^\unde{c}_1$ if $\Jc=J_{\mrs 1}$ or $\undetc=-\undet^\unde{c}_2$ if $\Jc=J_{\mrs 2}$ \\
$\bt_\mrt$ \> tangential contact traction, i.e.~tangential part of $\bt_\mrc$ \\
$T_k$ \> temperature at $\bx_k\in\sB_k$; unit:~[K] \\
$T_\mrc$ \> temperature of the interfacial contact medium at $\bx\in\sS_\mrc$; unit:~[K] \\
$[\![T]\!]$ \> $:=T_2-T_1$; temperature jump (or temperature gap) across the contact interface \\
$u_k$ \> mass-specific internal bulk energy at $\bx_k\in\sB_k$; unit:~[J/kg]  \\
$\unde{U}_k$ \> $=\rho_{0k}\,u_k$; internal bulk energy density per undeformed volume of $\sB_k$; unit:~[J/m$^3$] \\
$u_{12}$ \> internal interaction energy between $\bx_1\in\sS_1$ and $\bx_2\in\sS_2$; unit:~[J] \\
$u_\mrc$ \> bond-specific internal energy of the contact interface $\sS_\mrc$ (per bonding site); unit:~[J]  \\
$\unde{U_c}$ \> $= \Nc\,u_\mrc$; internal contact energy density per undeformed area; unit:~[J/m$^2$] \\
$\bbU$ \> total internal energy of the two-body system; unit:~[J] \\
$\bbU_\mrc$ \> internal energy of the contact interface $\sS_\mrc$; unit:~[J] \\
$\bv_k$ \> material velocity at $\bx_k\in\sB_k$; unit:~[m/s] \\
$[\![\bv]\!]$ \> $:=\bv_2-\bv_1$; velocity jump across the contact interface \\
$\bv$ \> $:=\bv_1$ in case $\bv_2=\bv_1$ at the common contact point $\bx_2=\bx_1\in\sS_\mrc$ \\
$\xi^\alpha$ \> $\alpha\in\{1,2\}$; curvilinear coordinates determining points on a surface; unit:~[m] or [1] \\
$\xi^\alpha_\mrp$ \> $\alpha\in\{1,2\}$; local surface coordinates defining the closest projection point $\bx_\mrp\in\sS_1$ \\
$\xi^\alpha_\mre$ \> reversible (elastic) part of $\xi^\alpha_\mrp$ associated with sticking contact \\
$\xi^\alpha_\mri$ \> irreversible (inelastic) part of $\xi^\alpha_\mrp$ associated with sliding contact \\
$\bx_k$ \> current position of a material point at time $t>0$ in body $\sB_k$; unit:~[m] \\
$\bX_k$ \> initial position of a material point in body $\sB_k$; unit:~[m] \\
$\bx_\mrc$ \> current position of a material point at time $t>0$ on contact surface $\sS_\mrc$; unit:~[m] \\
$\bx_\mrp$ \> $=\bx_1(\xi^\alpha_\mrp,t)\in\sS_1$; closest projection point of the projection $\bx_2\in\sS_2\rightarrow\sS_1$
\end{tabbing}

\section{Introduction}\label{s:intro}

Many applications in science and technology involve coupled interactions at interfaces.
An example is the mechanically sensitive chemical bonding appearing in adhesive joining, implant osseointegration and cell adhesion. 
Another example is the thermal heating arising in frictional contact, adhesive joining and electrical contacts.  
Further examples are the temperature-dependent mechanical contact conditions in melt-based production technologies, such as welding, soldering, casting and additive manufacturing.
A general understanding and description of these examples requires a general theory that couples chemical, mechanical, thermal and electrical contact.
Such a theory is developed here for the first three fields in the general framework of nonlinear continuum mechanics and irreversible thermodynamics.
Two cases are considered in the theory: touching contact and non-touching interactions.
While the former is dominating at large length scales, the latter provides a link to atomistic contact models.

There is a large literature body on coupled contact models. 
General approaches deal, however, with two-field and not with three-field contact coupling, such as is considered here.\\
General thermo-mechanical contact models have appeared in the early nineties, starting with the work of \citet{zavarise92} that combined a frictionless contact model with heat conduction.
Following that, \citet{johansson93} presented the full coupling for linear thermo-elasticity.
This was extended by \citet{wriggers94} to large deformations using an operator split technique for the coupling - a staggering scheme based on successively solving mechanical and thermal subproblems.
\citet{oancea97} extended this to a monolithical coupling formulation and proposed general constitutive models for thermo-mechanical contact.
Following these initial works, many thermo-mechanical contact studies have appeared, studying the extension to wear \citep{stromberg96,stupkiewicz99,molinari01}, rough-surface contact \citep{willner99} multiscale contact \citep{temizer10,temizer14,temizer16} and adhesion \citep{dittmann18}.
Beyond that, there have also been many advancements in the computational description of thermo-mechancial contact \citep{saracibar98,laursen99,stromberg99,pantuso00,adam02,xing02,bergman04,rieger04,hueber09,dittmann14,khoei18,seitz19}. \\ 
General chemo-mechanical contact models go back even further -- to the work of \citet{derjaguin75} that describes molecular, e.g.~van der Waals, adhesion between an elastic sphere and a half-space.
\citet{argento97} then extended this approach to a general surface formulation for interacting continua.
The work was then generalized to a nonlinear continuum mechanical contact formulation by \citet{sauer-phd} and \citet{sauer07b,sauer07a}. 
Subsequently, the framework was applied to the study of cell adhesion \citep{zeng10}, generalized to various surface interaction models \citep{spbc}, and combined with sliding friction models \citep{adhfric}.
In a strict sense, the chemo-mechanical contact models mentioned so far are not coupled models.
Rather, the chemical surface interaction is described by distance-dependent potentials.
Hence, the resulting problem is a single-field problem that only depends on deformation.
A second field variable representing the chemical contact state is not used. \\
This is different to the debonding model of \citet{fremond88}.
There a state variable is introduced in order to describe irreversible damage during debonding.
It is essentially a phenomenological debonding model, where the bond degrades over time following a first order ordinary differential equation (ODE). 
\citet{raous99} extended the model to sliding contact and implemented it within a finite element formulation.
Its finite element implementation is also discussed in \citet{wriggers-contact} in the framework of large deformations.
Subsequently the model has been extended to thermal effects \citep{bonetti09}, applied to multiscale contact \citep{wriggers08} and generalized to various constitutive models \citep{delpiero10}, among others.
Even though the Fr\'emond model is a coupled two-field model, it only describes debonding and not bonding.
It therefore does not provide a general link to chemical contact reactions. \\ 
The adhesion and debonding models mentioned above are similar to \textit{cohesive zone models} (CZMs).
Those propose phenomenological traction-separation laws for debonding, that are often derived from a potential, like the seminal model of \citet{xu93}, thus ensuring thermodynamic consistency.  
CZMs have been extended to thermo-mechanical
debonding through the works of \citet{hattiangadi04,willam04,fagerstrom08,ozdemir10,fleischhauer13} and \citet{esmaeili16}, among others.
Recently, CZMs have also been coupled to a hydrogen diffusion model in order to study fatigue \citep{busto17}.
CZMs usually have a damage/degradation part that sometimes follows from an evolution law, e.g.~see \citet{willam04}.
This makes them very similar to the \citet{fremond88} model. 
Like the Fr\'emond model, CZMs have not yet been combined with chemical contact reactions, and so this aspect is still absent in general contact models. \\ 
Adhesion models based on chemical bonding and debonding reactions have been developed by Bell and coworkers in the late seventies and early eighties in the context of cell adhesion \citep{bell78,bell84}.
Many subsequent works have appeared based on these models, for example to study substrate adhesion \citep{hammer87}, strip peeling \citep{dembo88}, nanoparticle endocytosis \citep{decuzzi07}, sliding contact \citep{deshpande08}, cell nanoindentation \citep{zhang08}, cell migration \citep{sarvestani09}, cell spreading \citep{sun09}, focal adhesion dynamics \citep{olberding10} and substrate compliance \citep{huang11}, among others.
Similar chemical bonding models have also been used to describe sticking and sliding friction, see \citet{srinivasan09}.
The Bell model is an ODE for the chemical reaction.
Assuming separation of chemical and mechanical time scales, this can be simplified into an algebraic equation that describes chemical equilibrium \citep{evans85}.
Even though Bell-like models have been combined with contact-induced deformations in several of the works mentioned above,
the contact formulations that have been considered in those works are not general continuum mechanical contact formulations as will be considered here.
\\
Another related topic is the field of \textit{tribochemistry} that is concerned with the growth of so-called \textit{tribofilms} during sliding contact. 
Chemical evolution models are used to describe the tribofilms in the framework of elementary contact models,
e.g. see \citet{andersson12} and \citet{ghanbarzadeh16}. 
There is also a review article on how mechanical stresses can affect chemical reactions at the molecular scale \citep{kochhar15}.
But none of these works use general continuum mechanical contact formulations. 
\\
Chemical contact reactions can be described by a state $\phi(\bx,t)$ that follows from an evolution law of the kind $\dot\phi = f(\phi)$.
Mathematically, they are thus similar to the description of contact ageing \citep{dieterich78,rice83,ruina83}, contact wear \citep{stromberg96,stromberg99,stupkiewicz99} and contact debonding \citep{fremond88,raous99}.
The thermodynamics is, however, very different.

None of the present chemo-mechanical models is a general contact model accounting for the general contact kinematics, balance laws and constitutive relations.
This motivates the development of such a formulation here.
To the best of our knowledge, it is the first general thermo-chemo-mechanical contact model that accounts for large-deformation contact and sliding, chemical bonding and debonding reactions, thermal contact and the full coupling of these fields. 
It is derived consistently from the general contact kinematics, balance laws and thermodynamics, introducing constitutive examples only at the end.
Its generality serves as a basis for computational formulations and later extensions such as membrane contact or thermo-electro-chemo-mechanical contact.

The novelties of the proposed formulation can be summarized as follows. The formulation \\[-8mm]
\begin{itemize}
\item derives a self-contained, fully coupled chemical, mechanical and thermal contact model, \\[-7mm]
\item accounts for general non-linear deformations and material behavior, \\[-7mm]
\item consistently captures all the coupling present in the interfacial balance laws, \\[-7mm]
\item highlights the similarities between chemical, mechanical and thermal contact \\[-7mm]
\item obtains the general material-independent constitutive contact relations, \\[-7mm]
\item provides several interface material models derived from the 2.~law of thermodynamics, \\[-7mm]
\item is illustrated by three elementary contact solutions. \\[-7mm]
\end{itemize}
Chemical reactions here are restricted to reactions across the contact interface. 
In this, two assumptions are made: (1) that the reaction rate is much higher than any sliding rate, and (2) that there is no tangential diffusion of bonds on the contact surface.
Reactions inside the bodies and on their free surfaces, that can also occur without contact, are not considered in this study.

The remainder of this paper is organized as follows:
Sec.~\ref{s:kine} presents the generalized continuum kinematics characterizing the mechanical, chemical and thermal contact behavior.
The kinematics are required in order to formulate the general balance laws that govern the coupled contact system in Sec.~\ref{s:bal}. 
Based on these laws the general constitutive relations are derived in Sec.~\ref{s:consti}.
Sec.~\ref{s:ex} then gives several coupled constitutive examples satisfying these relations. 
In order to illustrate these, Sec.~\ref{s:test} provides three analytical test cases for coupled contact.
The paper concludes with Sec.~\ref{s:concl}.

\section{Continuum contact kinematics}\label{s:kine}

This section introduces the kinematic variables characterizing the mechanical, thermal and chemical interaction between two deforming bodies $\sB_k$, $k =1,2$, following the established developments of continuum mechanics \citep{holzapfel} and contact mechanics \citep{laursen,wriggers-contact}.
The novelty here consists in their contrasting juxtaposition. %\\

\begin{figure}[ht]
\begin{center} \unitlength1cm
\begin{picture}(0,6.3)
\put(-7.5,-.3){\includegraphics[width=40mm]{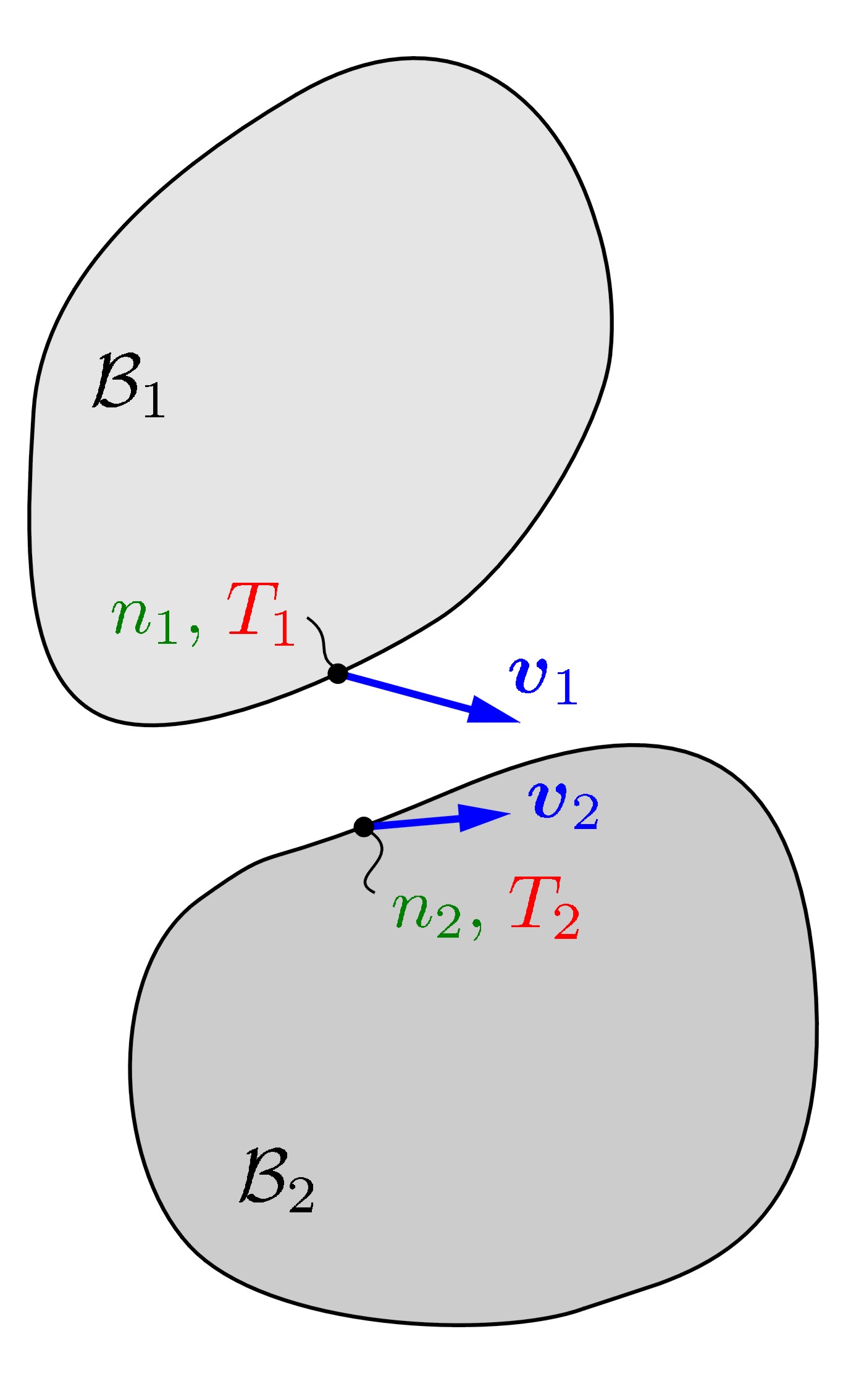}}
\put(-2,-.3){\includegraphics[width=40mm]{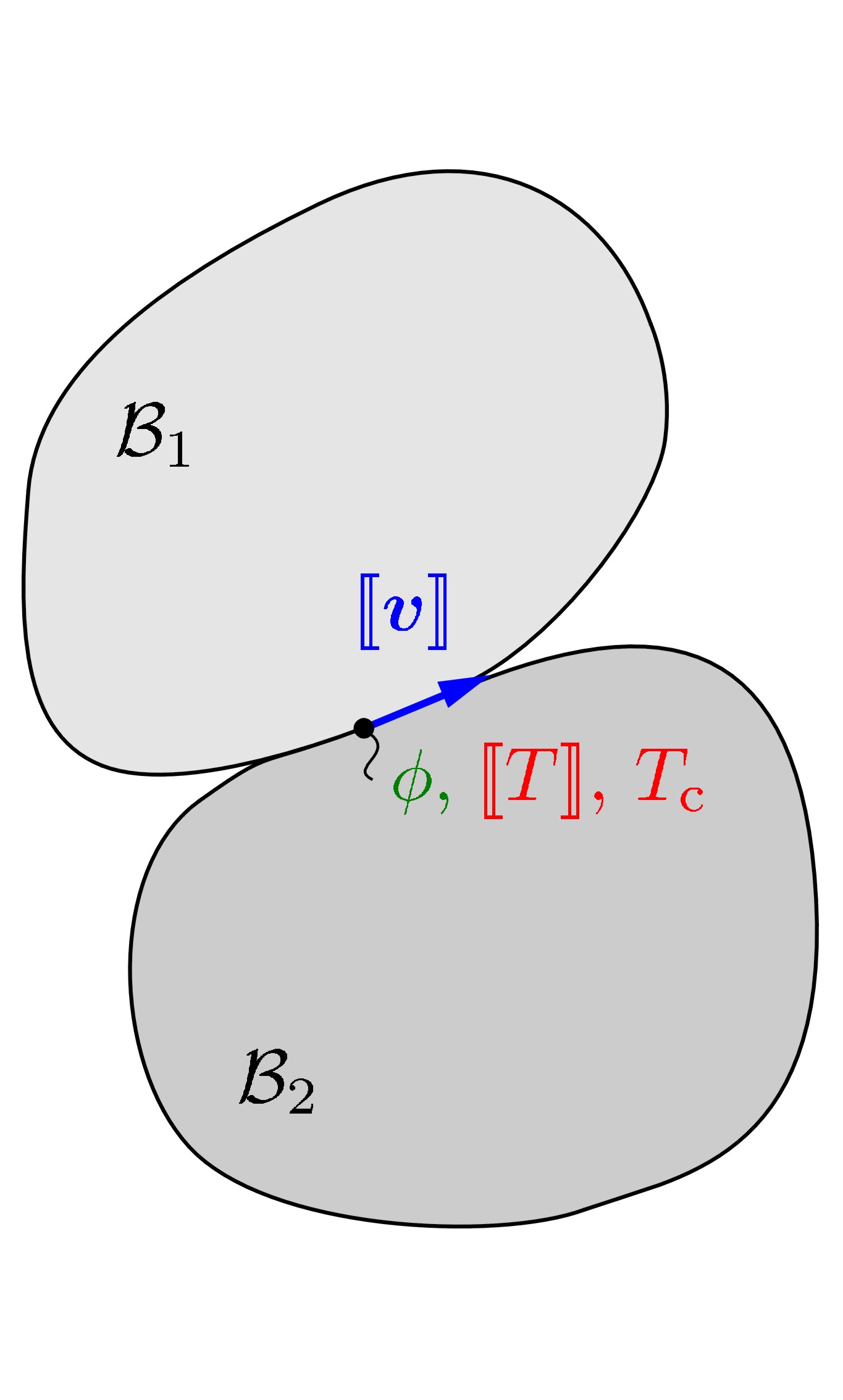}}
\put(3.5,-.3){\includegraphics[width=40mm]{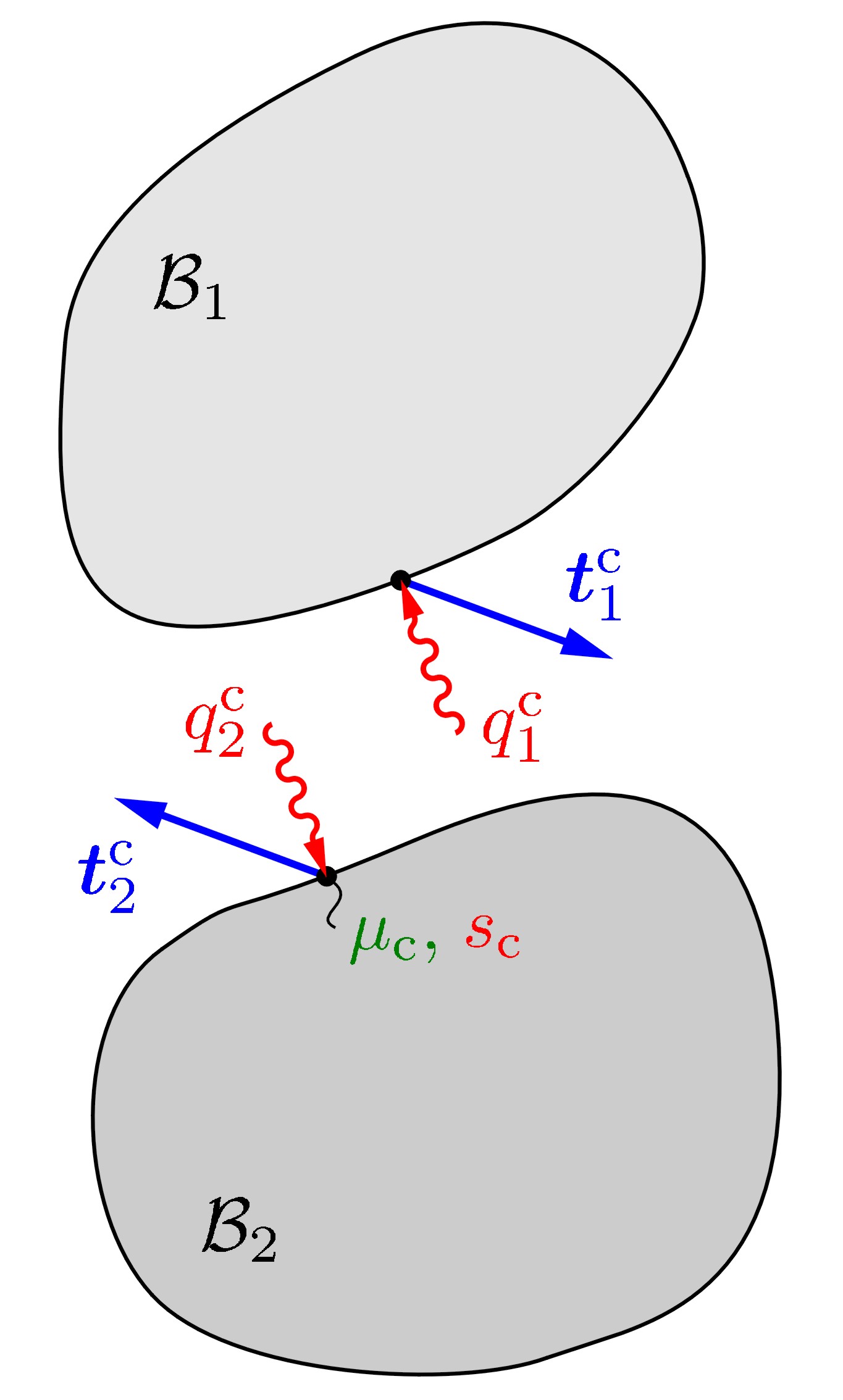}}
\put(-7.3,0){a.}
\put(-1.8,0){b.}
\put(3.6,0){c.}
\end{picture}
\caption{Continuum contact description: a.~bodies before contact; b.~bodies in contact; c.~free-body diagram for contact. Here, $n_k$, $\bv_k$ and $T_k$ denote the bonding site density, velocity and temperature, respectively, on the contact surface of body $\sB_k$ ($k=1,2$) at time $t$.
During contact, $\phi$ denotes the degree of bonding, $[\![\bv]\!]$ the velocity jump, $[\![T]\!]$ the temperature jump, $T_\mrc$ the contact temperature, $\bt^\mrc_k$ the contact tractions, $q^\mrc_k$ the heat influx, $\mu_\mrc$ the chemical contact potential and $s_\mrc$ the contact entropy. $T_\mrc$ and $s_\mrc$ are associated with an interfacial medium.} 
\label{f:cont}
\end{center}
\end{figure}
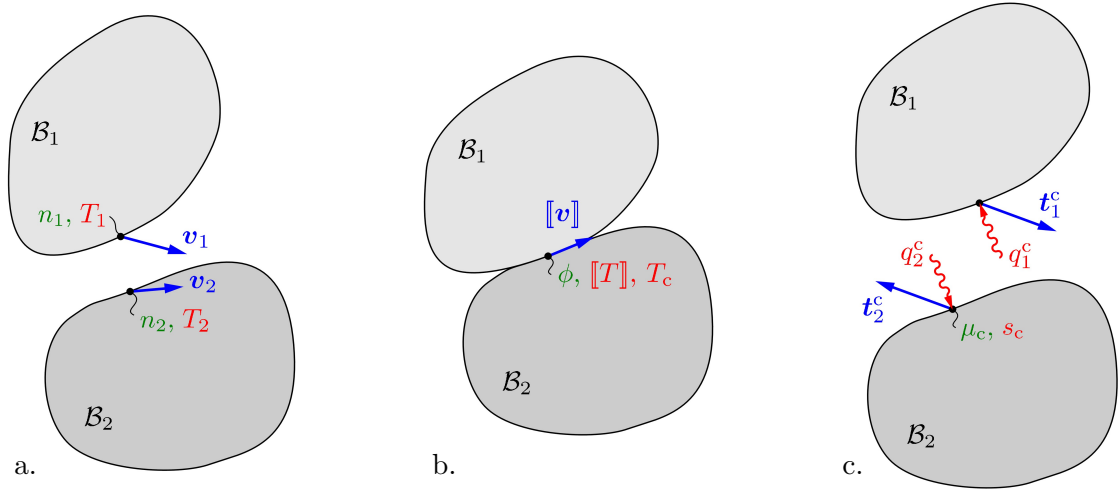

\begin{figure}[ht]
\begin{center} \unitlength1cm
\begin{picture}(0,5)
\put(-3.8,-.2){\includegraphics[height=52mm]{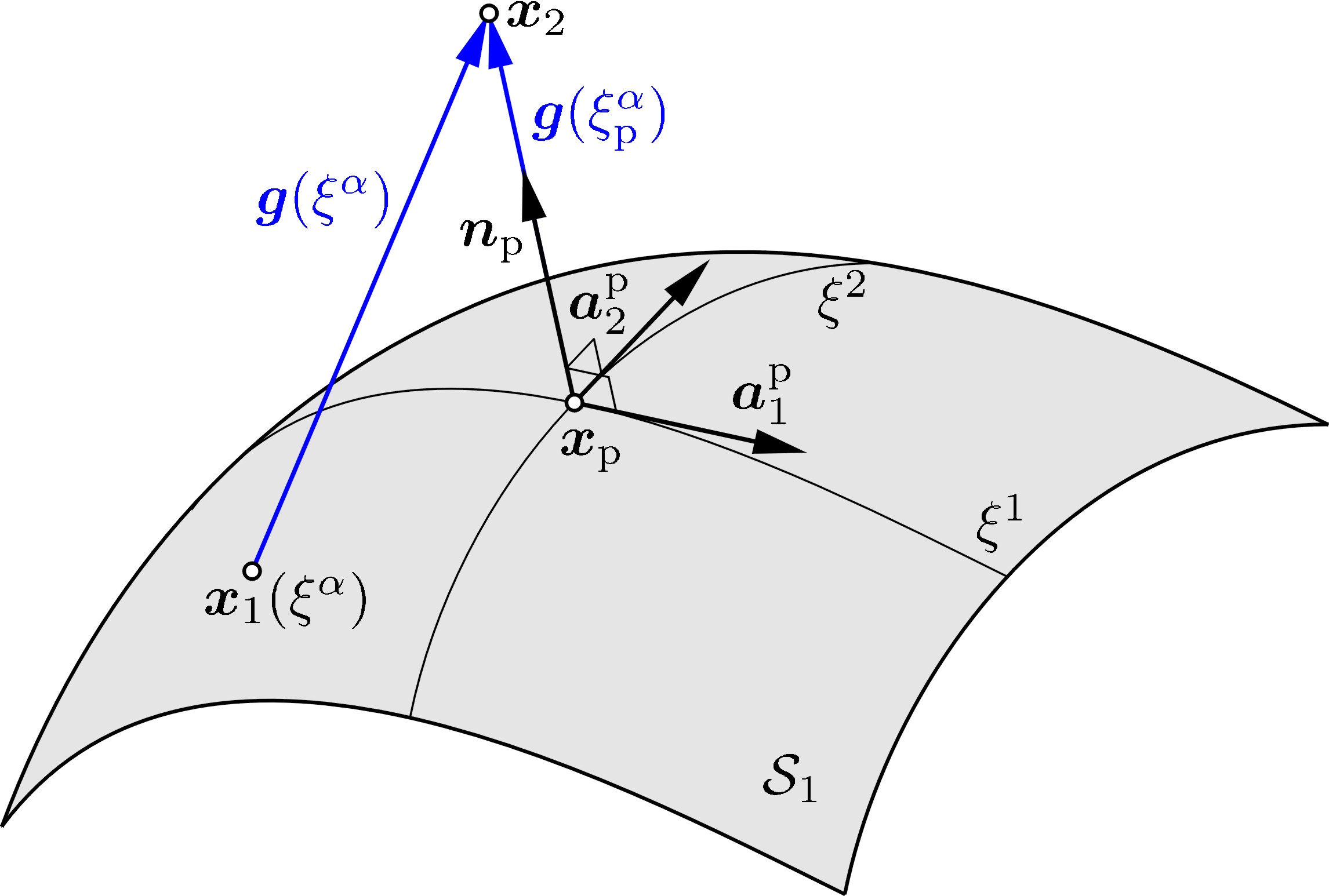}}
\end{picture}
\caption{Mechanical contact: Computation of the contact gap $\bg$ via a closest point projection of \textit{slave} point $\bx_2$ onto the \textit{master} surface $\sS_1$.}
\label{f:cpp}
\end{center}
\end{figure}

\begin{figure}[ht]
\begin{center} \unitlength1cm
\begin{picture}(0,4.5)
\put(-3.6,-.1){\includegraphics[height=45mm]{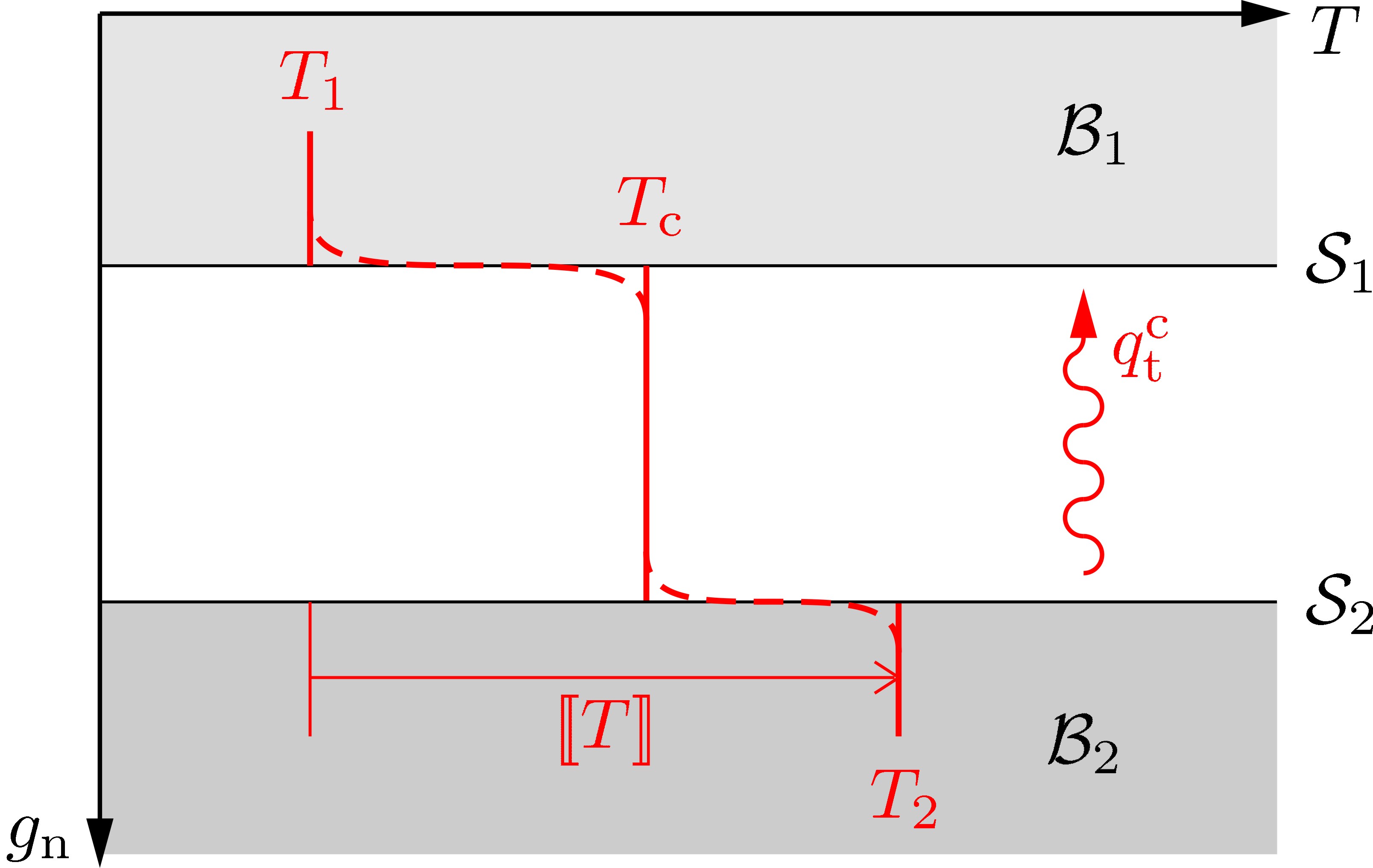}}
\end{picture}
\caption{Thermal contact:
Assumed temperature profile $T(g_\mrn)$ across the contact interface (along coordinate $g_\mrn$).
Thermal contact between two bodies is characterized by the temperature jump $[\![T]\!]:=T_2 - T_1$ and the contact temperature $T_\mrc$ associated with an interfacial medium.
$[\![T]\!]$ leads to the transfer heat flux $q^\mrc_\mrt = h [\![T]\!]$ identified in Sec.~\ref{s:consti_c}.} 
\label{f:Tcont}
\end{center}
\end{figure}

The primary field variables are the current mass density $\rho_k = \rho_k(\bX_{\!k},t)$, current surface bonding site density $n_k=n_k(\bX_{\!k},t)$, current position $\bx_k = \bx_k(\bX_{\!k},t)$, current velocity $\dot\bx_k =: \bv_k = \bv_k(\bX_{\!k},t)$ and current temperature $T_k = T_k(\bX_{\!k},t)$ that are all functions of space and time.
Here, the spatial dependency is expressed through the initial position $\bX_k = \bx_k\big|_{t\,=\,0}$, and the dot denotes the material time derivative
\eqb{l}
\dot{...} = \ds\pa{...}{t}\Big|_{\bX_k\,=\,\mathrm{fixed}}\,.
\label{e:mtd}\eqe
The two bodies can come into contact and interact mechanically, thermally and chemically on their common contact surface $\sS_\mrc := \partial\sB_1\cap\partial\sB_2$, see Fig.~\ref{f:cont}. 
In case of long-range interaction, like van der Waals adhesion, they may also interact when not in direct contact. 
In this case, interaction is considered to take place on the (non-touching) surfaces regions $\sS_k\subset\partial\sB_k$.\footnote{In principle, long-range interaction can also take place in the volume. The present theory can be extended straightforwardly to account for such interactions. Alternatively, one may also project them onto the surface following \citet{sauer09b}.}
%-----------------------------------------------------------------
%-----------------------------------------------------------------
\\
The deformation within each body is characterized by the deformation gradient
\eqb{l}
\bF_k := \ds\pa{\bx_k}{\bX_k}\,,
\eqe
from which various strain measures can be derived, such as the Green-Lagrange strain tensor
\eqb{l}
\bE_k := \big(\bF_k^\mrT\bF_k-\bone \big)/2\,.
\eqe
The deformation generally contains elastic and inelastic contributions that lead to the additive strain decomposition\footnote{Even though the additive decomposition of the strain tensor is motivated from small deformations, it also applies to large deformations, where a multiplicative decomposition of the deformation gradient in the form $\bF = \bF_\mre\bF_\mri$ is usually considered. In this case $\bE^\mre = \big(\bF^\mrT\bF - \bF_\mri^\mrT\bF_\mri\big)/2$ and $\bE^\mri = \big(\bF_\mri^\mrT\bF_\mri-\bone\big)/2$.}
\eqb{l}
\bE_k = \bE_k^\mre + \bE_k^\mri\,.
\label{e:Esplit}\eqe
An example for an inelastic strain is thermal expansion.
From $\bF_k$ also follows the quantity
\eqb{l}
J_k := \det\bF_k\,,
\eqe
which governs the local volume change 
\eqb{l}
\dif v_k = J_k\,\dif V_k
\label{e:dv}\eqe
at $\bx_k\in\sB_k$ between undeformed and deformed configuration. 
Similarly, the local surface area change
\eqb{l}
\dif a_k = J_{\mrs k}\,\dif A_k
\label{e:da}\eqe
at $\bx_k\in\sS_k$ is governed by the quantity
\eqb{l}
J_{\mrs k} := \det_\mrs \bF_k\,,
\eqe
where $\det_\mrs(...)$ is the surface determinant on $\sS_k$, e.g.~see \citet{FeFi}. 
\\
Time dependent deformation is characterized by the symmetric velocity gradient
\eqb{l}\
\bD_k :=  \big(\nabla\bv_k + \nabla\bv_k^\mrT \big)/2\,,
\label{e:Dk}\eqe
also known as the rate of deformation tensor.
The velocity also gives rise to the identities
\eqb{l}
\ds\frac{\dot J_k}{J_k} = \divz\bv_k 
\label{e:dJk}\eqe
and
\eqb{l}
\ds\frac{\dot J_{\mrs k}}{J_{\mrs k}} = \divz_{\!\mrs}\,\bv_k\,, 
\label{e:dJs}\eqe
where $\divz_{\!\mrs}(...)$ is the surface divergence on $\sS_k$.
\\
Mechanical interaction between the two bodies is characterized by the gap vector
\eqb{l}
\bg := \bx_2 - \bx_1\,,
\label{e:g}\eqe
that is defined for all pairs $\bx_1\in\sS_1$ and $\bx_2\in\sS_2$.
Two cases have to be distinguished:
i.~Both $\bx_1$ and $\bx_2$ are material points -- a setting that is suitable for describing long-range interactions; or more generally: 
ii.~One of the points is not necessarily a material point -- a setting that arises when using a closest point projection suitable for short-range interactions.  
In the latter case, one surface, say $\sS_1$, is designated the so-called \textit{master surface}, while the other, $\sS_2$, is designated the so-called \textit{slave surface} \citep{hallquist85}.
The master surface is then used to define normal and tangential contact contributions.  
Therefore the master point $\bx_1$ in \eqref{e:g} is determined as the closest point projection of slave point $\bx_2$ onto master surface $\sS_1$.
Parameterizing $\sS_1$ by the curvilinear surface coordinates $\xi^\alpha$, $\alpha\in\{1,\,2\}$, the closest projection point is given by
\eqb{l}
\bx_\mrp := \bx_1(\xi^\alpha_\mrp,t)\,,
\label{e:xp}\eqe
where $\xi^\alpha_\mrp$ is the value of $\xi^\alpha$ that solves the minimum distance problem
\eqb{l}
\bg\cdot\ba_\alpha = 0\,.
\eqe
Here $\ba_\alpha := \partial\bx_1/\partial\xi^\alpha$ ($\alpha=1,2$) are tangent vectors of $\sS_1$, which at $\bx_\mrp$ are denoted $\ba_\alpha^\mrp$, i.e.~$\ba_\alpha^\mrp := \ba_\alpha(\xi^\alpha_\mrp,t)$.
Likewise $\bn_\mrp = \bn_1(\xi^\alpha_\mrp,t)$ is the surface normal at $\bx_\mrp$.
The closest point projection is illustrated in Fig.~\ref{f:cpp}.
%-----------------------------------------------------------------
%-----------------------------------------------------------------
The gap can be decomposed into its normal and tangential parts
\eqb{l}
g_\mrn := \bg\cdot\bn\,,\quad
\bg_\mrt := \bg - g_\mrn\,\bn\,,
\label{e:gnt}\eqe
where the latter is zero at $\bx_\mrp$.
Hence, for contact $\bg = g_\mrn\,\bn_\mrp$ and $g_\mrn=0$, so that
\eqb{l}
\dot\bg = \dot g_\mrn\,\bn_\mrp\,. 
\label{e:gd}\eqe
Based on \eqref{e:g} and \eqref{e:xp}, the material time derivative of $\bg$ can also be written as
\eqb{l}
\dot\bg = \dot\bx_2 - \dot\bx_1 - \dot\xi^\alpha_\mrp\,\ba_\alpha^\mrp\,,
\label{e:dbg}\eqe
where, in agreement with definition \eqref{e:mtd}, $\dot\bx_1 = \partial\bx_1/\partial t|_{\xi^\alpha_\mrp\,=\,\mathrm{fixed}}$ and $\dot\xi^\alpha_\mrp = \partial\xi^\alpha_\mrp/\partial t|_{\bX_2\,=\,\mathrm{fixed}}$.
The latter part 
\eqb{l}
\Lg :=  \dot\xi^\alpha_\mrp\,\ba_\alpha^\mrp \,,
\label{e:Lg}\eqe
(with summation implied over $\alpha$ and often denoted $\sL_v\bg_\mrt$) is equal to the \textit{Lie derivative} of the tangential gap vector  
\citep{wriggers-contact}.
Likewise \eqref{e:gd} is equal to the Lie derivative of the normal gap vector.
$\Lg$ is zero when $\xi^\alpha_\mrp$ is fixed (s.t.~$\dot\xi^\alpha_\mrp=0$), i.e.~when $\bx_1$ is a material point as in case i above.
In case ii, $\Lg$ is equal to the relative tangential velocity between the two surfaces. 
Thus, $\dot\xi^\alpha_\mrp=0$ also denotes the case of tangential sticking, while $\dot\xi^\alpha_\mrp\neq0$ characterizes tangential sliding. 
The sliding motion is irreversible and accumulates over time.
Combining \eqref{e:gd} and \eqref{e:dbg}, one can identify the velocity jump
\eqb{l}
[\![\bv]\!] := \bv_2 - \bv_1 = \dot g_\mrn\,\bn_\mrp + \Lg = \mathring\bg\,,
\label{e:vsplitdef}\eqe
and see that it is the Lie derivative of the total gap vector in case ii.
Even during sticking, it can be advantageous (see remark~\ref{r:22}) to allow for (small) motion that is reversible upon unloading.
This leads to the tangential velocity decomposition 
\eqb{l}
\dot\xi^\alpha_\mrp = \dot\xi^\alpha_\mre + \dot\xi^\alpha_\mri\,, 
\label{e:xisplit}\eqe
where $\dot\xi^\alpha_\mre$ captures the reversible (elastic) motion during sticking, while $ \dot\xi^\alpha_\mri$ captures the irreversible (inelastic) motion during sliding.
Plugging \eqref{e:xisplit} into \eqref{e:vsplitdef} yields
\eqb{l}
[\![\bv]\!] =  \Lge+ \Lgi \,,
\label{e:vsplit}\eqe 
where $\Lge := \dot g_\mrn\,\bn_\mrp + \dot\xi^\alpha_\mre\,\ba^\mrp_\alpha$ characterizes normal contact and tangential sticking, while $\Lgi := \dot\xi^\alpha_\mri\,\ba^\mrp_\alpha$ characterizes tangential sliding.
Decomposition \eqref{e:vsplit} is analogous to decomposition \eqref{e:Esplit}, making the contact kinematics analogous to the kinematics of the bodies.
It contains Eq.~\eqref{e:vsplitdef} for the case $\dot\xi^\alpha_\mre = 0$.
Expressions for $\dot g_\mrn$ and $\dot\xi^\alpha_\mrp$ can be found e.g.~in \citet{wriggers-contact}.
\\
Thermal contact is characterized by the temperature jump between surface points $\bx_1$ and $\bx_2$,
\eqb{l}
[\![T]\!] := T_2 - T_1\,,
\eqe
also denoted as the thermal gap \citep{saracibar98}, and the contact temperature $T_\mrc$ that is associated with an interfacial medium, e.g.~a lubricant or wear particles, see Fig.~\ref{f:Tcont}.
It is assumed that this medium is very thin, such that $T_\mrc$ is constant through the thickness of the medium and can only vary along the surface. 
%-----------------------------------------------------------------
%-----------------------------------------------------------------
\\
In order to characterize chemical contact, the bonding state variable $\phi=\phi(\bx,t)$ is introduced on the contact surface, as discussed in the following section.
It varies between $\phi=0$ for no bonding and $\phi=1$ for full bonding, and can be associated with a chemical gap (e.g. defined as $1-\phi$).
\\
Tab.~\ref{t:conj} summarizes the kinematic contact variables and their corresponding kinetic counterparts that will be discussed in later sections. 
%------------------------------------------------------------------
\begin{table}[h]
\centering
\begin{tabular}{|l|l|l|}
  \hline
  field & kinematic variable & kinetic variable \\[0mm] \hline 
  & & \\[-3.5mm]
  chemical & bonding state $\phi$ & chemical contact potential $\mu_\mrc$ \\[.5mm]
  mechanical & velocity jump $[\![\bv]\!]$ & contact traction $\bt_\mrc$ \\[.5mm]
  thermal (gap) & temperature jump $[\![T]\!]$ & contact heat influx $q_\mrc$ \\[.5mm]
  thermal (medium) & contact temperature $T_\mrc$ & contact entropy $s_\mrc$ \\[.5mm]
   \hline
\end{tabular}
\caption{Energy-conjugated contact pairs.}
\label{t:conj}
\end{table}
%------------------------------------------------------------------

\begin{remark}\label{r:21}
Apart from the tangential sticking constraint $\dot\xi^\alpha_\mrp=0$, there is also the normal contact constraint $g_\mrn\geq0$.
Normal and tangential contact are usually treated separately in
most friction algorithms, but there are also unified approaches directly based on the gap vector $\bg$, e.g.~see \citet{wriggers03} and \citet{duong18}.
Such an approach is also taken in the following sections.
\end{remark}

\begin{remark}\label{r:22}
There are reasons for relaxing the contact constraints $g_\mrn\geq0$ and $\dot\xi^\alpha_\mrp=0$.
One is to use a penalty regularization, which is typically simpler to implement than the exact constraint enforcement.  
Another is to use the elastic gap $\bg_\mre$ to capture the elastic deformation of (microscale) surface asperities during contact.
In both cases $\bg_\mre$ is nonzero (but typically small) during contact.
\end{remark}

\begin{remark}\label{r:24}
The designation into \textit{master} and \textit{slave} surfaces introduces a bias in the contact formulation that can affect the accuracy and robustness of computational methods.
The bias can be removed if alternating \textit{master/slave} designations are used, as is done for 3D friction in \citet{jones06} and \citet{spbf}.
\end{remark}

\begin{remark}\label{r:25}
As seen above, the Lie derivatives $\mathring\bg$ and $\mathring\bg_e$ only contain the relative changes of the normal and tangential gap.
They do not contain the basis changes $\dot\bn_\mrp$ and $\dot\ba^\mrp_\alpha$.
This makes the Lie derivative objective and a suitable quantity for the constitutive modeling \citep{laursen}, see Sec.~\ref{s:poti}.
\end{remark}

\section{Balance laws}\label{s:bal}

This section derives the chemical, mechanical and thermal balance laws for a generally coupled two-body system.
The resulting equations turn out to be in the same form as the known relations for chemical reactions \citep{prigogine} and thermo-mechanical contact \citep{laursen}.
The novelty here lies in establishing their coupling and highlighting the similarities between the chemical, mechanical and thermal contact equations.
The derivation is based on the following three mathematical ingredients.
The first is Reynold's transport theorem for volume integrals,
\eqb{l}
\ds\pad{}{t}\int_{\sB_k}...\,\dif v_k = \int_{\sB_k}\Big(\dot{(...)} + \divz\bv_k\,(...) \Big)\,\dif v_k\,.
\label{e:vRey}\eqe
It follows from substituting \eqref{e:dv} on the left, using the product rule, and then applying \eqref{e:dJk}.
Applied to surfaces, Reynold's transport theorem simply adapts to
\eqb{l}
\ds\pad{}{t}\int_{\sS_k}...\,\dif a_k = \int_{\sS_k}\Big(\dot{(...)} +  \divz_{\!\mrs}\,\bv_k\,(...) \Big)\,\dif a_k\,.
\label{e:sRey}\eqe
It follows from substituting \eqref{e:da} on the left, using the product rule, and then applying \eqref{e:dJs}.
\\
The second ingredient is the divergence theorem,
\eqb{l}
\ds\int_{\partial\sB_k}...\,\bn_k\,\dif a_k = \int_{\sB_k} \divz(...)\,\dif v_k\,,
\label{e:divtheo}\eqe
where $\bn_k$ is the outward normal vector of boundary $\partial\sB_k$.
The third ingredient is the localization theorem,
\eqb{l}
\ds\int_\sP ...\,\dif v = 0 \quad \forall\,\sP\subset\sB \quad\Leftrightarrow\quad ... = 0 \quad\forall\,\bx\in\sB\,,
\label{e:Loc}\eqe
that can be equally applied to surface integrals.

Some of the field quantities appearing in the following balance laws can be expressed per mass, bonding site, current volume, current area, reference volume or reference area. 
The notation distinguishing these options is summarized in Tab.~\ref{t:not}.
%------------------------------------------------------------------
\begin{table}[h]
\centering
\begin{tabular}{|l|l|l|l|l|}
  \hline
  quantity & per mass/bond & per curr.~vol./area & per ref.~vol./area & in total \\[0mm] \hline 
  & & & & \\[-3.5mm]
  internal energy & $u_k$, $u_\mrc$ & $U_k$, $U_\mrc$ & $\unde{U}_k$, $\unde{U_c}$ & $\bbU$ \\[.5mm]
  Helmholtz free energy & $\psi_k$, $\psi_\mrc$ & $\Psi_k$, $\Psi_\mrc$ & $\unde{\Psi}_k$, $\unde{\Psi_c}$ & -- \\[.5mm]
  entropy & $s_k$, $s_\mrc$ & $S_k$, $S_\mrc$ & $\unde{S}_k$, $\unde{S_c}$ & $\bbS$ \\[.5mm]
  reaction rate & $r_\mrc$ & $R_\mrc$ & $\unde{R_c}$ & -- \\[.5mm]
  chemical potential & $\mu_\mrc$ & $M_\mrc$ & $\unde{M_c}$ & -- \\[.5mm]
  body force / traction & $\bexk$ & $\bar\bt_k$, $\bt_\mrc$ & $\bar\undet_k$, $\undetc$ & -- \\[.5mm]
  heat source / flux & $\rexk$ & $\bar\bq_k$, $q_\mrc$ & $\unde{\bar{q}}_k$, $\unde{q_c}$ & -- \\[.5mm]
   \hline
\end{tabular}
\caption{Notation for various quantities: $\bullet_k$ are bulk quantities expressed per mass or volume of body $k$, while $\bullet_\mrc$ are contact surface quantities expressed per bond(ing site) or area. 
All $a_k$ ($a=u,\,\psi,\,s$) satisfy $A_k = \rho_k\,a_k$ and $\unde{A}_k = J_k\,A_k = \rho_{k0}\,a_k$, while 
all $a_\mrc$ ($a=u,\,\psi,\,s,\,r,\,\mu$) satisfy $A_\mrc = \nc\,a_\mrc$ and $\unde{A_c} = \Jc\,A_\mrc = \Nc\,a_\mrc$.
Not all combinations are required here: the reaction rate, chemical potential, contact tractions and contact heat flux are only defined on the contact surface, while the body force and heat source are only defined in the bulk.}
\label{t:not}
\end{table}
%------------------------------------------------------------------

\subsection{Conservation of mass and bonding sites}\label{s:mass}

Assuming no mass sources, the mass balance of each body is given by the statement
\eqb{l}
\ds\pad{}{t}\int_{\sP_k}\rho_k\,\dif v_k = 0 \quad\forall\,\sP_k\subset\sB_k\,.
\eqe
Applying \eqref{e:vRey} and \eqref{e:Loc}, this leads to the local balance law
\eqb{l}
\dot\rho_k + \rho_k\,\divz\bv_k = 0 \quad\forall\,\bx_k\in\sB_k\,.
\label{e:mk}\eqe
Due to \eqref{e:dJk}, this ODE is solved by $\rho_k = \rho_{0k}/J_k$, where $\rho_{0k}:=\rho_{k}\big|_{t=0}$ is the initial mass density.

\begin{figure}[h]
\begin{center} \unitlength1cm
\begin{picture}(0,4.6)
\put(-6.9,0){\includegraphics[height=45mm]{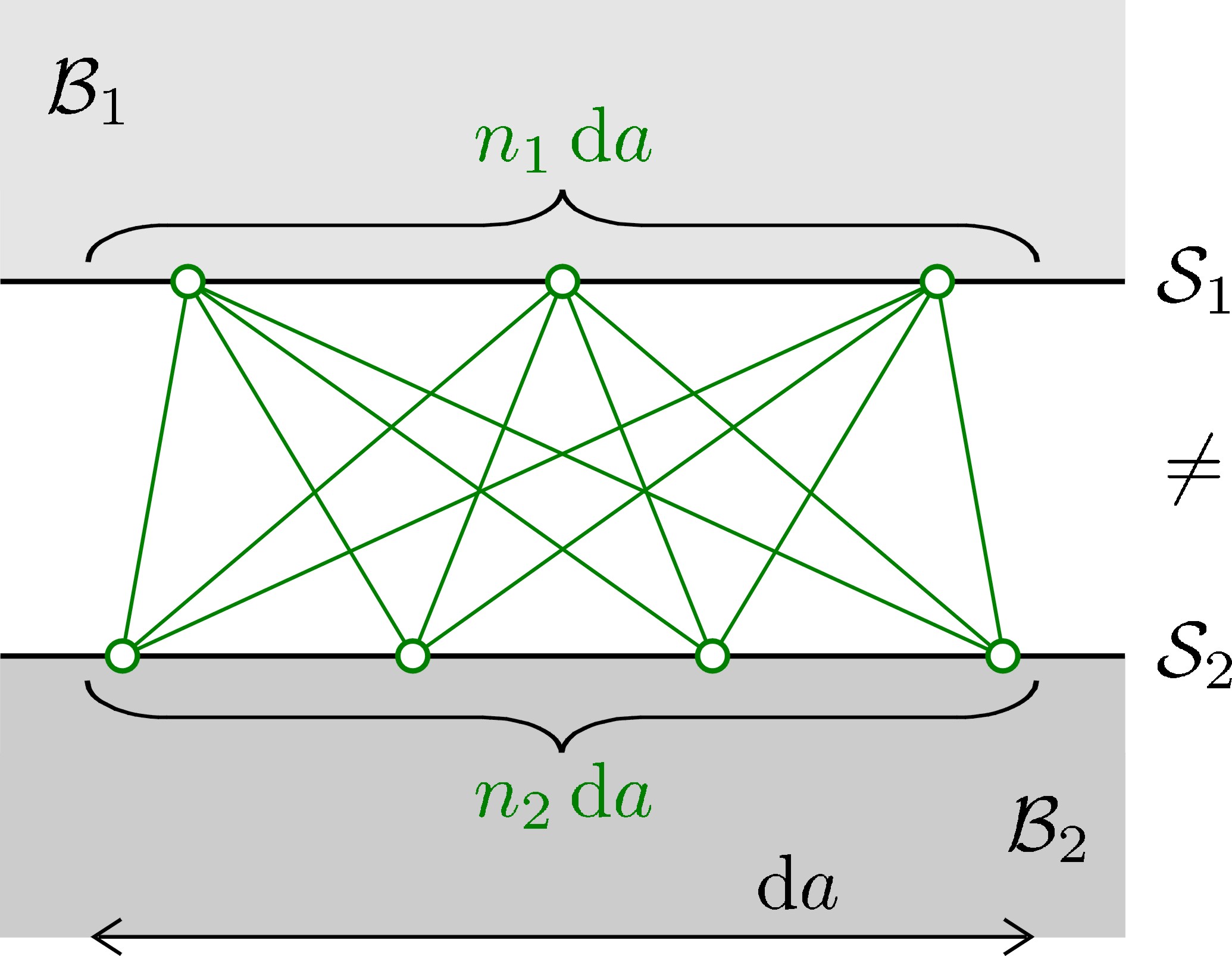}}
\put(1.1,0){\includegraphics[height=45mm]{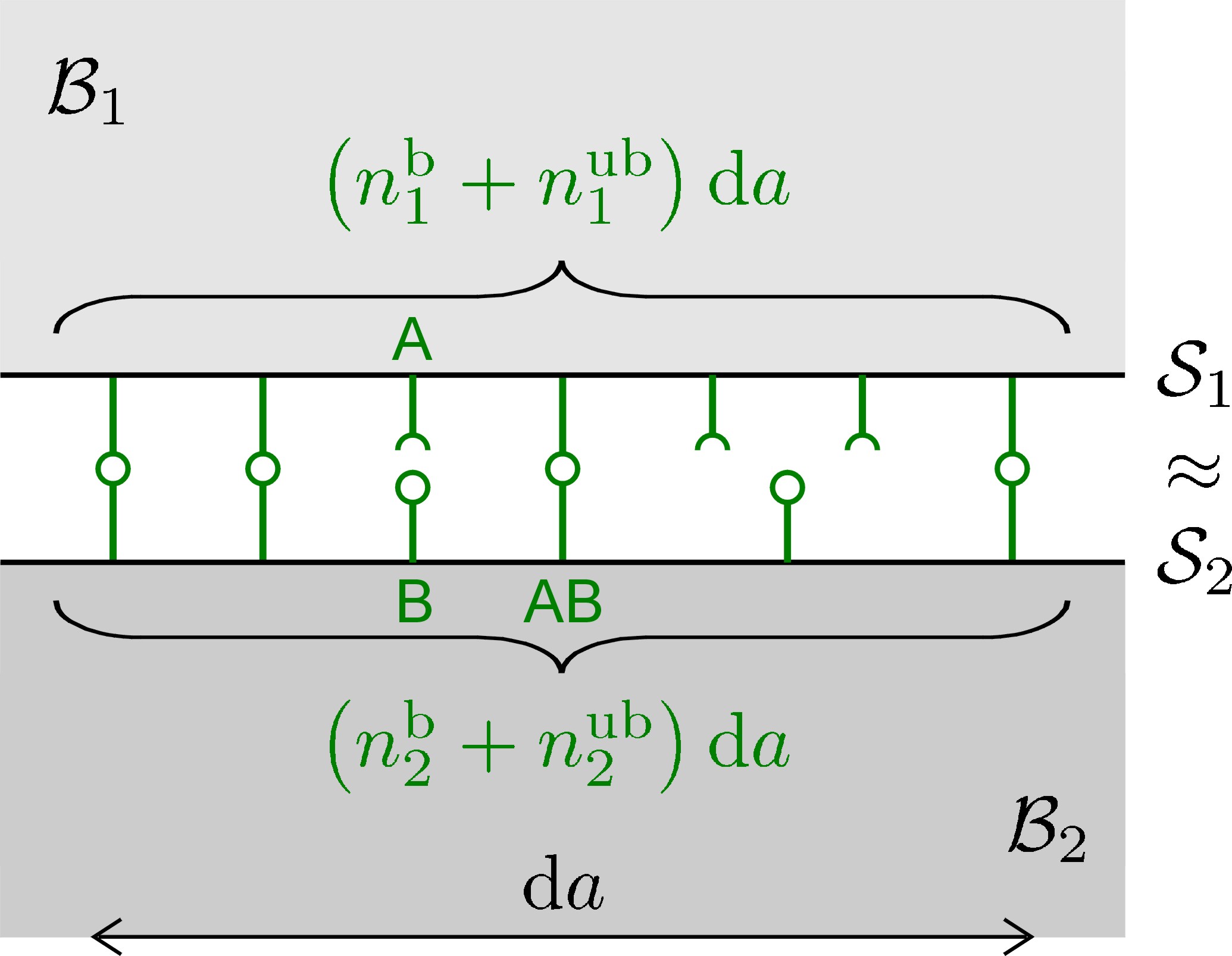}}
\put(-7.3,0){a.}
\put(.7,0){b.}
\end{picture}
\caption{Chemical contact interactions: 
a.~Long-range interactions between all bonding sites (e.g.~molecules) of two non-touching surfaces. 
Here $n_1\,\dif a = 3$ and $n_2\,\dif a = 4$.
b.~Short-range bonding reactions between neighboring bonding sites $\unde{A}$ and $\unde{B}$ of two close surfaces. 
Here $n^\mrb_1\,\dif a = n^\mrb_2\,\dif a =4$ bonds have formed, while $n^{\mru\mrb}_1\,\dif a=3$ and $n^{\mru\mrb}_2\,\dif a=2$ sites are unbonded.} 
\label{f:phicont}
\end{center}
\end{figure}

In order to model chemical bonding, the interacting surfaces are considered to have a certain bonding site density $n_k$ (the number of bonding sites per current area) composed of bonded sites and unbonded sites, i.e.
\eqb{l}
n_k = n_k^{\mrb} + n_k^{\mru\mrb}\,.
\eqe
The number of bonding sites is considered to be conserved, i.e.
\eqb{l}
\ds\pad{}{t}\int_{\sP_k}n_k\,\dif a_k = 0 \quad\forall\,\sP_k\subset\sS_k\,,
\label{e:nk0}\eqe
implying
\eqb{l}
\dot n_k +  n_k\,\divz_{\!\mrs}\,\bv_k = 0 \quad\forall\,\bx_k\in\sS_k\,,
\label{e:nk}\eqe
due to \eqref{e:sRey} and \eqref{e:Loc}.
Due to \eqref{e:dJs}, this ODE is solved by $n_k = N_k/J_{\mrs k}$, where $N_k:=n_{k}\big|_{t=0}$ is the initial bonding site density.
Two cases will be considered in the following, see Fig.~\ref{f:phicont}:\\
(a) Long-range interaction\footnote{Long-range interactions are understood to be interactions between non-touching surfaces $\sS_1\neq\sS_2$ here. 
These include electrostatic and van-der-Waals interactions, even if the latter are classified as short-range in other works.} (between non-touching surfaces), where each bonding site on one surface can interact with all other bonding sites on the other surface.\\
(b) Short-range interactions (between very close or even touching surfaces), where each bonding site on one surface can only bond to a single bonding site on the other surface.
%-----------------------------------------------------------------
%-----------------------------------------------------------------
\\
In the first case we assume $n_k^{\mru\mrb}=0$ such that $n_k = n_k^\mrb$.
Then no further equation is needed for $n_k^\mrb$. 
An example are van-der-Waals interactions described by the Lennard-Jones potential discussed in Sec.~\ref{s:LJ}.
In the second case, discussed in the remainder of this section, further equations are needed for $n^\mrb_k$.
Since, in this case, the total number of bonded sites is the same on both surfaces, we must have
\eqb{l}
\boxed{\ds\int_{\sS_1} n_1^\mrb\,\dif a_1 = \int_{\sS_2} n_2^\mrb\,\dif a_2}\,.
\label{e:netbond}\eqe
If the two surfaces are very close, $\sS_1\approx\sS_2$ can be replaced by a common reference surface $\sS_\mrc$.
Eq.~\eqref{e:netbond} then implies that $n^\mrb_1 = n^\mrb_2 =: n_\mrb$ due to localization (since \eqref{e:netbond} is still true for any subregion $\sP_\mrc\subset\sS_\mrc$).
This case is considered in the following.
\\
The evolution of $n_\mrb$ is described by the chemical reaction
\eqb{l}
%\mathrm{A + B \rightleftharpoons AB}\,, \\
\unde{A + B \rightleftharpoons AB}\,,
\eqe
where $\unde{AB}$ denotes the bond, while $\unde{A}$ and $\unde{B}$ are its components on the two surfaces prior to bonding, see Fig.~\ref{f:phicont}b. 
The bonding reaction is characterized by the reaction rate
\eqb{l}
R_\mrc = \overrightarrow{R_\mrc} - \overleftarrow{R}_\mrc\,,
\label{e:Rc}\eqe
composed of the bonding reaction rate $\overrightarrow{R_\mrc}$ (forward reaction) and the debonding reaction rate $\overleftarrow{R}_\mrc$ (backward reaction).
An example for these is discussed in Sec.~\ref{s:reac}.
Given $R_\mrc$, the balance law for the bonded sites (considering no surface diffusion) then follows as \citep{sahu17}
\eqb{l}
\ds\pad{}{t}\int_{\sP_\mrc}n_\mrb\,\dif a = \int_{\sP_\mrc}R_\mrc\,\dif a \quad\forall\,\sP_\mrc\subset\sS_\mrc\,,
\eqe
which gives
\eqb{l}
\dot n_\mrb + n_\mrb\,\divz_{\!\mrs}\,\bv_\mrc = R_\mrc \quad\forall\,\bx_\mrc\in\sS_\mrc\,,
\label{e:nb}\eqe
where $\bv_\mrc:=\bv_1=\bv_2$ is the common velocity required to enable chemical bonding.
For the unbonded sites on the two surfaces, we have the two balance laws
\eqb{l}
\ds\pad{}{t}\int_{\sP_k}n_k^{\mru\mrb}\,\dif a_k = -\int_{\sP_k}R_\mrc\,\dif a_k \quad\forall\,\sP_k\subset\sS_k\,,\quad k=1,2\,,
\eqe
leading to
\eqb{l}
\dot n_k^{\mru\mrb} + n_k^{\mru\mrb}\,\divz_{\!\mrs}\,\bv_k = -R_\mrc \quad\forall\,\bx_k\in\sS_k\,,
\label{e:nub}\eqe
where still $\bv_1=\bv_2$ (as long as $n_\mrb>0$).
Summing \eqref{e:nb} and \eqref{e:nub} leads to \eqref{e:nk}, and so one equation (for each $k$) is redundant.
\\
For convenience, the non dimensional phase field 
\eqb{l}
\phi := \ds\frac{n_\mrb}{\nc}\,,\quad 0\leq\phi\leq1\,,
\label{e:phi_def}\eqe
is introduced, where $\nc$ is the reference value for the bonding site density, either picked as $\nc=n_1$, in case surface 1 is taken as reference, or $\nc=n_2$, in case surface 2 is taken as reference.\footnote{Both choices are useful: The first implies using the \textit{master surface} $\sS_1$, already used as a reference for mechanical contact, as reference surface for chemical contact. 
On the other hand, the second choice is advantageous in computations, where usually the \textit{slave surface} $\sS_2$ is used for the integration of the weak form.}
Then
\eqref{e:nb} can be rewritten into
\eqb{l}
\boxed{\nc\,\dot\phi = R_\mrc} \quad \forall\,\bx_\mrc\in\sS_\mrc\,,
\label{e:phi}\eqe
using Eq.~\eqref{e:nk}.
Eq.~\eqref{e:phi} is the evolution law for the chemical contact state. 

\begin{remark}\label{r:31}
As we are focusing on solids, the derivation of \eqref{e:phi} assumes no surface mobility (or diffusion) of the bonds. In the context of fluidic membranes, the surface mobility of bonds is usually accounted for, e.g.~see \citet{brochard02} and \citet{freund04}.
\end{remark}

\begin{remark}\label{r:32}
In the derivation leading up to Eq.~\eqref{e:nk}, $\bv_1=\bv_2$ has been used.
But it suffices to assume that the time scale of chemical reactions is much smaller than the timescale of sliding, such that the two reacting surfaces can be assumed stationary w.r.t.~one another.
\end{remark}

\subsection{Momentum balance}

The linear momentum balance for the entire two-body system is given by
\eqb{l}
\ds\sum_{k=1}^2\pad{}{t}\int_{\sB_k}\rho_k\,\bv_k\,\dif v_k = \sum_{k=1}^2\bigg[\int_{\sB_k}\rho_k\,\bexk\,\dif v_k + \int_{\partial_t\sB_k}\bar\bt_k\,\dif a_k \bigg]\,,
\label{e:tmom}\eqe
where $\bexk$ and $\bar\bt_k$ are prescribed body forces and surface tractions. 
The latter are prescribed on the Neumann boundary $\partial_t\sB_k\subset\partial\sB_k$ that is disjoint from contact surface $\sS_k$.
The bonding events, described by Eq.~\eqref{e:phi}, are not considered to affect the momentum of the system.
\\
If the two bodies are cut apart, the additional contact interaction traction $\bt_k^\mrc$ needs to be taken into account on the interface (see Fig.~\ref{f:cont}).
The individual momentum balance for every part $\sP_k\subset\sB_k$ ($k=1,2$) then reads
\eqb{l}
\ds\pad{}{t}\int_{\sP_k}\rho_k\,\bv_k\,\dif v_k = \int_{\sP_k}\rho_k\,\bexk \,\dif v_k + \int_{\partial\sP_k}\bt_k\,\dif a_k 
\quad\forall\,\sP_k\subset\sB_k\,,
\label{e:tmomk}\eqe
where $\bt_k$ is the traction on surface $\partial\sP_k$ that contains the cases
\eqb{llll}
\bt_k \is \bar\bt_k & $on $\partial_t\sB_k\,, \\[1mm]
\bt_k \is \bt^\mrc_k & $on $\sS_k\,.
\label{e:tk}\eqe
Applying \eqref{e:vRey}, \eqref{e:divtheo} and \eqref{e:Loc} leads to the corresponding local form
\eqb{l}
\boxed{\rho_k\,\dot\bv_k = \divz\bsig_{\!k} + \rho_k\,\bexk} \quad\forall\,\bx_k\in\sB_k\,,
\label{e:sf_mom}\eqe
where $\bsig_{\!k}$ is the Cauchy stress at $\bx_k$ that is defined by the formula
\eqb{l}
\bsig_{\!k}\,\bn_k = \bt_k\,.
\label{e:Cauchy}\eqe
Using \eqref{e:tmomk}, Eq.~\eqref{e:tmom} implies
\eqb{l}
\boxed{\ds\int_{\sS_1}\bt_1^\mrc\,\dif a_1 + \int_{\sS_2}\bt_2^\mrc\,\dif a_2 = \mathbf{0}}\,.
\label{e:netforce}\eqe
This simply states that the net contact forces are in equilibrium.
It is the corresponding statement to Eq.~\eqref{e:netbond} for mechanical contact.
If the two surfaces touch, i.e.~$\sS_1\approx\sS_2=:\sS_\mrc$ and $\dif a_1\approx \dif a_2$, \eqref{e:netforce} implies that
\eqb{l}
\boxed{\bt^\mrc_1=-\bt^\mrc_2=:\bt_\mrc}\quad\forall\bx_\mrc\in\sS_\mrc\,.
\label{e:sf_moc}\eqe
Hence there is no jump in the contact traction. 
Such a jump only arises if an interface stress (e.g.~surface tension) is considered.

The angular momentum balances for the individual bodies and the entire two-body system can be written down analogously to \eqref{e:tmomk} and
\eqref{e:tmom}, respectively.
As long as no distributed body and surface moments are considered, the angular momentum balance of each body has the well-known consequence $\bsig_{\!k}^\mrT = \bsig_{\!k}$ \citep{chadwick}, while global angular momentum balance implies
\eqb{l}
\boxed{\ds\int_{\sS_1}\bx_1\times\bt_1^\mrc\,\dif a_1 + \int_{\sS_2}\bx_2\times\bt_2^\mrc\,\dif a_2 = \mathbf{0}}\,,
\label{e:netmom}\eqe
analogously to \eqref{e:netforce}.
This statement is relevant for separated surfaces ($\sS_1\neq\sS_2$), but in the case of touching surfaces ($\sS_1\approx\sS_2$), it leads to the already known traction equivalence \eqref{e:sf_moc} at common contact points $\bx_1=\bx_2$.

\begin{remark}\label{r:33}
In \eqref{e:sf_moc} the \textit{master surface} $\sS_1$ is taken as reference surface for defining the reference traction $\bt_\mrc$, which is commonly done in mechanical contact formulations \citep{wriggers-contact}. 
In case of chemical and thermal contact, however, the two possible choices $n_\mrc=n_1$ or $n_\mrc=n_2$ for the reference bonding site density introduced in \eqref{e:phi_def} are maintained in this treatment.
\end{remark}

\subsection{Energy balance}

The energy balance for the entire two-body system is
\eqb{l}
\ds\pad{\bbE}{t} 
= \sum_{k=1}^2\bigg[\int_{\sB_k}\rho_k\,\rexk\, \dif v_k + \int_{\partial_q\sB_k}\bar q_k\,\dif a_k  
+ \int_{\sB_k}\rho_k\,\bv_k\cdot\bexk\,\dif v_k + \int_{\partial_t\sB_k}\bv_k\cdot\bar\bt_k\,\dif a_k \bigg]\,,
\label{e:tenb}\eqe
where the total energy in the system,
\eqb{l}
\bbE = \bbK + \bbU\,, 
\eqe
is given by the kinetic energy
\eqb{l}
\bbK = \ds\frac{1}{2}\int_{\sB_1}\rho_1\,\bv_1\cdot\bv_1\,\dif v_1 + \frac{1}{2}\int_{\sB_2}\rho_2\,\bv_2\cdot\bv_2\,\dif v_2
\eqe
and the internal energy
\eqb{l}
\bbU = \ds\int_{\sB_1}\rho_1\,u_1\,\dif v_1 + \int_{\sB_2}\rho_2\,u_2\,\dif v_2 + \bbU_\mrc\,,
\eqe
which accounts for the individual energies $u_k=u_k(\bx_k)$ in $\sB_k$ and the contact energy $\bbU_\mrc$.
$\bbU_\mrc$ can describe long range surface interactions, as discussed in Remark~\ref{r:36}, or it can correspond to the energy of a third medium residing in the contact interface, e.g.~a thin film of lubricants or wear particles.
In the latter case, assuming a sufficiently thin film, $\bbU_\mrc$ can be expressed as the surface integral 
\eqb{l}
\bbU_\mrc :=  \ds\int_{\sS_\mrc}\nc\,u_\mrc\,\dif a\,,
\label{e:Uc}\eqe
where $u_\mrc$ is the contact energy per bonding site on contact surface $\sS_\mrc$.\footnote{where -- as in~Eq.~\eqref{e:phi_def} -- either surface 1 or surface 2 can be used as reference surface for defining $\nc$, see also footnote~5.}
Further, $\rexk$ and $\bar q_k$ in \eqref{e:tenb} denote external heat sources in $\sB_k$ and external heat influxes on $\partial_q\sB_k\subset\partial\sB_k$, respectively.
The Neumann boundary $\partial_q\sB_k$ is considered disjoint from contact surface $\sS_k$.
An external heat source on the interface is not required here.
As will be seen later, the present setup already accounts for the heat from interfacial friction and interfacial reactions.
\\
If the two bodies are cut apart, the additional contact heat influx $q_k^\mrc$ needs to be taken into account on the interface (see Fig.~\ref{f:cont}), leading to the individual energy balance for each body
\eqb{lll}
\ds\pad{}{t}\int_{\sP_k}\rho_k\,e_k\, \dif v_k \is \ds\int_{\sP_k}\rho_k\,\rexk\, \dif v_k + \int_{\partial\sP_k}q_k\,\dif a_k \\[4mm]
\plus \ds\int_{\sP_k}\bv_k\cdot\rho_k\,\bexk\,\dif v_k + \int_{\partial\sP_k}\bv_k\cdot\bt_k\,\dif a_k
\quad \forall\,\sP_k\subset\sB_k\,,
\label{e:enbk}\eqe
where $e_k := u_k + \bv_k\cdot\bv_k/2$ and $\bt_k$ satisfies \eqref{e:tk}. 
Further, $q_k$ is the heat influx on surface $\partial\sP_k$ that contains the cases
\eqb{llll}
q_k \is \bar q_k & $on $\partial_q\sB_k\,, \\[1mm]
q_k \is q^\mrc_k & $on $\sS_k\,.
\eqe
Introducing the heat flux vector $\bq_k$ at $\bx_k\in\sB_k$ that is defined by Stokes formula
\eqb{l}
-\bq_k\cdot\bn_k = q_k \,,
\eqe
and using \eqref{e:Cauchy}, the divergence theorem \eqref{e:divtheo} can be applied to obtain
\eqb{l}
\ds\int_{\partial\sP_k}q_k\,\dif a_k = -\int_{\sP_k}\divz\bq_k\, \dif v_k
\eqe
and
\eqb{l}
\ds\int_{\partial\sP_k}\bv_k\cdot\bt_k\,\dif a_k = \int_{\sP_k}\big(\bv_k\cdot\divz\bsig_{\!k} + \bsig_{\!k}:\bD_k\big)\, \dif v_k\,,
\eqe
where $\bD_k$ is the symmetric velocity gradient from \eqref{e:Dk}.
Using \eqref{e:vRey}, \eqref{e:Loc} and \eqref{e:sf_mom}, the local form of \eqref{e:enbk} thus becomes 
\eqb{l}
\boxed{\rho_k\,\dot u_k = \rho_k\,\rexk - \divz\bq_k + \bsig_{\!k}:\bD_k} \quad\forall \bx_k\in\sB_k\,.
\label{e:sf_en}\eqe
With this and Eq.~\eqref{e:sf_mom}, the combined balance statement \eqref{e:tenb} implies
\eqb{l}
\boxed{\ds
\dot\bbU_\mrc + \int_{\sS_1}\big(q_1^\mrc+\bt_1^\mrc\cdot\bv_1\big)\,\dif a_1 + \int_{\sS_2}\big(q_2^\mrc+\bt_2^\mrc\cdot\bv_2\big)\,\dif a_2 = 0 }\,,
\label{e:neten}\eqe
which is the corresponding statement to Eqs.~\eqref{e:netbond} and \eqref{e:netforce} for thermal contact.
Here
\eqb{l}
\dot\bbU_\mrc =  \ds\int_{\sS_\mrc}\nc\,\dot u_\mrc\,\dif a\,,
\eqe
follows from \eqref{e:Uc} for bonding site conservation \eqref{e:nk0}.
If the two surfaces are very close or even touch, i.e. $\sS_1\approx\sS_2=:\sS_\mrc$ and $\dif a_1\approx \dif a_2$ (i.e.~for very thin interfacial media), this implies that (since $\sS_\mrc$ can be considered an arbitrary subregion of the contact surface)
\eqb{l}
\boxed{\nc\,\dot u_\mrc + q_1^\mrc + q_2^\mrc + \bt_\mrc\cdot(\bv_1-\bv_2) = 0} \quad\forall \bx_\mrc\in\sS_\mrc\,,
\label{e:sf_enc}\eqe
due to \eqref{e:sf_moc}.
Eq.~\eqref{e:sf_enc} states that the energy rate $\dot u_\mrc$ and the mechanical contact power $\bt_\mrc\cdot(\bv_1-\bv_2)$ cause the heat influxes $q_k^\mrc$ on $\sS_k$. 
Eq.~\eqref{e:sf_enc} is equivalent to Eq.~(6.63) of \citet{laursen}.\footnote{In \citet{laursen} the contact traction $\bt_\mrc$ and heat fluxes $q_k^\mrc$ are defined with opposite sign, and \eqref{e:sf_enc} is written per unit reference area. 
Further, $u_\mrc$ does not contain chemical energy in \citet{laursen}.}

\begin{remark}\label{r:34}
Using \eqref{e:vsplit}, the contact power becomes
\eqb{l}
\bt_\mrc\cdot(\bv_1-\bv_2) = \bt_\mrc\cdot\Lge + \bt_\mrc\cdot\Lgi\,,
\eqe
showing that it contains elastic and inelastic contributions associated with sticking and sliding, respectively.
The first one is associated with energy stored in the contact interface (defined by \eqref{e:psic}, later). 
This energy vanishes for an exact enforcement of sticking ($\bg_\mre=\mathbf{0}$). 
The second contribution causes dissipation, which is also zero in the case of sticking (since $\Lgi=\mathbf{0}$) and in case of frictionless sliding (since $\bt_\mrc\perp\Lgi$). 
In principle, the second contribution could also contain stored energy.
In the context of plasticity such a stored energy occurs for hardening \citep{rosakis00}.
But this is not considered here.
\end{remark}

\begin{remark}\label{r:35}
Note, that contrary to the contact tractions governed by \eqref{e:sf_moc}, the contact heat flux according to \eqref{e:sf_enc} can have a jump across the interface. 
This has also been recently explored by \citet{javili14a}.
\end{remark}

\begin{remark}\label{r:36}
For long-range interactions between two surfaces, $\bbU_\mrc$ can be written as\footnote{e.g.~by coarse-graining the molecular interactions across the two surfaces \citep{sauer07a}}
\eqb{l}
\bbU_\mrc := \ds\int_{\sS_1}\int_{\sS_2}n_1\,n_2\,u_{12}\,\dif a_2\,\dif a_1\,,
\label{e:Uclr}\eqe
where $u_{12}=u_{12}(\bx_1,\bx_2)$ is an interaction energy defined between points on the two surfaces.
For bonding site conservation \eqref{e:nk0} this leads to 
\eqb{l}
\dot\bbU_\mrc = \ds\int_{\sS_1}\int_{\sS_2}n_1\,n_2\,\dot u_{12}\,\dif a_2\,\dif a_1\,,
\eqe
in \eqref{e:neten}.
Localization in the form of \eqref{e:sf_enc} is not possible for long-range interaction so that Eq.~\eqref{e:neten} then remains the only governing equation.
\end{remark}

\subsection{Entropy balance and the second law of thermodynamics}

The entropy balance for the entire two-body system can be written as$^6$
\eqb{l}
\ds\pad{\bbS}{t} = \sum_{k=1}^2\bigg[\int_{\sB_k}\rho_k\,\big(\eta^\mre_k+\eta^\mri_k\big)\, \dif v_k + \int_{\partial_q\sB_k}\bar{\tilde q}_k\,\dif a_k \bigg]
+ \int_{\sS_\mrc} \nc\,\eta^\mri_\mrc\,\dif a\,,
\label{e:tentb}\eqe
where the total entropy in the system,
\eqb{l}
\bbS := \ds\int_{\sB_1}\rho_1\,s_1\,\dif v_1 + \int_{\sB_2}\rho_2\,s_2\,\dif v_2 
+ \int_{\sS_\mrc}\nc\,s_\mrc\,\dif a\,,
\label{e:Stot}\eqe
accounts for the individual entropies $s_k$ in $\sB_k$ and the contact entropy $s_\mrc$ that is associated with an interfacial medium.
Further, $\eta_k^\mre$ and $\eta_k^\mri$ are the external and internal entropy production rates in $\sB_k$, $\bar{\tilde q}_k$ is the entropy influx on the heat flux boundary $\partial_q\sB_k$, and $\eta^\mri_\mrc$ is the internal entropy production rate of the interface. 
An external entropy production rate is not needed for the interface, as long as no heat source is considered in the interface.
Like $u_\mrc$ in \eqref{e:Uc}, $s_\mrc$ and $\eta^\mri_\mrc$ are bond-specific.
\\
If the two bodies are cut apart, the additional contact entropy influx $\tilde q_k^\mrc$ needs to be taken into account on the interface, leading to the individual entropy balances ($k=1,2$)
\eqb{l}
\ds\pad{}{t}\int_{\sP_k}\rho_k\,s_k\,\dif a = \int_{\sP_k}\rho_k\,\big(\eta^\mre_k+\eta^\mri_k\big)\, \dif v + \int_{\partial\sP_k}\tilde q_k\,\dif a 
\quad \forall\,\sP_k\subset\sB_k\,,
\label{e:entb}\eqe
where $\tilde q_k$ is the entropy influx on surface $\partial\sP_k$ that contains the cases
\eqb{llll}
\tilde q_k \is \bar{\tilde q}_k & $on $\partial_q\sB_k\,, \\[1mm]
\tilde q_k \is \tilde q^\mrc_k & $on $\sS_k\,.
\eqe
Introducing the entropy flux $\tilde\bq_k$ in body $\sB_k$ defined by
\eqb{l}
-\tilde\bq_k\cdot\bn_k = \tilde q_k \,, 
\eqe
theorems \eqref{e:vRey}, \eqref{e:divtheo} and \eqref{e:Loc} can be applied to \eqref{e:entb} to give the local form
\eqb{l}
\boxed{\rho_k\,\dot s_k = \rho_k\,\eta^\mre_k + \rho_k\,\eta^\mri_k - \divz\tilde\bq_k} \quad\forall \bx_k\in\sB_k\,.
\label{e:sf_ent}\eqe
Plugging this equation into the total entropy balance \eqref{e:tentb} implies
\eqb{l}
\boxed{ 
\ds\int_{\sS_\mrc} \nc\,\big(\dot s_\mrc-\eta^\mri_\mrc\big)\,\dif a + \int_{\sS_1}\tilde q_1^\mrc\,\dif a_1 + \int_{\sS_2}\tilde q_2^\mrc\,\dif a_2 
= 0}\,.
\label{e:netent}\eqe
This is the corresponding entropy statement to Eqs.~\eqref{e:netbond}, \eqref{e:netforce} and \eqref{e:neten}. 
If the two surfaces touch, i.e.~$\sS_1\,\approx\sS_2=:\sS_\mrc$, it implies that
\eqb{l}
\boxed{\nc\,\dot s_\mrc = n_1\,\eta_\mrc^\mri - \tilde q_1^\mrc - \tilde q_2^\mrc} \quad\forall \bx_\mrc\in\sS_\mrc\,.
\label{e:sf_entc}\eqe
The second law of thermodynamics states that
\eqb{l}
\eta^\mri_k \geq 0\,,\quad \eta^\mri_\mrc \geq 0\,.
\label{e:2law}\eqe
In the absence of chemical contact, \eqref{e:sf_entc} then becomes equivalent to Eq.~(6.66) of \citet{laursen}.

\section{General constitution}\label{s:consti}

This section derives the general constitutive equations for the two bodies and their contact interface -- as they follow from the internal energy and the second law of thermodynamics. 
In general, the internal energy is a function of the mechanical, chemical and thermal state of the system that is characterized by the deformation, the bonding state and the entropy.
Introducing the Helmholtz free energy, the thermal state can be characterized by the temperature instead of the entropy.
The derivation uses the framework of general irreversible thermodynamics established in chemistry \citep{onsager31a,onsager31b,prigogine,degroot} and mechanics \citep{coleman64} following the coupled thermo-mechanical treatment of \citet{laursen} and the coupled chemo-mechanical treatment of \citet{sahu17}. 
The novelty here is the extension to coupled thermo-chemo-mechanical contact leading to the establishment of the interfacial thermo-chemo-mechanical energy balance in Eq.~\eqref{e:qc_m2}, which is then discussed in detail.

\subsection{Thermodynamic potentials}\label{s:poti}

For each $\sB_k$, the Helmholtz free energy (per mass),
\eqb{l}
\psi_k = u_k - T_k\,s_k\,,
\label{e:psi_def}\eqe
is introduced as the chosen thermodynamic potential. 
Thus,
\eqb{l}
T_k\,\dot s_k = \dot u_k - \dot\psi_k - \dot T_k\,s_k\,.
\label{e:dsk}\eqe
Inserting \eqref{e:sf_en}, then leads to
\eqb{l}
T_k\,\rho_k\,\dot s_k = \bsig_{\!k}:\bD_k + \rho_k\,\rexk - \divz\bq_k - \rho_k\,\big(\dot\psi_k + \dot T_k\,s_k\big)\,.
\label{e:dsk2}\eqe
For the interface, the Helmholtz free energy (per bonding site) is introduced as
\eqb{l}
\psi_\mrc = u_\mrc - T_\mrc\,s_\mrc\,,
\label{e:psic_def}\eqe
where $T_\mrc$ is the interface temperature introduced in Sec.~\ref{s:kine} (see Fig.~\ref{f:cont}).
Thus,
\eqb{l}
T_\mrc\,\dot s_\mrc = \dot u_\mrc - \dot\psi_\mrc - \dot T_\mrc\,s_\mrc\,.
\label{e:dsc}\eqe
Inserting \eqref{e:sf_enc}, then leads to
\eqb{l}
\nc\,T_\mrc\,\dot s_\mrc = -\bt_\mrc\cdot(\bv_1-\bv_2) - q_1^\mrc - q_2^\mrc - \nc\,\big(\dot\psi_\mrc + \dot T_\mrc\,s_\mrc\big)\,.
\label{e:dsc2}\eqe
In this study, the Helmholtz free energy within $\sB_k$ is considered to take the functional form
\eqb{l}
\psi_k = \psi_k\big(\bE^\mre_k,T_k\big)\,,
\label{e:psik}\eqe
where $\bE^\mre_k$ is the elastic Green-Lagrange strain tensor introduced by Eq.~\eqref{e:Esplit}.
For the interface, the Helmholtz free energy is considered to take the form
\eqb{l}
\psi_\mrc = \psi_\mrc\big(\bg_\mre,T_\mrc,\phi\big)\,,
\label{e:psic}\eqe
where $\bg_\mre = g_\mrn\,\bn_\mrp + \xi^\alpha_\mre\,\ba^\mrp_\alpha$ is the elastic part of the gap as introduced by Eq.~\eqref{e:vsplit}.
From Eqs.~\eqref{e:psik} and \eqref{e:psic} follow
\eqb{l}
\dot\psi_k = \ds\pa{\psi_k}{\bE^\mre_k}:\dot \bE^\mre_k + \pa{\psi_k}{T_k}\,\dot T_k
\label{e:dpsik}\eqe
and
\eqb{l}
\dot\psi_\mrc = \ds\pa{\psi_\mrc}{\bg_\mre}\cdot\Lge + \pa{\psi_\mrc}{T_\mrc}\,\dot T_\mrc + \pa{\psi_\mrc}{\phi}\,\dot\phi \,.
\label{e:dpsic}\eqe
In \eqref{e:dpsic} $\Lge = \dot g_\mrn\,\bn_\mrp + \dot\xi^\alpha_\mre\,\ba^\mrp_\alpha$ appears instead of $\dot\bg_\mre$ (cf.~\citet{laursen}, Eq.~(6.72)), since $\psi_\mrc$ should be objective and hence not depend on the surface basis, i.e.~$\partial\psi_\mrc/\partial\bn_\mrp$ and $\partial\psi_\mrc/\partial\ba_\alpha^\mrp$ vanish.
The following subsections proceed to derive the general constitutive equations for the two bodies and their interaction based on $\psi_k$ and $\psi_\mrc$.

\subsection{Constitutive equations for the two individual bodies}

Inserting \eqref{e:sf_ent} and \eqref{e:dsk2} into (\ref{e:2law}.1), gives
\eqb{l}
\rho_k\,\eta^\mri_k = \rho_k\bigg(\ds\frac{\rexk}{T_k}-\eta_k^\mre\bigg) - \divz\bigg(\frac{\bq_\mrk}{T_k}-\tilde\bq_\mrk\bigg) - \frac{\bq_\mrk\cdot\grad T_k}{T_k^2}
+ \frac{\bsig_{\!k}:\bD_k}{T_k} - \frac{\rho_k}{T_k}\big(\dot\psi_k+\dot T_k\,s_k\big) \geq 0\,.
\eqe
Since this is true for any $\rexk$, $\bq_k$, $T_k$, $\bv_k$ and $\dot\psi_k$, we can identify the relations
\eqb{l} 
\eta_k^e = \ds\frac{\rexk}{T_k}~~$and$~~\tilde\bq_k = \ds\frac{\bq_k}{T_k}
\label{e:entflux}\eqe
for the external entropy source and the entropy flux.
We are then left with the well know dissipation inequality
\eqb{l}
\rho_k\,\eta^\mri_k =  \ds\frac{\bsig_{\!k}:\bD_k}{T_k} - \frac{\bq_\mrk\cdot\grad T_k}{T_k^2} - \frac{\rho_k}{T_k}\big(\dot\psi_k+\dot T_k\,s_k\big) \geq 0\,,
\eqe
which, in view of \eqref{e:Esplit}, \eqref{e:dpsik} and 
\eqb{l}
\bsig_{\!k}:\bD_k = \bS_{\!k}:\dot\bE_k/J_k\,, 
\label{e:sigD}\eqe
where $\bS_k := J_k\,\bF_{\!k}^{-1}\bsig_{\!k}\,\bF_{\!k}^{-\mrT}$ is the second Piola-Kirchhoff stress, becomes
\eqb{l}
T_k\,\rho_k\,\eta^\mri_k =  \ds\frac{1}{J_k}\bigg(\bS_{\!k}-\pa{\unde{\Psi}_{k}}{\bE^\mre_k}\bigg):\dot\bE^\mre_k - \rho_k\,\bigg(s_k+\pa{\psi_k}{T_k}\bigg)\dot T_k + \bsig_k:\bD_k^\mri - \frac{\bq_\mrk\cdot\grad T_k}{T_k}  \geq 0\,.
\label{e:2lawa}\eqe
Here $\unde{\Psi}_{k} := J_k\,\rho_k\,\psi_k = \rho_{k0}\,\psi_k$ is the Helmholtz free energy per reference volume, and 
$\bD_k^\mri$ is the inelastic part of $\bD_k$ that is related to $\bE_k^\mri$ via \eqref{e:sigD}.
Since \eqref{e:2lawa} is true for any $\bE_k$ and $T_k$, we find the constitutive relations
\eqb{l}
\boxed{\bS_k = \ds\pa{\unde{\Psi}_k}{\bE^\mre_k}}\,,\quad
\boxed{s_k = -\ds\pa{\psi_k}{T_k}}\,,\quad
\boxed{\bsig_{\!k}\cdot\bD^\mri_k\geq0}\,,\quad
\boxed{\bq_k\cdot\grad T_k\leq 0}\,.
\label{e:cons_k}\eqe
Inserting \eqref{e:dpsik} and \eqref{e:cons_k} back into \eqref{e:dsk2} then leads to the entropy evolution equation
\eqb{l}
T_k\,\rho_k\,\dot s_k = \bsig_{\!k}:\bD^\mri_k + \rho_k\,\rexk - \divz\bq_k\,.
\label{e:sk_evol}\eqe
Here, the term $ \bsig_{\!k}:\bD^\mri_k + \rho_k\,\rexk$ can be understood as an effective heat source.
\\
The equations in \eqref{e:cons_k} are the classical constitutive equations for thermo-mechanical bodies. 
They happen to be very similar to the contact equations obtained in the following section.

\begin{remark}\label{r:41}
The above derivation considers elastic and inelastic strain rates following from \eqref{e:Esplit}, which are analogous counterparts to the elastic and inelastic velocity jump, see \eqref{e:vsplit}, required for a general contact description.
On the other hand, a decomposition of stress $\bsig_{\!k}$ is not considered, which corresponds to a Maxwell-like model for the elastic and inelastic behavior.
For other constitutive models, that are not considered here, elastic and inelastic stress contributions can appear and need to be separated.
\end{remark}

\begin{remark}\label{r:41a}
Apart from constitutive equations \eqref{e:cons_k} and the corresponding field equations \eqref{e:sf_mom} and \eqref{e:sk_evol}, one also needs to determine the strain decomposition of \eqref{e:Esplit}.
Hence another equation is needed, e.g.~an evolution law for the inelastic strain $\bE_k^\mri$.
A simple example is thermal expansion, where $\bE_k^\mri$ follows from temperature $T_k$.
Likewise, an evolution law for the inelastic gap $\bg_\mri$ will be needed, e.g.~see \citet{wriggers-contact}.
\end{remark}

\subsection{Constitutive equations for the contact interface}\label{s:consti_c}

Next we examine the case that there is a common contact interface $\sS_1 \approx \sS_2=\sS_\mrc$. 
Inserting \eqref{e:sf_entc} and \eqref{e:dsc2} into (\ref{e:2law}.2) and using (\ref{e:entflux}.2), we find
\eqb{l}
\nc\,\eta^\mri_\mrc = \ds\bigg(\frac{1}{T_1}-\frac{1}{T_\mrc}\bigg)q_1^\mrc + \ds\bigg(\frac{1}{T_2}-\frac{1}{T_\mrc}\bigg)q_2^\mrc
+ \frac{\bt_\mrc\cdot(\bv_2-\bv_1)}{T_\mrc} 
- \frac{\nc}{T_\mrc}\big(\dot\psi_\mrc+\dot T_\mrc\,s_\mrc\big) \geq 0\,.
\eqe
Further inserting \eqref{e:vsplit} and \eqref{e:dpsic}, and introducing the nominal contact traction $\undetc:=\Jc\,\bt_\mrc$, we find
\eqb{lll}
\nc\,\eta^\mri_\mrc \is \ds\bigg(\frac{1}{T_1}-\frac{1}{T_\mrc}\bigg)q_1^\mrc + \ds\bigg(\frac{1}{T_2}-\frac{1}{T_\mrc}\bigg)q_2^\mrc
+ \frac{1}{T_\mrc\,\Jc}\bigg(\undetc-\pa{\unde{\Psi}_\mrc}{\bg_\mre}\bigg)\cdot\Lge 
+ \frac{\bt_\mrc\cdot\Lgi}{T_\mrc} \\[4mm]
\mi \ds\frac{\nc}{T_\mrc}\bigg(s_\mrc+\pa{\psi_\mrc}{T_\mrc}\bigg)\dot T_\mrc - \frac{\nc}{T_\mrc}\,\mu_\mrc\,\dot\phi \geq 0\,,
\label{e:Ei12}\eqe
where $\unde{\Psi_c} := \Jc\,\nc\,\psi_\mrc = \Nc\,\psi_\mrc$ and where
\eqb{l}
\mu_\mrc := \ds\pa{\psi_\mrc}{\phi}
\label{e:muc}\eqe
denotes the chemical potential associated with the interface reactions.
Here, $\Jc$ is the reference value for the area stretch, chosen analogously to $\nc$ in \eqref{e:phi_def}.
Since inequality \eqref{e:Ei12} is true for any $T_k$, $\Lge$, $\Lgi$ and $\phi$, we find the constitutive relations
\eqb{l}
\boxed{\undetc = \ds\pa{\unde{\Psi_c}}{\bg_\mre}}\,,\quad
\boxed{\bt_\mrc\cdot\Lgi\geq 0}
\label{e:cons_tc}\eqe
for the contact tractions,
\eqb{l}
\boxed{s_\mrc = -\ds\pa{\psi_\mrc}{T_\mrc}}
\label{e:cons_sc}\eqe
for the interfacial entropy,
\eqb{l}
\boxed{\mu_\mrc\,R_\mrc\leq 0}
\label{e:cons_muc}\eqe
for the reaction rate $R_\mrc$ (from \eqref{e:phi}), and
\eqb{lll}
\ds\bigg(\frac{1}{T_1}-\frac{1}{T_\mrc}\bigg)q_1^\mrc + \ds\bigg(\frac{1}{T_2}-\frac{1}{T_\mrc}\bigg)q_2^\mrc \geq 0\,,
\eqe
for the contact heat fluxes.
Multiplying by $T_1$, $T_2$ and $T_\mrc$ (that are all positive), the last statement can be rewritten into
\eqb{lll}
T_2\big(T_\mrc-T_1\big)\,q_1^\mrc + T_1\big(T_\mrc-T_2\big)\,q_2^\mrc \geq 0\,.
\label{e:qc_inq}\eqe
Since this has to be satisfied for any $q^\mrc_k$, setting either $q^\mrc_2=0$ or $q^\mrc_1=0$ yields the two separate conditions ($k=1,2$)
\eqb{l}
\boxed{\big(T_\mrc-T_k\big)\,q_k^\mrc \geq 0}\,,
\label{e:qkinq}\eqe
for $q_k^\mrc$. 
They are, for example, satisfied for the simple and well-known linear heat transfer law
\eqb{l}
q_k^\mrc = h_k\,\big(T_\mrc-T_k\big)\,,
\label{e:qc_k1}\eqe
where the constant $h_k \geq 0$ is the heat transfer coefficient between body $\sB_k$ and the interfacial medium.
Introducing the mean influx into $\sB_k$,
\eqb{l}
q_\mrm^\mrc := \ds\frac{q_1^\mrc+q_2^\mrc}{2}\,,
\label{e:qm_def}\eqe
and the transfer flux from $\sB_2$ to $\sB_1$ (see Fig.~\ref{f:Tcont}),
\eqb{l}
q_\mrt^\mrc := \ds\frac{q_1^\mrc-q_2^\mrc}{2}\,,
\label{e:qt_def}\eqe
such that $q_1^\mrc=q_\mrm^\mrc+q_\mrt^\mrc$ and $q_2^\mrc=q_\mrm^\mrc-q_\mrt^\mrc$, one can also rewrite inequality \eqref{e:qc_inq} into
\eqb{lll}
\big((T_1+T_2)T_\mrc - 2T_1T_2\big)q_\mrm^\mrc + \big(T_2-T_1\big)T_\mrc\,q_\mrt^\mrc \geq 0\,.
\label{e:qc_inq2}\eqe
Since this also has to be true for $q_\mrm^\mrc=0$, we find the further condition
\eqb{l}
\boxed{\big(T_2-T_1\big)\,q_\mrt^\mrc \geq 0}\,,
\label{e:qtinq}\eqe
which is, for example, satisfied by
\eqb{l}
q_\mrt^\mrc = h\,\big(T_2-T_1\big)\,,
\label{e:qc_t1}\eqe
where the constant $h \geq 0$ is the heat transfer coefficient between bodies $\sB_1$ and $\sB_2$.
\\
From \eqref{e:sf_enc} we can further find that the mean influx is given by
\eqb{l}
q^\mrc_\mrm = \ds\frac{1}{2}\big( \bt_\mrc\cdot(\bv_2-\bv_1) - \nc\,\dot u_\mrc \big)\,,
\eqe
which in view of \eqref{e:vsplit}, \eqref{e:dsc}, \eqref{e:dpsic}, (\ref{e:cons_tc}.1) and \eqref{e:cons_sc} becomes
\eqb{l}
\boxed{q^\mrc_\mrm = \ds\frac{1}{2}\big(\bt_\mrc\cdot\Lgi - \mu_\mrc\,R_\mrc - \nc\,\dot s_\mrc\,T_\mrc \big)}\,,
\label{e:qc_m2}\eqe
i.e.~the mean heat influx is caused by mechanical dissipation (from friction), chemical dissipation (from reactions), and entropy changes at the interface.
The three terms are composed of the conjugated pairs identified in table \ref{t:conj}.
While the first two terms are always positive (due to (\ref{e:cons_tc}.2) and \eqref{e:cons_muc}) and thus lead to an influx of heat into the bodies, the third term can be positive or negative. 
It thus allows for a heat flux from the bodies into the interface, where the heat is stored as internal energy, see the example in Sec.~\ref{s:test2b}.
Defining the contact enthalpy per bonding site $e_\mrc$ from
\eqb{l}
n_\mrc\,e_\mrc := n_\mrc\,u_\mrc - \bt_\mrc\cdot\bg\,,
\eqe
which is the logical extension of the classical enthalpy definition\footnote{$e_k = u_k + p_k V_k$, where $p_k$ is the pressure and $V_k$ the volume of body $\sB_k$}, one finds
\eqb{l}
q_\mrm^\mrc = -\ds\frac{n_\mrc}{2}\dot e_\mrc\big|_{\undetc\,=\,\mathrm{fixed}}\,,
\eqe
based on the Lie derivative of \eqref{e:vsplitdef}.
Thus the mean heat influx is proportional to the enthalpy change at constant nominal contact traction.
Following the classical definition \citep{laidler96}, the contact reactions can thus be classified according to
\eqb{l}
q^\mrc_\mrm \left\{\begin{array}{ll} 
> 0 & $exothermic contact reaction,$ \\  
= 0 & $isothermic contact reaction,$ \\
< 0 & $endothermic contact reaction,$
\end{array}\right.
\eqe
in case there is no heat coming from mechanical dissipation ($\bt_\mrc\cdot\Lgi=0$).
Corresponding examples are given in Sec.~\ref{s:test2}.

\begin{remark}\label{r:43}
Interface equation \eqref{e:qc_m2} is analogous to the bulk equation \eqref{e:sk_evol}. 
While \eqref{e:sk_evol} describes the entropy evolution in each body, \eqref{e:qc_m2} describes the entropy evolution of the contact interface.
In case of transfer law \eqref{e:qc_k1}, this evolution law explicitly follows from \eqref{e:qm_def} and \eqref{e:qc_m2} as
\eqb{l}
n_\mrc\,T_\mrc\,\dot s_\mrc = \bt_\mrc\cdot\Lgi - \mu_\mrc\,R_\mrc - (h_1+h_2)\,T_\mrc + h_1T_1 + h_2T_2\,.
\label{e:Tc_evolv}\eqe
Like \eqref{e:sk_evol}, it can be rewritten as an evolution law for the temperature using the entropy-temperature relationship, i.e.~\eqref{e:cons_sc}.
The analogy between \eqref{e:sk_evol} and \eqref{e:qc_m2} is not complete, as \eqref{e:sk_evol} contains an explicit heat source, while \eqref{e:qc_m2} contains a chemical dissipation term.
In principle, an explicit heat source can also be considered on the interface, while a chemical reaction can also be considered in the bulk. 
This would lead to a complete analogy between \eqref{e:qc_m2} and \eqref{e:sk_evol}.
In the absence of chemical reactions, \eqref{e:qc_m2} resorts to the thermo-mechanical energy balance found in older works, see \citet{johansson93,stromberg96,oancea97}.
\end{remark}

\begin{remark}\label{r:42}
A special case is $s_\mrc=0$ $\forall$ $t$ (for example because there is no interfacial medium),
see the example in Sec.~\ref{s:test2a}. 
In this case $\dot s_\mrc = 0$, such that \eqref{e:qc_m2} becomes $2q^\mrc_\mrm = \bt_\mrc\cdot\Lgi - \nc\,\mu_\mrc\,\dot\phi$, which is not an evolution law for $T_\mrc$ anymore, but just an expression for the energy influx into $\sB_1$ and $\sB_2$.
\end{remark}

\begin{remark}\label{r:44}
According to Eq.~\eqref{e:qc_m2}, equal energy flows from the interface into bodies $\sB_1$ and $\sB_2$. 
The transfer heat flux $q^\mrc_t$ then accounts for the possibility that the bodies heat up differently. 
Alternatively, as considered in the works of \citet{saracibar98}, \citet{pantuso00} and \citet{xing02}, factors can be proposed for the relative contributions going into the two bodies. 
\citet{pantuso00} also account for a loss of the interface heat to other bodies, like an interfacial gas.
\end{remark}

\begin{remark}\label{r:45}
The nominal traction $\undetc := \Jc\,\bt_\mrc$ is not a physically attained traction. 
Only the traction $\bt_\mrc = \nc\,\partial\psi_\mrc/\partial\bg_\mre$ is.
Since $\nc\neq\nc(\bg_\mre)$, one can also write $\bt_\mrc = \partial\Psi_\mrc/\partial\bg_\mre$, for $\Psi_\mrc := \nc\,\psi_\mrc$.
\end{remark}

\begin{remark}\label{r:46}
Multiplying \eqref{e:qc_m2} by the area change $\Jc$ yields
\eqb{l}
\unde{q}^\mrc_\mrm := \Jc\,q^\mrc_\mrm = \ds\frac{1}{2}\big(\undetc\cdot\Lgi - \unde{M_c}\,\dot\phi - \dot{\unde{S}}_\unde{c}\,T_\mrc \big)\,,
\label{e:qc_m2a}\eqe
where $\unde{S_c}:=\Nc\,s_\mrc$ and $\unde{M_c}:=\Nc\,\mu_\mrc$ are the contact entropy and chemical potential per reference area. 
Due to $\unde{\Psi_c} := \Nc\,\psi_\mrc$, \eqref{e:muc} and \eqref{e:cons_sc}, they follow directly from
\eqb{l}
\unde{S_c} = -\ds\pa{\unde{\Psi_c}}{T_\mrc}\,,\qquad \unde{M_c} = \ds\pa{\unde{\Psi_c}}{\phi}\,.
\label{e:SMuc}\eqe
\end{remark}

\begin{remark}\label{r:47}
Alternatively to transfer laws for $q^\mrc_1$ and $q^\mrc_2$, such as \eqref{e:qc_k1}, one can also propose laws for $q^\mrc_\mrt$ and $q^\mrc_\mrm$.
Examples are given by \eqref{e:qc_t1} and 
\eqb{l}
q_\mrm^\mrc = h_\mrm\bigg(T_\mrc - \ds\frac{2\,T_1T_2}{T_1+T_2}\bigg)\,,
\eqe
for some $h_\mrm\geq0$, as they satisfy condition \eqref{e:qc_inq2}.
\end{remark}

\begin{remark}\label{r:48}
In case $q^\mrc_\mrm = 0$ and relation \eqref{e:qc_k1} holds, $T_\mrc$ explicitly follows from \eqref{e:qm_def} as
\eqb{l}
T_\mrc = \ds\frac{h_1T_1 + h_2T_2}{h_1 + h_2}\,.
\eqe
One can then find the well known relation $h^{-1} = h_1^{-1} + h_2^{-1}$ from \eqref{e:qt_def} and \eqref{e:qc_t1}, which also has to be true for $q^\mrc_\mrm \neq 0$ in case all $h_k$ are constants.
\end{remark}

\begin{remark}\label{r:49}
In case of perfect thermal contact $T_\mrc = T_1 = T_2$, the fluxes $q^\mrc_1$ and $q^\mrc_2$ become the unknown Lagrange multipliers to the thermal contact constraints $T_1=T_\mrc$ and $T_1=T_\mrc$.
Alternatively, $q^\mrc_\mrt$ and $q^\mrc_\mrm$ can be used as the Lagrange multipliers to the constraints $T_2 = T_1$ and $T_\mrc = (T_1 + T_2)/2$.
\end{remark}

\subsection{Constitutive equations for non-touching (non-dissipative) interactions}\label{s:lr}

n case of non-touching (long-range) interactions, the two surfaces remain distinct, i.e.~$\sS_1\neq\sS_2$. 
For simplification we consider that the surfaces have the uniform temperatures $T_1$ and $T_2$, and that the space between between $\sS_1$ and $\sS_2$ has no mass, temperature or entropy, i.e.~$s_\mrc\equiv0$. 
Further, since the surfaces don't touch, no bonding can take place, i.e.~$\phi\equiv0$.
The Helmholtz free energy of the interface then is
\eqb{l}
\psi_{12} = u_{12}\,,
\eqe
so that $\dot u_{12} = \dot\psi_{12}$.
The Helmholtz free energy is now considered to take the form
\eqb{l}
\psi_{12} = \psi_{12}\big(\bx_1,\bx_2)\,,
\eqe
such that
\eqb{l}
\dot\psi_{12} = \ds\pa{\psi_{12}}{\bx_1}\cdot\bv_1 + \pa{\psi_{12}}{\bx_2}\cdot\bv_2\,.
\label{e:dpsi12}\eqe
Inserting this into \eqref{e:neten} yields
\eqb{l}
\ds\int_{\sS_1}\bigg(\bt_1^\mrc+n_1\pa{\psi^\mrc_2}{\bx_1}\bigg)\cdot\bv_1\,\dif a_1 
+ \ds\int_{\sS_2}\bigg(\bt_2^\mrc+n_2\pa{\psi^\mrc_1}{\bx_2}\bigg)\cdot\bv_2\,\dif a_2 
+ \int_{\sS_1}q^\mrc_1\,\dif a_1
+ \int_{\sS_2}q^\mrc_2\,\dif a_2
= 0
\eqe
where
\eqb{l}
\psi^\mrc_1 := \ds\int_{\sS_1}n_1\,\psi_{12}\,\dif a_1~,\quad
\psi^\mrc_2 := \ds\int_{\sS_2}n_2\,\psi_{12}\,\dif a_2\,.
\label{e:psi_1+2}\eqe
Noting that $\bv_k$ and $q_k^\mrc$ are arbitrary and can be independently taken as zero, we find
\eqb{l}
\boxed{\bt_1^\mrc = -n_1\ds\pa{\psi^\mrc_2}{\bx_1}}~,\quad
\boxed{\bt_2^\mrc = -n_2\ds\pa{\psi^\mrc_1}{\bx_2}}~,
\label{e:bt_lr}\eqe
for the interaction tractions,
and
\eqb{l}
\ds\int_{\sS_1}q_1^\mrc\,\dif a_1 + \ds\int_{\sS_2}q_2^\mrc\,\dif a_2 = 0 \,,
\label{e:netq}\eqe
for the interaction heat fluxes.
At the same time \eqref{e:netent} and (\ref{e:2law}.2) yield
\eqb{l}
\ds\int_{\sS_1}\frac{q_1^\mrc}{T_1}\,\dif a_1 + \ds\int_{\sS_2}\frac{q_2^\mrc}{T_2}\,\dif a_2 \geq 0 \,.
\eqe
If the temperature is constant, one can multiply this by $T_1T_2$ and insert \eqref{e:netq} to get
\eqb{l}
\boxed{(T_2 - T_1) \ds\int_{\sS_1}q_1^\mrc\,\dif a_1 \geq 0} \,,
\eqe
which is the corresponding statement to \eqref{e:qtinq} for long-range interactions.
The traction laws in \eqref{e:bt_lr} are identical to those obtained from a variational principle \citep{spbc}.

\section{Constitutive examples for contact}\label{s:ex}

This section lists examples for the preceding constitutive equations for contact and long-range interaction 
accounting for the coupling between mechanical, thermal and chemical fields. 
The examples are based on the material parameters defined in Tab.~\ref{t:mat}. 
While many of the examples are partially known, their thermo-chemo-mechanical coupling has not yet been explored in detail.
%------------------------------------------------------------------
\begin{table}[h]
\centering
\begin{tabular}{|l|l|l|}
  \hline
  symbol & name & unit \\[0.5mm] \hline 
  & & \\[-3.5mm]
  $\unde{E_n}$, $\unde{E_t}$ & normal \& tangential contact stiffness & N/m$^3$ \\[.5mm] 
  $\unde{C_c}$ & interfacial contact heat capacity at $T_0$ & J/(K\,m$^2)$  \\[.5mm] 
  $\unde{K_c}$ & bond energy density & J/m$^2$\\[.5mm]
  $T_0$ & reference temperature & K \\[.5mm]
  $t_\mrn^\mathrm{max}$, $t_\mrt^\mathrm{max}$ & normal \& tangential bond strength & N/m$^2$\\[.5mm]
  $g_0$, $\psi_0$ & reference distance \& reference energy & m, J \\[.5mm]
  $\mu_0$, $\mu$ & static and dynamic friction coefficient & 1 \\[.5mm]
  $\eta$ & dynamic interface viscosity & N\,s/m$^3$ \\[.5mm]
  $h$ & heat transfer coefficient & J/(K\,s\,m$^2$) \\[.5mm]
  $C_\mrr$ & reaction rate constant & 1/(J s m$^2$) \\[.5mm] 
   \hline
\end{tabular}
\caption{Material parameters for the contact interface.}
\label{t:mat}
\end{table}
%------------------------------------------------------------------

\subsection{Contact potential}\label{s:exco}

A simple example for the contact potential (per unit reference area) is the quadratic function
\eqb{l}
\unde{\Psi_c}(\bg_\mre,T_\mrc,\phi) = \ds\frac{1}{2}\,\bg_\mre\cdot\undeEc\,\bg_\mre 
- \frac{\unde{C_c}}{2T_0}(T_\mrc-T_0)^2 
+ \frac{\unde{K_c}}{2} (\phi-1)^2 \,,
\label{ex:psi_c}\eqe
with\footnote{where $\bn=\bn_1$ is the surface normal of the master body}
\eqb{l}
\undeEc = \unde{E_n}\, \bn\otimes\bn + \unde{E_t}\, \bi\,,
\label{e:bveps}\eqe
where
\eqb{l}
\bi := \bone - \bn\otimes\bn
\eqe
is the surface identity on $\sS_\mrc$. 
Here, $\unde{E_n}$ and $\unde{E_t}$ denote the normal and tangential contact stiffness (e.g.~according to a penalty regularization), $\unde{C_c}$ denotes the contact heat capacity (at $T_\mrc=T_0$) and $\unde{K_c}$ denotes the bond energy. 
\eqref{ex:psi_c} corresponds to an extension of the models of \citet{johansson93,stromberg96} and \citet{oancea97} to chemical bonding.
The parameters $\unde{E}_\bullet$, $\unde{C_c}$ and $\unde{K_c}$ are defined here per undeformed surface area.
They can be constant or depend on the other contact state variables, i.e.~$\unde{E}_\bullet = \unde{E}_\bullet(T_\mrc,\phi)$, $\unde{C_c} = \unde{C_c}(\bg_\mre,\phi)$ and $\unde{K_c} = \unde{K_c}(\bg_\mre,T_\mrc)$.
If they are constant, (\ref{e:cons_tc}.1) and \eqref{e:SMuc} yield the contact traction, entropy and chemical potential (per reference area)
\eqb{lll}
\undetc \is \undeEc\,\bg_\mre\,, \\[1mm]
\unde{S_c} \is \unde{C_c}\,\big(T_\mrc/T_0-1\big)\,, \\[1mm]
\unde{M_c} \is \unde{K_c}\,(\phi-1)\,.
\label{e:tsmu}\eqe
If $\unde{E}_\bullet$, $\unde{C_c}$ and $\unde{K_c}$ are not constant, further terms are generated from (\ref{e:cons_tc}.1) and \eqref{e:SMuc}.
An example is given in \eqref{e:Psi_bt}.

\begin{remark}\label{r:51}
As noted in Remark~\ref{r:34}, the first term in \eqref{ex:psi_c} is zero if the sticking constraint is enforced exactly.
In that case, the elastic gap $\bg_\mre$ is zero, while $\undeEc$ approaches infinity.
If there is no mass associated with the contact interface, its heat capacity $\unde{C_c}$, and hence the second term in \eqref{ex:psi_c}, is also zero. 
On the other hand, non-zero $\bg_\mre$ can be used to capture the deformation of surface asperities during contact (see Remark~\ref{r:22}), while non-zero $\unde{C_c}$ can be used to capture the heat capacity of trapped wear debris \citep{johansson93}.
\end{remark}

\begin{remark}\label{r:52}
The last part in \eqref{ex:psi_c} corresponds to a classical surface energy.  
In the unbonded state, ($\phi=0$) the free surface energy is $\unde{K_c}/2$.
\end{remark}

\begin{remark}\label{r:53}
Choice \eqref{ex:psi_c} has minimum energy at full bonding ($\phi=1$). 
Since $0\leq\phi\leq1$, $\mu_\mrc \leq 0$ follows for $\unde{K_c}>0$. 
Thus $R_\mrc\geq0$ due to \eqref{e:cons_muc}, which leads to $\dot\phi\geq0$ due to \eqref{e:phi}.
Hence, according to \eqref{e:phi} and \eqref{ex:psi_c}, the bonding state $\phi$ is monotonically increasing over time.
Then only mechanical debonding, illustrated by the examples in Secs.~\ref{s:exde} and \ref{s:test3}, leads to a decrease in $\phi$.
Chemical debonding, on the other hand, requires a modification of \eqref{ex:psi_c}, see following remark.
\end{remark}

\begin{remark}\label{r:54}
One can change the last term of \eqref{ex:psi_c} into
\eqb{l} 
\ds\frac{\unde{K_c}}{2} (\phi-\phi_\mathrm{eq})^2 \,,
\label{e:phieq}\eqe
such that the minimum energy state is at $\phi=\phi_\mathrm{eq}$.
This case implies that chemical equilibrium is a balance of bonding and debonding reactions, like in the model of \citet{bell78}.
An example for this is given in Sec.~\ref{s:reac}.  
Like $\unde{K_c}$, the  parameter $0\leq\phi_\mathrm{eq}\leq1$ can be a function of $\bg_\mre$ and $T_\mrc$.
\end{remark}

\begin{remark}\label{r:55}
The contact traction $\bt_\mrc = \undetc/\Jc$ can be decomposed into the contact pressure $p_\mrc = -\bn\cdot\bt_\mrc$ and tangential contact traction $\bt_\mrt =\bi\,\bt_\mrc$ (such that $\bt_\mrc = -p_\mrc\,\bn + \bt_\mrt$).
From \eqref{e:bveps}-(\ref{e:tsmu}.1) and $\epsilon_\bullet := \unde{E}_\bullet/\Jc$ thus follow $p_\mrc = -\epsilon_\mrn\,g_\mrn$ and $\bt_\mrt = \epsilon_\mrt\,\bg_{\mathrm{et}}$, where $g_\mrn$ and $\bg_{\mathrm{et}}$ are the normal and tangential parts of $\bg_\mre$ following from~\eqref{e:gnt}.
\end{remark}

\begin{remark}\label{r:56}
The normal contact contribution in \eqref{ex:psi_c} and (\ref{e:tsmu}.1) is only active up to a debonding limit.
Beyond that it becomes inactive, i.e.~by setting $\unde{E_n}=0$.
This is discussed in the following section.
\end{remark}

\subsection{Adhesion / debonding limit}\label{s:exde}

\begin{figure}[h]
\begin{center} \unitlength1cm
\begin{picture}(0,5.7)
\put(-7.9,.3){\includegraphics[height=49mm]{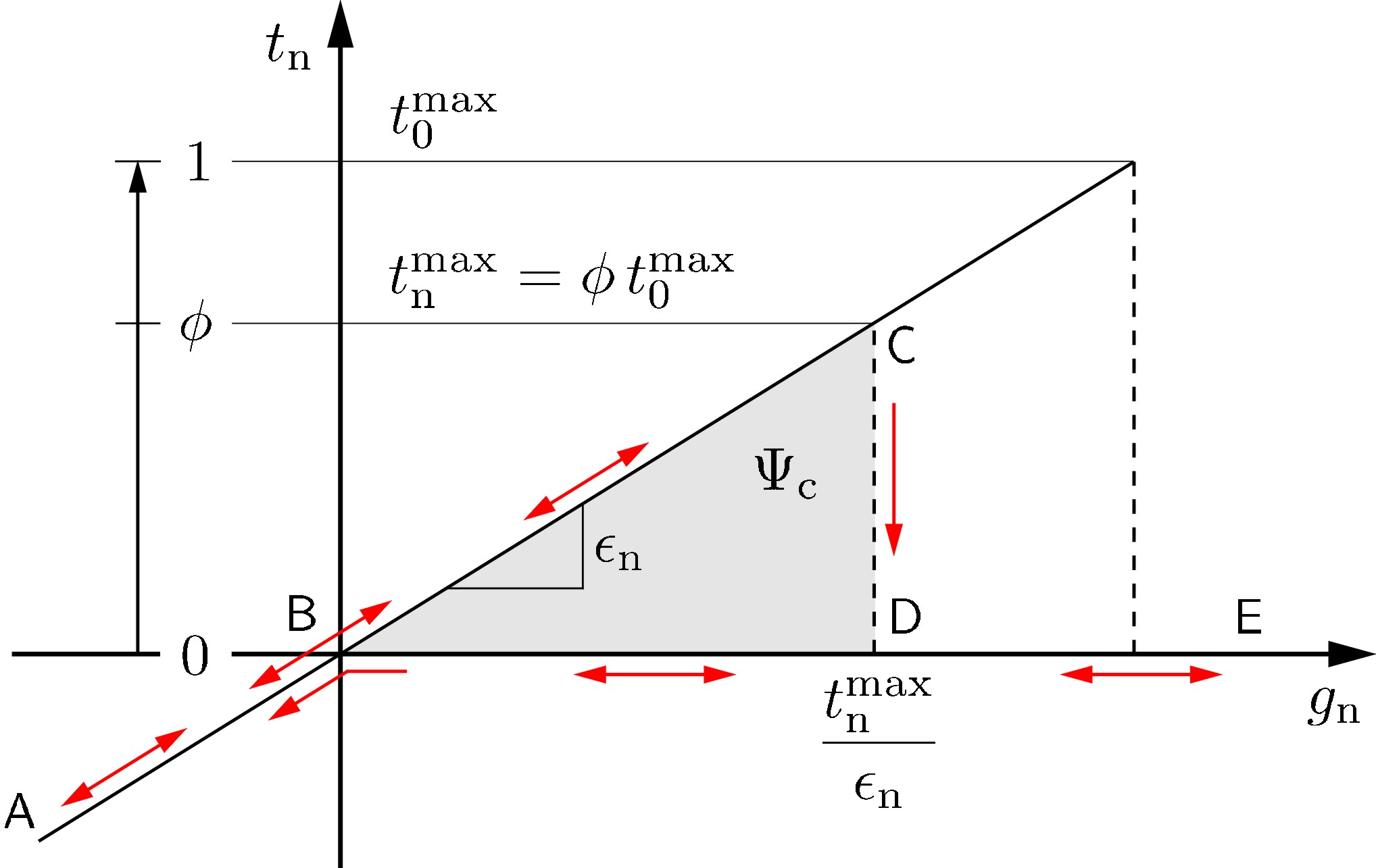}}
\put(0.15,-.1){\includegraphics[height=58mm]{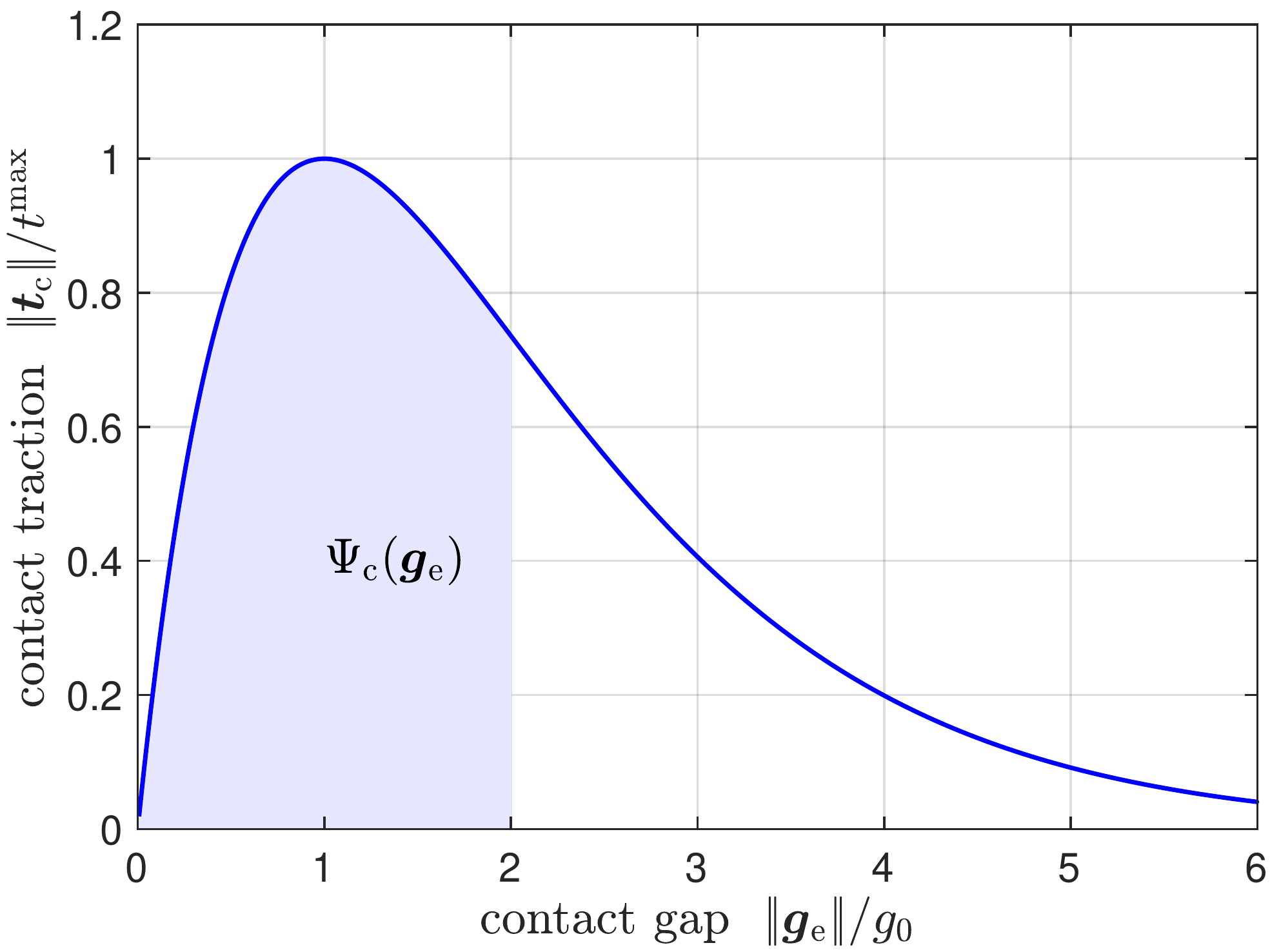}}
\put(-7.9,-.1){a.}
\put(0.9,-.1){b.}
\end{picture}
\caption{Normal contact behavior: a.~Irreversible debonding according to (\ref{e:tsmu}.1) and \eqref{e:debond}; b.~Reversible adhesion according to \eqref{ex:psi_czm}; Figure b is adopted from \citet{compadh}.}
\label{f:debond}
\end{center}
\end{figure}

An adhesion or debonding limit implies a lower bound on the contact pressure $p_\mrc$, i.e.
\eqb{l}
-p_\mrc \leq t_\mrn^\mathrm{max}\,,
\label{e:debond}\eqe
where the tensile limit (or bond strength) $t_\mrn^\mathrm{max}$ can depend on the contact state, i.e.~$t_\mrn^\mathrm{max} =  t_\mrn^\mathrm{max}(T_\mrc,\phi)$.
An example is
\eqb{l}
t_\mrn^\mathrm{max} = \phi\,\ds\frac{T_0}{T_\mrc}\,t_0^\mathrm{max}\,.
\eqe
According to this, the bond strength is increasing with $\phi$ and decreasing with $T_\mrc$.
\\
Once the limit is reached, sudden debonding occurs, resulting in $\phi=0$.\footnote{Debonding can also be modeled as compliant, such that the irreversible path $\unde{C}\rightarrow\unde{D}$ has a finite negative slope.} 
This makes debonding (mechanically) irreversible (unless $t_\mrn^\mathrm{max}=0$).
Fig.~\ref{f:debond}a shows a graphical representation of this.
%-----------------------------------------------------------------
%-----------------------------------------------------------------

Debonding can also be described in the context of cohesive zone models.
An example is the exponential cohesive zone model
\eqb{l}
\Psi_\mrc = -\ds t^\mathrm{max}\,\big(g_\mre+g_0\big)\,\exp\!\bigg(1-\frac{g_\mre}{g_0}\bigg)\,,\quad g_\mre := \norm{\bg_\mre}\,,
\label{ex:psi_czm}\eqe
where $t^\mathrm{max}$ and $g_0$ are material parameters that can depend on $T_\mrc$ and $\phi$. 
Eq.~\eqref{ex:psi_czm} is an adaption of the model of \citet{xu93}.
It leads to the contact traction
\eqb{l}
\bt_\mrc = -\ds t^\mathrm{max}\,\exp\!\bigg(1-\frac{g_\mre}{g_0}\bigg)\,\frac{\bg_\mre}{g_0}\,,
\label{ex:bt_czm}\eqe
according to (\ref{e:cons_tc}.1) and Remark~\ref{r:45}.
Eq.~\eqref{ex:bt_czm} replaces traction law (\ref{e:tsmu}.1) and its limit \eqref{e:debond}.
It is illustrated in Fig.~\ref{f:debond}b.
Traction law \eqref{ex:bt_czm} is reversible unless $t^\mathrm{max}$ is a function of $\phi$ that drops to zero beyond some $g_\mre$. 
\\
Model \eqref{ex:psi_czm} is similar to adhesion models for non-touching contact discussed next.

\subsection{Non-touching adhesion}\label{s:LJ}

An example for non-touching contact interactions according to Sec.~\ref{s:lr} is the interaction potential
\eqb{l}
\psi_{12}(\bar g_k) = \ds\psi_0\,\bigg[\frac{1}{45}\bigg(\frac{g_0}{\bar g_k}\bigg)^{\!\!10}\!-\frac{1}{3}\bigg(\frac{g_0}{\bar g_k}\bigg)^{\!\!4}\,\bigg]\,,\quad \bar g_k := \norm{\bx_k-\bx_\ell} > 0\,,
\label{e:psi12ex}\eqe
where $\psi_0$ and $g_0$ are constants and where either $k=1$ and $\ell=2$ or $k=2$ and $\ell=1$.
From \eqref{e:psi_1+2} and \eqref{e:bt_lr} then follows 
\eqb{l}
\bt_k^\mrc(\bx_k) = -n_k\,\ds\int_{\sS_\ell}n_\ell\,\pa{\psi_{12}}{\bx_k}\,\dif a_\ell\,,
\label{e:bt_lr2}\eqe
with
\eqb{l}
-\ds\pa{\psi_{12}}{\bx_k} = \frac{\psi_0}{g_0}\,\bigg[\frac{10}{45}\bigg(\frac{g_0}{\bar g_k}\bigg)^{\!\!11}\!-\frac{4}{3}\bigg(\frac{g_0}{\bar g_k}\bigg)^{\!\!5}\,\bigg]\bar\bg_k\,, \quad \bar\bg_k := \frac{\bx_k-\bx_\ell}{\bar g_k}\,.
\eqe
These expressions are valid for general surface geometries and arbitrarily long-range interactions. 
For short-range interactions between locally flat surfaces, these expressions can be integrated analytically to give \citep{spbc}
\eqb{l}
\bt_k^\mrc(\bx_k) = 2\pi\,n_1\,n_2\,g_\mrn\,\psi_{12}(g_\mrn)\,\bn_\mrp\,,
\label{e:bt_sr}\eqe
where $g_\mrn$ is the normal gap between point $\bx_k$ and surface $\sS_\ell$ and $\bn_\mrp$ is the surface normal of $\sS_\ell$ at $\bx_\mrp$. 

\begin{remark}\label{r:57}
The surface interaction potential \eqref{e:psi12ex} can be derived from the classical Lennard-Jones potential for volume interactions \citep{sauer07a,sauer09b}.
\end{remark}

\begin{remark}\label{r:58}
Example \eqref{e:psi12ex} is only valid for separated bodies ($\bar g_k > 0$). 
Formulation \eqref{e:bt_lr2} can however also be used for other potentials $\psi_{12}$ that admit penetrating bodies (with negative distances).
Examples, such as a penalty-type contact formulation, are given in \citet{spbc}.
Computationally, \eqref{e:bt_lr2} leads to the so-called \textit{two-half-pass contact algorithm}, which is thus a thermodynamically consistent algorithm.
\end{remark}

\begin{remark}\label{r:59}
Eq.~\eqref{e:bt_lr2} also applies to the Coulomb potential for electrostatic interactions \citep{spbc}.
However, in order to account for the full electro-mechanical coupling, the present theory needs to be extended.
\end{remark}

\begin{remark}\label{r:510}
We note again that for non-touching contact, the tractions $\bt_1^\mrc$ and $\bt_2^\mrc$ generally only satisfy global contact equilibrium \eqref{e:netforce}, but not local contact equilibrium \eqref{e:sf_moc}.
\end{remark}

\begin{remark}\label{r:511}
Eq.~\eqref{e:bt_sr} is a pure normal contact model that does not produce local tangential contact forces.
Tangential contact forces only arise globally when $\bt_k^\mrc$ acts on rough surfaces. 
\end{remark}

\subsection{Sticking limit}

\begin{figure}[h]
\begin{center} \unitlength1cm
\begin{picture}(0,5.8)
\put(-5.5,-.3){\includegraphics[height=61mm]{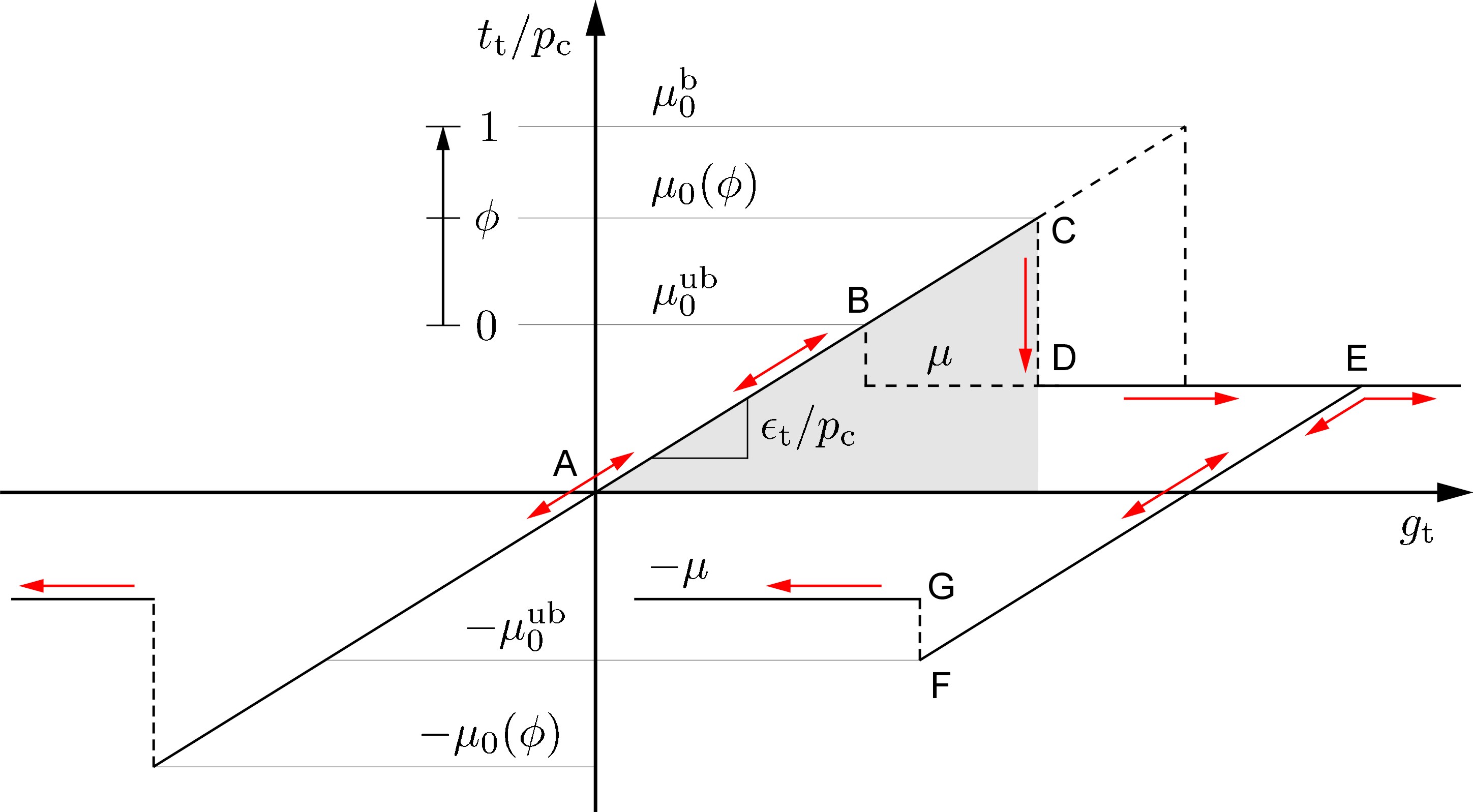}}
\end{picture}
\caption{Tangential contact behavior: Sticking, debonding and sliding according to \eqref{e:tt_max}-\eqref{e:mu0} and \eqref{e:dry_fric}. The latter two are irreversible.}
\label{f:debont}
\end{center}
\end{figure}

Similar to the debonding limit \eqref{e:debond}, a sticking limit implies a bound on the tangential traction $\bt_\mrt$, i.e.
\eqb{l}
\norm{\bt_\mrt} \leq t_\mrt^\mathrm{max}\,,
\label{e:tt_max}\eqe
where the limit value $t_\mrt^\mathrm{max}$ is generally a function of contact pressure, temperature and bonding state, i.e.
$t_\mrt^\mathrm{max} =  t_\mrt^\mathrm{max}(p_\mrc,T_\mrc,\phi)$. 
It is reasonable to assume that it is monotonically increasing with $p_\mrc$ and $\phi$ and decreasing with $T_\mrc$, e.g. 
\eqb{l}
t_\mrt^\mathrm{max} = \mu_0\,p_\mrc\,,
\eqe
where 
\eqb{l}
\mu_0 = \big(\phi\,\mu_0^\mrb + (1-\phi)\,\mu_0^\mathrm{ub}\big)\ds\frac{T_0}{T_\mrc}
\label{e:mu0}\eqe
is a temperature- and bonding state-dependent coefficient of sticking friction based on the constants $\mu_0^\mrb$ and $\mu_0^\mathrm{ub}$ describing the limits for full bonding and full unbonding, respectively.
Once the limit is reached, the bonds break ($\phi=0$) and tangential sliding occurs, as is discussed in the following section.
Fig.~\ref{f:debont} shows a graphical representation of this.
%-----------------------------------------------------------------
%-----------------------------------------------------------------

\subsection{Sliding models}

The simplest viscous friction model satisfying (\ref{e:cons_tc}.2) is the tangential traction model
\eqb{l}
\bt_\mrt = \bet\,\ds\frac{\Lgi}{g_\mrn}\,,
\label{e:visc_fric}\eqe
where $\bet$ is the positive definite dynamic viscosity tensor and $g_\mrn$ is the normal gap. 
In the isotropic case, $\bet = \eta\bone$.
\\
One of the simplest dry friction models satisfying (\ref{e:cons_tc}.2) for $p_\mrc\geq0$ is the Amontons-Coulomb law\footnote{Often just referred to as Amontons' law or Coulomb's law.}
\eqb{l}
\bt_\mrt = \bmu\,p_\mrc\ds\frac{\Lgi}{\norm{\Lgi}}\,,
\label{e:dry_fric}\eqe
where $\bmu$ is the positive definite coefficient tensor for sliding friction.
In the isotropic case, $\bmu = \mu\bone$.
For adhesion, where $p_\mrc\geq-t^\mathrm{max}_\mrn$, this extends to
\eqb{l}
\bt_\mrt = \bmu\,\big(p_\mrc+t_\mrn^\mathrm{max}\big)\ds\frac{\Lgi}{\norm{\Lgi}}\,.
\label{e:M-C}\eqe
In the adhesion literature this extension is often attributed to \citet{derjaguin34a}, whereas in soil mechanics it is usually referred to as Mohr-Coulomb's law.
Note that $\bmu\,t_\mrn^\mathrm{max}$ can be taken as a new constant.
The application of \eqref{e:M-C} to coupled adhesion and friction in the context of nonlinear 3D elasticity has been recently considered by \citet{adhfric,adhfric2}.

\begin{remark}\label{r:512}
A transition model between dry (rate-independent) and viscous (rate-dependent) sliding friction is Stribeck's curve, e.g.~see \citet{gelinck00}. 
\end{remark}

\begin{remark}\label{r:513}
The above sliding models can be temperature dependent. 
An example for a temperature dependent friction coefficient $\mu=\mu(T_\mrc)$ is given in \citet{laursen}.
Temperature dependent viscosity models are e.g.~discussed in \citet{roelands} and \citet{seeton06}.
\end{remark}

\begin{remark}\label{r:514}
The present formulation assumes fixed bonding sites that break during sliding friction.
Sliding friction is thus independent from $\phi$. 
Bonding sites can, however, also be mobile and hence stay intact during sliding.
Sliding friction may thus depend on $\phi$.
This requires a mobility model for the bonds, e.g.~following the work of \citet{freund04}.
\end{remark}

\begin{remark}\label{r:515}
The friction coefficient can be considered dependent on the surface deformation, as has been done by \citet{stupkiewicz01}.
\end{remark}

\begin{remark}\label{r:516}
Friction coefficients that are dependent on the sliding velocity and (wear) state have been considered in the framework of
rate and state friction models \citep{dieterich78,ruina83}.
Those models are usually based on an additive decomposition of the shear traction $\bt_\mrt$ instead of an additive slip decomposition as is used here.
\end{remark}

\begin{remark}\label{r:517}
Note that $\bt_\mrt$ causes the mechanical dissipation $\sD_\mathrm{mech} = \bt_\mrt\cdot\Lgi$ that leads to heating of the contact bodies
due to Eq.~\eqref{e:qc_m2}.
The mechanical dissipation is $\sD_\mathrm{mech} = \Lgi \cdot\bet\,\Lgi/g_\mrn$ for model \eqref{e:visc_fric} and $\sD_\mathrm{mech} = p_\mrc\,\Lgi \cdot\bmu\,\Lgi/\norm{\Lgi}$ for model \eqref{e:dry_fric}; see Sec.~\ref{s:test1} for an example.
\end{remark}

\subsection{Heat transfer}

The simplest heat transfer model satisfying condition \eqref{e:qtinq} is the linear relationship already given in \eqref{e:qc_t1}, where $h>0$ is the heat transfer coefficient, also referred to as the thermal contact conductance.  
Eq.~\eqref{e:qc_t1} is analogous to the mechanical flux model in (\ref{e:tsmu}.1). 
Similar to the mechanical case, $h\rightarrow\infty$ implies $T_2-T_1\rightarrow0$.
In general, $h$ can be a function of the contact pressure, gap or bonding state, i.e.~$h = h(p_\mrc,g_\mrn,\phi)$.
Various models have been considered in the past.
Those usually consider $h$ to be additively split into a contribution coming from actual contact and contributions coming from radiation and convection across a small contact gap.
\\
During actual contact ($g_\mrn=0$, $p_\mrc>0$) a simple model is the power law dependency on the contact pressure, 
\eqb{l}
h = h_0 + h_\mrc \bigg(\ds\frac{p_\mrc}{H}\bigg)^{\!q}\,,
\label{e:h_spot}\eqe
where $h_0$, $h_\mrc$, $H$ and $q$ are positive constants \citep{wriggers94,laschet04}. \\
This model is a simplification of the more detailed model of \citet{song87} that accounts for microscopic contact roughness.
Another model for $h$ is the model by \citet{mikic74}.
A new model has also been proposed recently by \citet{martins16}.
The dependency of \eqref{e:h_spot} on the bonding state can, for example, be taken as
\eqb{lllll}
h_0 \is \phi\,h_0^\mrb \plus (1-\phi)\,h_0^\mathrm{ub}\,, \\[1.5mm]
h_\mrc \is \phi\,h_\mrc^\mrb \plus (1-\phi)\,h_\mrc^\mathrm{ub}\,,
\eqe
i.e.~assuming that $h_0$ and $h_\mrc$ increase monotonically with $\phi$.
Here $h_0^\mrb$, $h_0^\mathrm{ub}$, $h_\mrc^\mrb$ and $h_\mrc^\mathrm{ub}$ are model constants.
\\
Out of contact ($g_\mrn>0$, $p_\mrc=0$, $\phi=0$), the heat tranfer depends on the contact gap. 
Considering small $g_\mrn$, \citet{laschet04} propose an exponential decay of $h$ with $g_\mrn$ according to
\eqb{l}
h = h_\mathrm{rad} + h_\mathrm{gas} + (h_0 - h_\mathrm{rad} - h_\mathrm{gas})\exp(-C_\mathrm{trans}\,g_\mrn)\,,
\eqe
where 
\eqb{l}
h_\mathrm{rad} = \ds\frac{c_\mathrm{rad}\,(T_1^2+T_2^2)\,(T_1+T_2)}{\varepsilon_1^{-1}+\varepsilon_2^{-1}-1}
\eqe
and
\eqb{l}
h_\mathrm{gas} = \ds\frac{k_\mathrm{gas}(T_\mrc)}{g_\mrn+g_0(T_\mrc)}
\eqe
correspond to the heat transfer across the gap due to radiation and gas convection, respectively.
Here, $C_\mathrm{trans}$, $c_\mathrm{rad}$, $\varepsilon_1$ and $\varepsilon_2$ are constants, while $k_\mathrm{gas}$ and $g_0$ depend on the contact temperature~$T_\mrc$.
\\
We note that all the above models are consistent with the 2.~law as long as $h>0$.

\subsection{Bonding reactions}\label{s:reac}

The simplest reaction rate model satisfying condition \eqref{e:cons_muc} is the linear relationship,
\eqb{l}
R_\mrc = -C_\mrr\,\mu_\mrc\,,
\label{e:R}\eqe
where the constant $C_\mrr\geq0$ can for example be a function of the contact temperature and pressure, i.e.~$C_\mrr = C_\mrr(p_\mrc,T_\mrc)$.
Writing $C_\mrr = \nc\,c_\mrr$, the reaction equation in \eqref{e:phi} thus becomes $\dot\phi = -c_\mrr\,\mu_\mrc$. 
Alternatively (and in consistency with \eqref{e:cons_muc}), $c_\mrr\geq0$ can be taken as a constant.

Another, more sophisticated example that satisfies \eqref{e:cons_muc} is the exponential relationship \citep{sahu17},
\eqb{l}
R_\mrc = C_\mrr\,R\,T_\mrc\,\bigg( 1 - \ds\exp\frac{\mu_\mrc}{R\,T_\mrc} \bigg)\,,
\eqe
where $R$ is the universal gas constant.

\begin{remark}\label{r:518}
Plugging example (\ref{e:tsmu}.3) and \eqref{e:phi_def} into \eqref{e:R} yields $R_\mrc = c_\mrr\,k_\mrc\,(\nc-n_\mrb)$ with $k_\mrc=\unde{K_c}/\Nc$, which is a pure forward reaction (see also Remark~\ref{r:53}).
$\overrightarrow{k}:=c_\mrr\,k_\mrc$ is then the forward reaction rate coefficient.
An example on how this could depend on the contact gap is given in \citet{sun09}.
\end{remark}

\begin{remark}\label{r:519}
On the other hand, modification \eqref{e:phieq} leads to $\unde{M_c} = \unde{K_c}\,(\phi-\phi_\mathrm{eq})$ instead of (\ref{e:tsmu}.3).
Substituting $\unde{K_c} = \overrightarrow{\unde{K_c}} + \overleftarrow{\unde{K_c}} $ and $\phi_\mathrm{eq} = \overrightarrow{\unde{K_c}}/\unde{K_c}$ then leads to 
\eqb{l}
\unde{M_c} = \overrightarrow{\unde{K_c}} \,(\phi-1) + \overleftarrow{\unde{K_c}}\,\phi\,,
\eqe
such that \eqref{e:R} leads to \eqref{e:Rc} with the reaction rates
\eqb{llllll}
\overrightarrow{R_\mrc} \is \overrightarrow{k}\,(\nc-n_\mrb)\,, \quad & \overrightarrow{k} \dis c_\mrr\,\overrightarrow{\unde{K_c}}/\Nc\,, \\[1mm]
\overleftarrow{R_\mrc} \is \overleftarrow{k}\,n_\mrb\,, & \overleftarrow{k} \dis c_\mrr\,\overleftarrow{\unde{K_c}}/\Nc\,,
\eqe
for chemical bonding (forward reaction) and debonding (backward reaction), respectively.
The case in Remark~\ref{r:518} is recovered for the backward reaction rate coefficient $\overleftarrow{k}=0$.
\end{remark}

\section{Elementary contact test cases}\label{s:test}

This section presents the analytical solution of three new elementary contact test cases:  
thermo-mechanical sliding, thermo-chemical bonding and thermo-chemo-mechanical debonding.
Such test cases are for example useful for the verification of a computational implementation.
Another test case -- for thermo-mechanical normal contact -- can be found in \citet{wriggers94}.
All test cases consider two blocks with initial height $H_1$ and $H_2$ brought into contact. 
The energy and temperature change in the contacting bodies is then computed 
assuming instantaneous heat transfer through the bodies ($\bq_k=\mathbf{0}$), 
no heat source in the bulk ($\rexk=0$), 
no boundary heat flux apart from $q^\mrc_\mrm$ ($\bar q_k = 0$ \& $q^\mrc_\mrt=0$),
equal temperature within the bodies ($T_1 = T_2 = T_\mrc$),
quasi-static deformation ($\bD_k=\mathbf{0}$), 
homogeneous deformation ($\bE_k\neq\bE_k(\bx_k)$), 
uniform contact conditions ($\bg\neq\bg(\bx_\mrc)$),
and fixed contact area ($J_{sk}=1$), 
such that one can work with the Helmholtz free %energies (per unit contact area) $\Psi_k = H_k\,\rho_k\,\psi_k$ and 
contact energy $\Psi_\mrc = \nc\,\psi_\mrc$.\footnote{$\Jc=1$ leads to $\unde{U_c} = U_\mrc$, $\unde{S_c} = S_\mrc$, $\unde{\Psi_c} = \Psi_\mrc$, etc.} 
%where $H_k$ is the current height associated with body $\sB_k$.
Under these conditions the temperature change within each body is governed by the energy balance
\eqb{l}
%\dot U_k = q_\mrm^\mrc\,,
\rho_{k0}\,H_k\,\dot u_k = q_\mrm^\mrc\,,
\label{e:Uk}\eqe
that follows directly from integrating \eqref{e:sf_en} over the reference volume, writing $\dif V_k = H_k\, \dif A_k$ on the left hand side and applying the divergence theorem on the right. %dividing by the contact area. \\
Considering the bulk energy
\eqb{l}
%\Psi_k = \Psi_{\mathrm{mech}} - \ds\frac{C_k}{2T_0}\big(T_k-T_0\big)^2\,,
\psi_k = \psi_{\mathrm{mech}} - \ds\frac{c_k}{2T_0}\big(T_k-T_0\big)^2\,,
\eqe
%where $C_k := H_k\,\rho_k\,c_k$ is the heat capacity per unit contact area, leads to
where $c_k$ is the heat capacity per unit mass, leads to
\eqb{l}
%U_k = U_\mathrm{mech} + \ds\frac{C_k}{2T_0}\big(T_k^2-T_0^2\big)\,,
u_k = u_\mathrm{mech} + \ds\frac{c_k}{2T_0}\big(T_k^2-T_0^2\big)\,,
\label{e:Uk0}\eqe
due to \eqref{e:psi_def} and \eqref{e:cons_k}.
%For quasi-static deformations then follows $\dot U_k = C_k\,T_k\,\dot T_k/T_0$.
For quasi-static deformations then follows $\dot u_k = c_k\,T_k\,\dot T_k/T_0$.
%It is further assumed that both bodies heat up equally such that  $T_1 = T_2 = T_\mrc$.
Inserting this into \eqref{e:Uk} %and using $\rho_k\,h_k = \rho_{k0}\,H_k$ (since $J_{sk}=1$), 
then gives
\eqb{l}
%\ds\frac{C_k}{T_0}\,T_k\,\dot T_k = q_\mrm^\mrc \,,
\ds\frac{C_k}{T_0}\,T_\mrc\,\dot T_\mrc = q_\mrm^\mrc \,,
\label{e:Tk}\eqe
where $C_k := H_k\,\rho_{k0}\,c_k$ is the heat capacity per unit contact area satisfying $C_1=C_2$.

\subsection{Sliding thermodynamics}\label{s:test1}

\begin{figure}[h]
\begin{center} \unitlength1cm
\begin{picture}(0,5.7)
\put(-7.9,.4){\includegraphics[height=51mm]{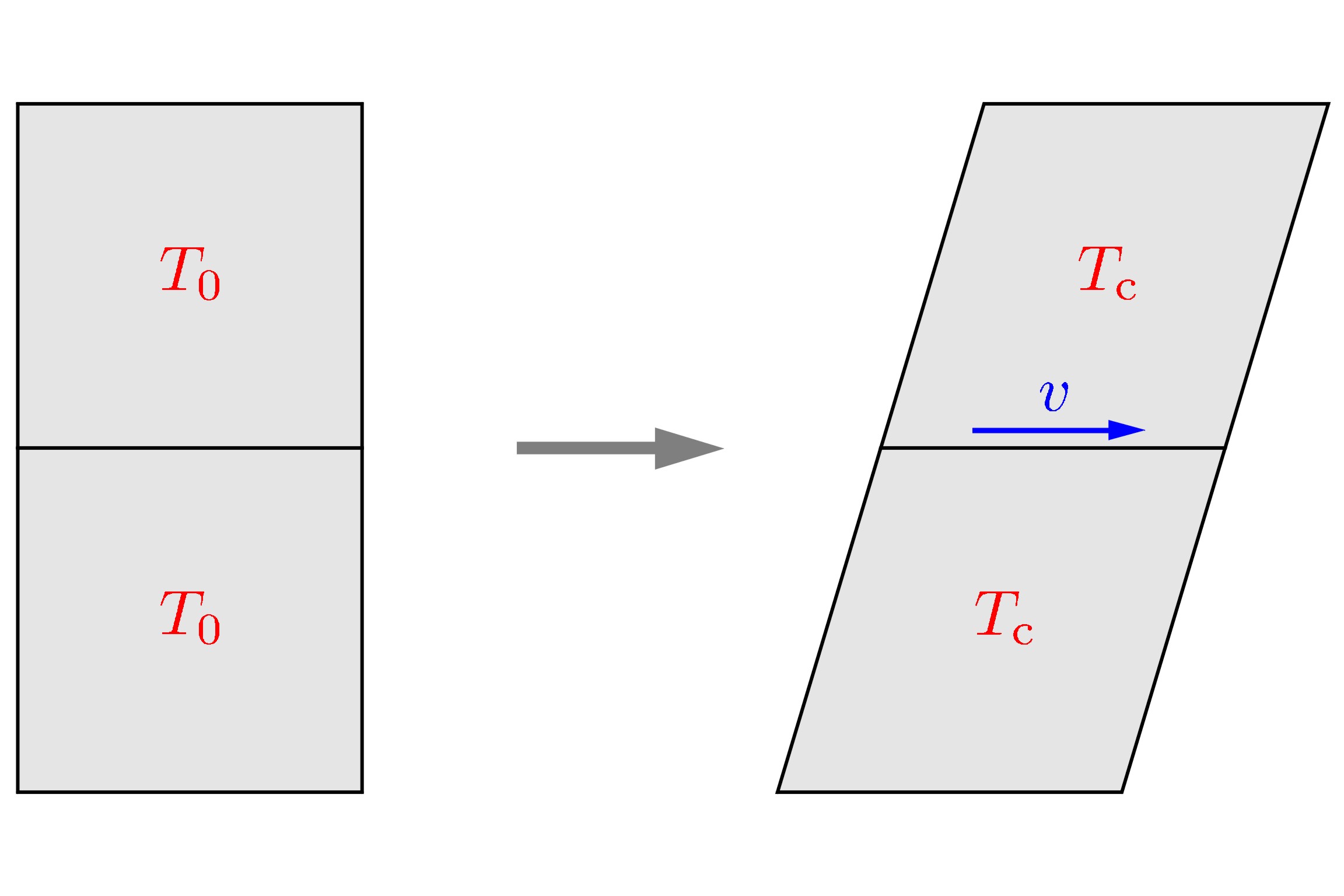}}
\put(0.15,-.1){\includegraphics[height=58mm]{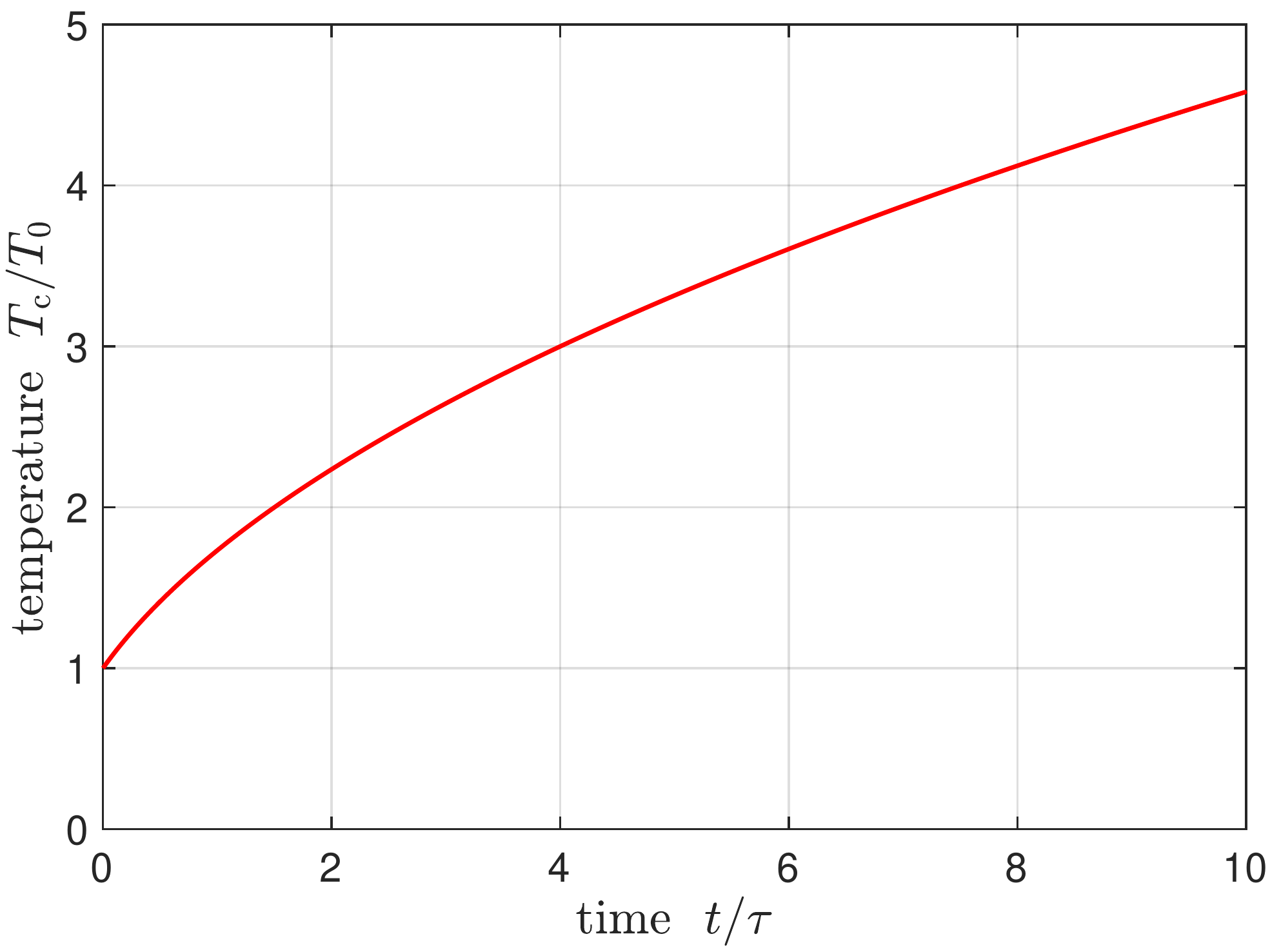}}
\put(-7.85,-.1){a.}
\put(.35,-.1){b.}
\end{picture}
\caption{Sliding thermodynamics: a.~model problem; b.~temperature rise during sliding}
\label{f:ex1}
\end{center}
\end{figure}

The first test case illustrates thermo-mechanical coupling %of sliding 
by calculating the temperature rise due to mechanical sliding. %at velocity $v$.
The test case is illustrated in Fig.~\ref{f:ex1}a: Two blocks initially at mechanical rest and temperature $T_0$ are subjected to steady sliding with the relative sliding velocity magnitude $v:=\norm{\Lgi}$. 
%-----------------------------------------------------------------
%-----------------------------------------------------------------
Following \citet{stromberg96} and \citet{oancea97}, the contact energy
\eqb{l}
\Psi_\mrc = \ds\frac{1}{2}\,\bg_\mre\cdot\bveps\,\bg_\mre - \frac{C_\mrc}{2T_0}\big(T_\mrc-T_0\big)^2
\label{e:Psi_t1}\eqe
is used with constant $\bveps$ and $C_\mrc$.
Eq.~\eqref{e:Psi_t1} leads to the entropy given in (\ref{e:tsmu}.2).
The mean influx, given by \eqref{e:qc_m2} and \eqref{e:qc_m2a}, then becomes
\eqb{l}
2q_\mrc^\mrm = \sD_\mathrm{mech} - \ds\frac{C_\mrc}{T_0}\,T_\mrc\,\dot T_\mrc\,,
\eqe
where $\sD_\mathrm{mech} = \eta v^2/g_\mrn$, according to model \eqref{e:visc_fric} and $\sD_\mathrm{mech} = \mu p v$ according to model \eqref{e:dry_fric}.
During stationary sliding ($v=$ const.) the mechanical deformation is time-independent.
From \eqref{e:Tk} thus follows
\eqb{l}
T_\mrc\,\dot T_\mrc = \ds\frac{T_0^2}{\tau}\,,\quad$with$~~\tau := \ds\frac{C_1 + C_2 + C_\mrc}{\sD_\mathrm{mech}}\,T_0\,,
\eqe
where $\tau$ is the time scale of the temperature rise.
Integrating this from the initial condition $T_\mrc(0) = T_0$ leads to the temperature rise 
\eqb{l}
T_\mrc(t) = T_0\,\sqrt{1+ 2\,t/\tau}\,,
\eqe
which is shown in Fig.~\ref{f:ex1}b.

\subsection{Bonding thermodynamics}\label{s:test2}

\begin{figure}[h]
\begin{center} \unitlength1cm
\begin{picture}(0,5.7)
\put(-7.9,.4){\includegraphics[height=51mm]{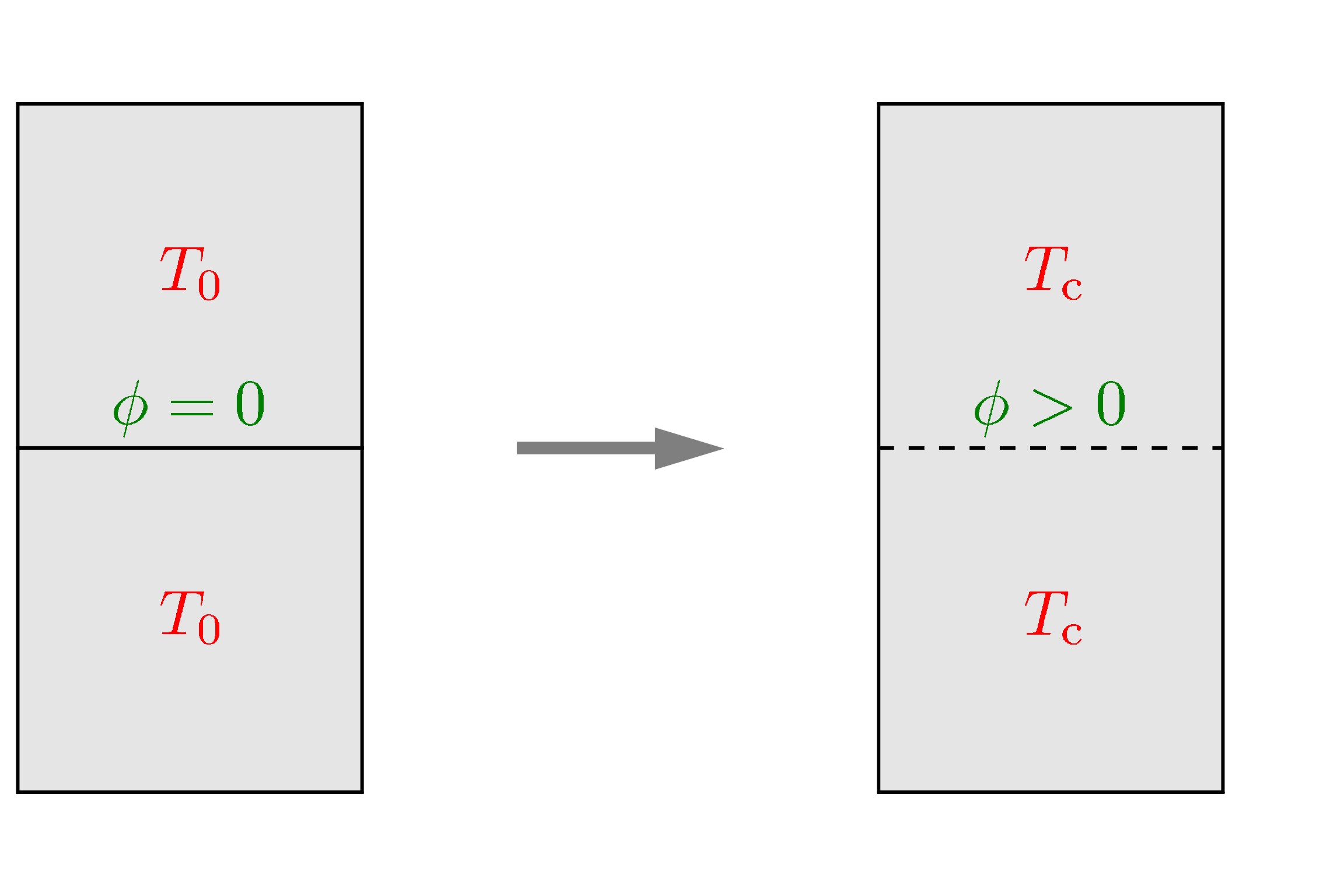}}
\put(0.15,-.1){\includegraphics[height=58mm]{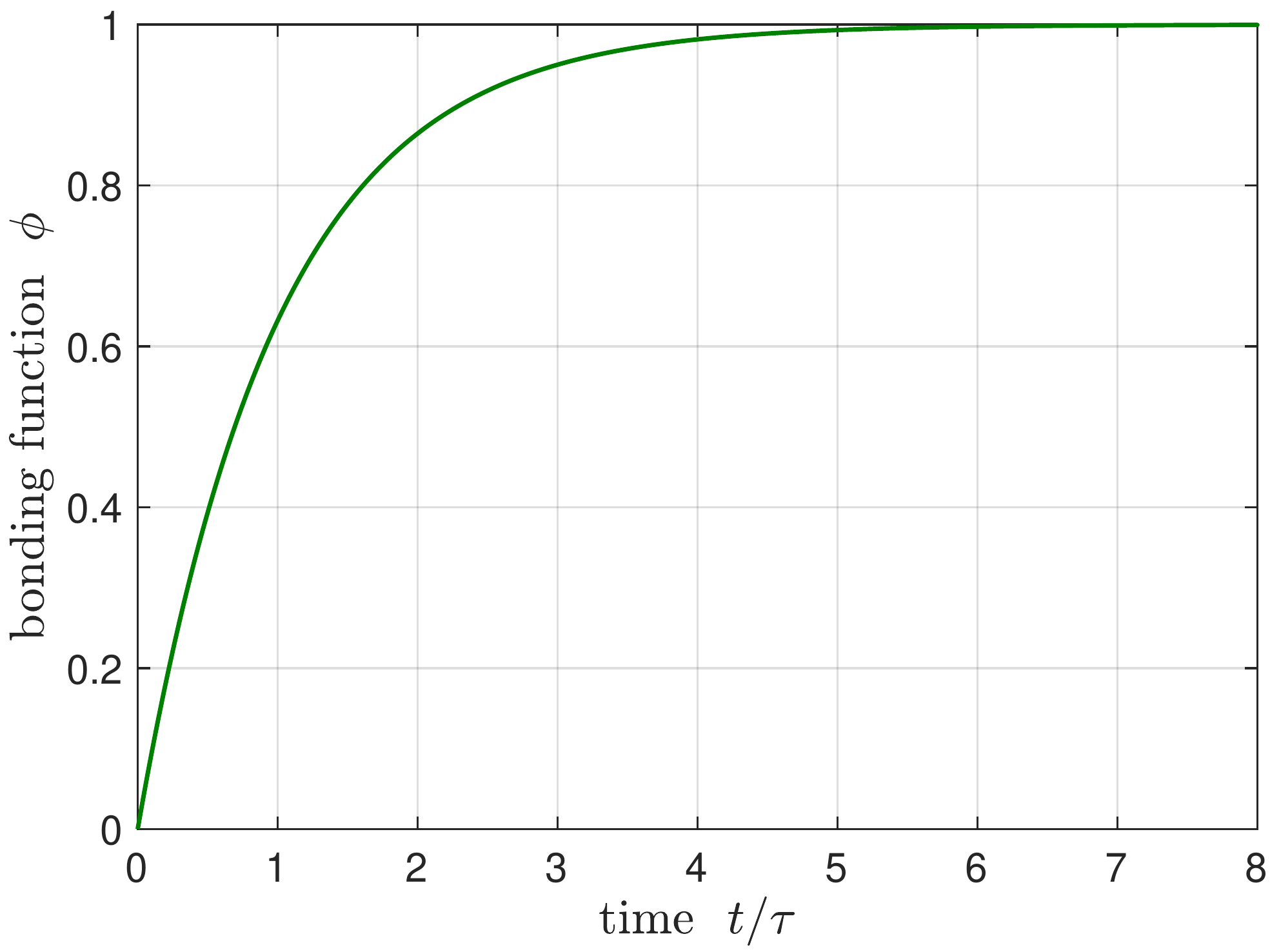}}
\put(-7.85,-.1){a.}
\put(.35,-.1){b.}
\end{picture}
\caption{Bonding thermodynamics: a.~model problem; b.~bonding function $\phi(t)$.}
\label{f:ex2a}
\end{center}
\end{figure}

The second test case illustrates thermo-chemical coupling by calculating the temperature change due to chemical bonding.
The test case is illustrated in Fig.~\ref{f:ex2a}a: Two blocks initially unbonded and at temperature $T_0$ are bonding and changing temperature. 
%-----------------------------------------------------------------
%-----------------------------------------------------------------
Now, the contact energy
\eqb{l}
\Psi_\mrc = \ds\frac{K_\mrc(T_\mrc)}{2}(\phi-1)^2\,,\quad$with$~~K_\mrc(T_\mrc) = K_0 + K_2\ds\frac{T_\mrc^2}{T_0^2}\,,
\label{e:Psi_bt}\eqe
is used, which, due to \eqref{e:SMuc}, leads to the chemical potential and entropy
\eqb{lll}
M_\mrc \is K_\mrc(T_\mrc)\,(\phi-1)\,, \\[3mm]
S_\mrc \is -\ds\frac{K_2\,T_\mrc}{T_0^2}(\phi-1)^2\,.
\eqe
This in turn leads to the internal energy
\eqb{l}
U_\mrc = \ds\frac{1}{2}(\phi-1)^2\bigg(K_0 - K_2\ds\frac{T_\mrc^2}{T_0^2}\bigg)\,,
\label{e:Ucex}\eqe
and the bonding ODE
\eqb{l}
\dot\phi = \ds\frac{1}{\tau}(1-\phi)\,,\quad $with$~~\tau = \ds\frac{\nc}{c_\mrr(T_\mrc)\,K_\mrc(T_\mrc)}\,,
\label{e:dphi_ex}\eqe
according to \eqref{e:phi}, \eqref{e:psic_def} and \eqref{e:R}.
ODE \eqref{e:dphi_ex} can only be solved with knowledge about $T_\mrc = T_\mrc(t)$.

\subsubsection{Exothermic bonding}\label{s:test2a}

\begin{figure}[h]
\begin{center} \unitlength1cm
\begin{picture}(0,5.7)
\put(-7.95,-.1){\includegraphics[height=58mm]{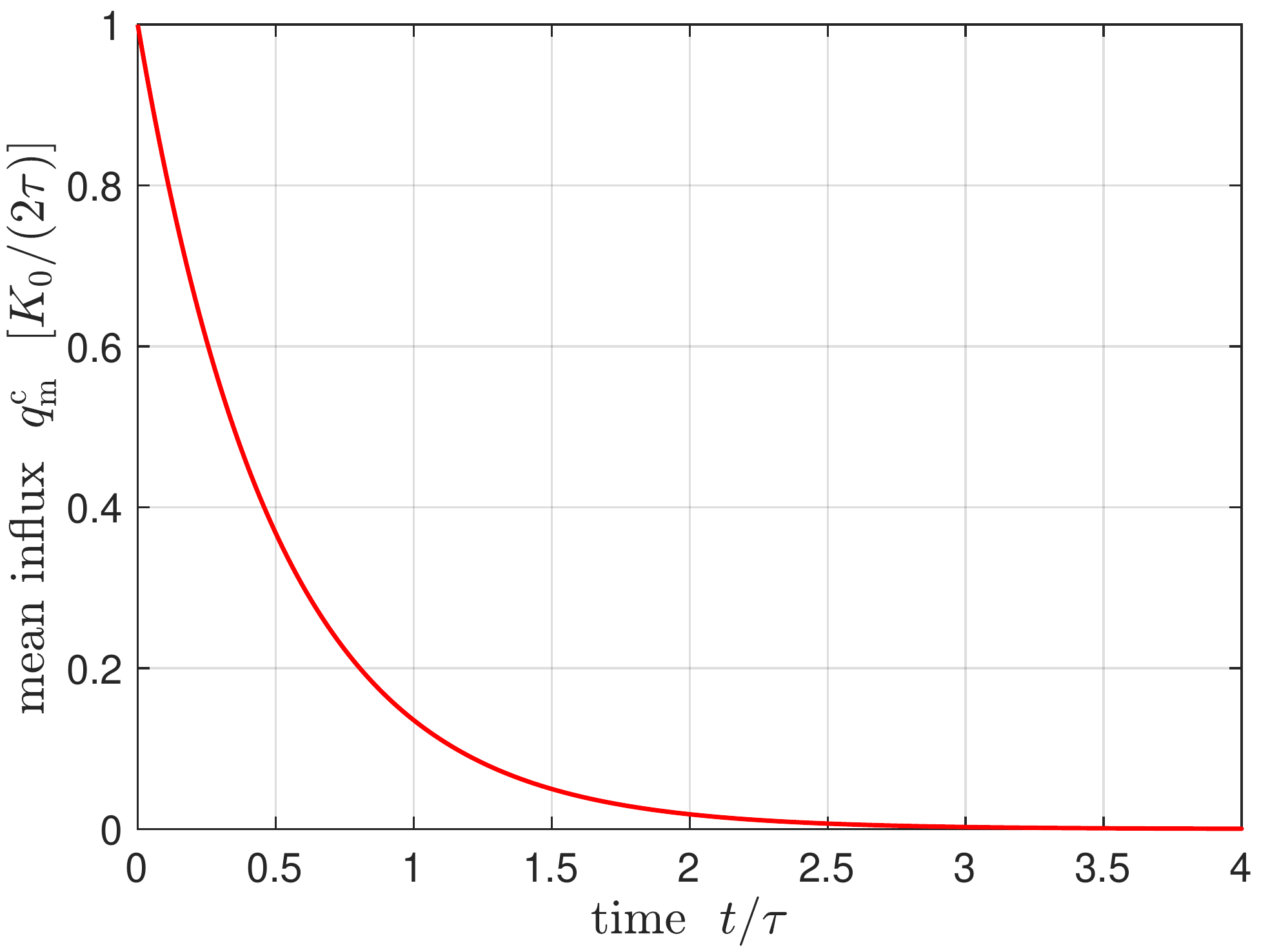}}
\put(0.15,-.1){\includegraphics[height=58mm]{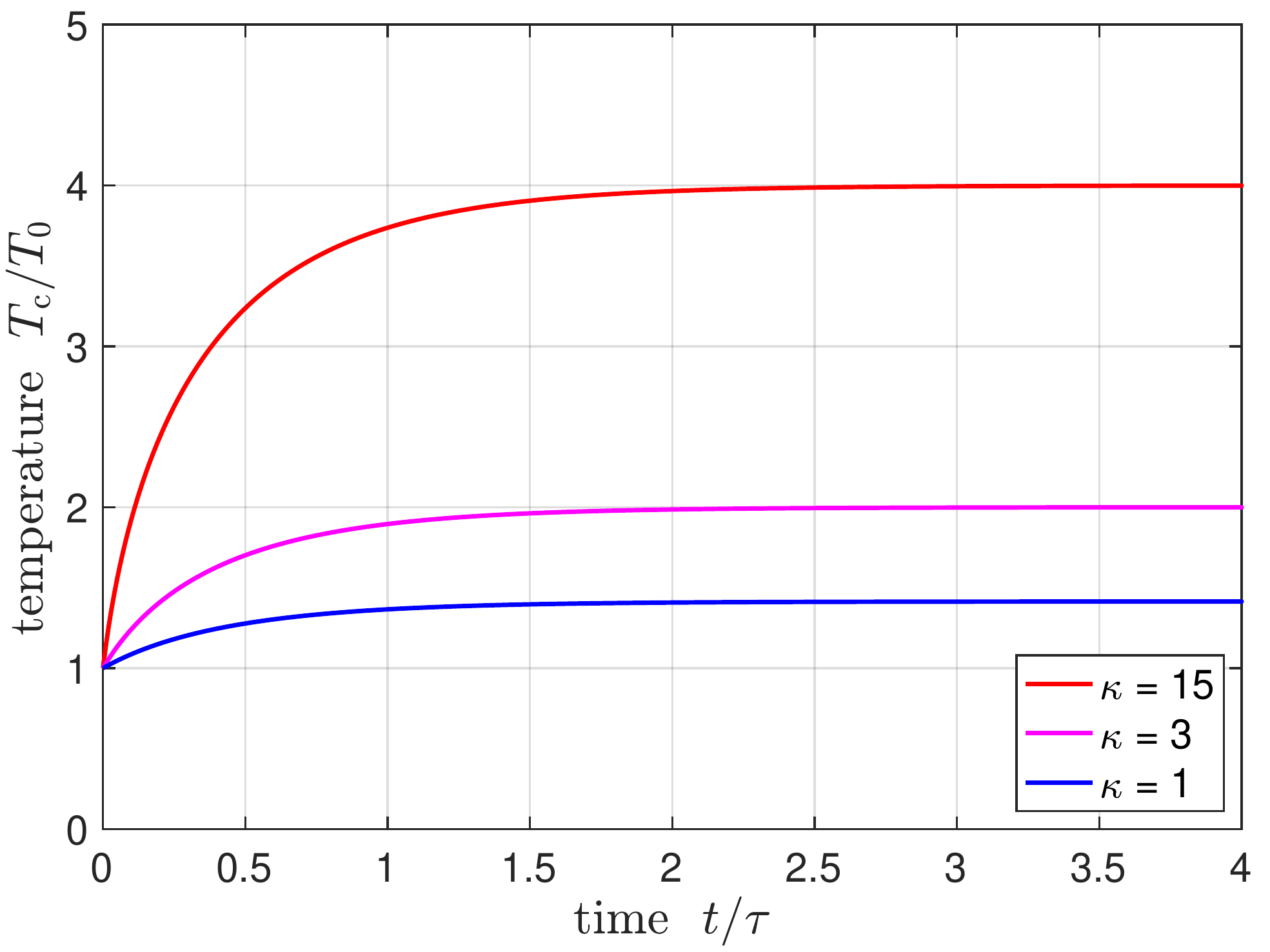}}
\put(-7.85,-.1){a.}
\put(.35,-.1){b.}
\end{picture}
\caption{Bonding thermodynamics: a.~heat influx $q_\mrm^\mrc(t)$ and b.~temperature rise $T_\mrc(t)$ during exothermic bonding}
\label{f:ex2b}
\end{center}
\end{figure}

Exothermic bonding occurs for $K_2=0$ (which implies $U_\mrc = \Psi_\mrc$).
Considering $c_\mrr=$ const., the  bonding ODE becomes independent of $T_\mrc$ and can be integrated from the initial condition $\phi(0) = 0$ to give
\eqb{l}
\phi(t) = 1 - e^{-t/\tau}\,,
\label{e:phi_ex}\eqe
where $\tau = n_1/(c_\mrr K_0)$ is the time scale of the exothermic bonding reaction.
Solution \eqref{e:phi_ex} is shown in Fig.~\ref{f:ex2a}b.
The mean influx, given in \eqref{e:qc_m2} and \eqref{e:qc_m2a}, then becomes
\eqb{l}
q_\mrm^\mrc = \ds\frac{K_0}{2\tau}\,e^{-2t/\tau}\,.
\eqe
It satisfies the exothermic condition $q_\mrm^\mrc > 0$ and is shown in Fig.~\ref{f:ex2b}a.
From \eqref{e:Tk} thus follows
\eqb{l}
C_k\,T_\mrc\,\dot T_\mrc = \ds\frac{K_0\,T_0}{2\tau}\,e^{-2t/\tau}\,,
\eqe
which can be integrated from the initial condition $T_\mrc(0) = T_0$ to give the temperature rise 
\eqb{l}
T_\mrc(t) = T_0\sqrt{1+\kappa\,\Big(1-e^{-2t/\tau}\Big)}\,,\quad $with$~~\kappa := \ds\frac{K_0}{2\,T_0\,C_k}\,,
\eqe
shown in Fig.~\ref{f:ex2b}b for various values of $\kappa$.
%-----------------------------------------------------------------
%-----------------------------------------------------------------

\subsubsection{Endothermic bonding}\label{s:test2b}

\begin{figure}[h]
\begin{center} \unitlength1cm
\begin{picture}(0,5.7)
\put(-7.95,-.1){\includegraphics[height=58mm]{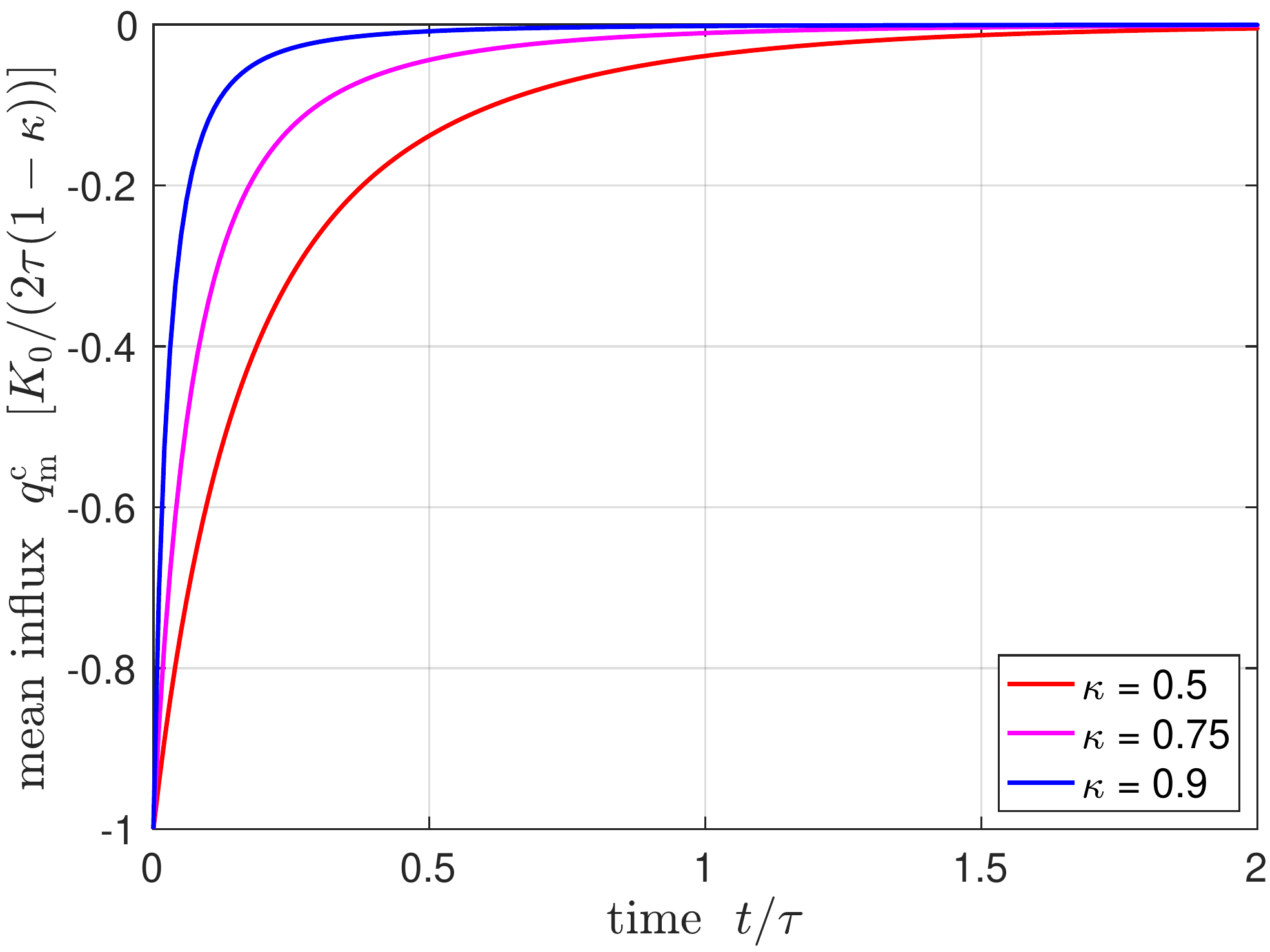}}
\put(0.15,-.1){\includegraphics[height=58mm]{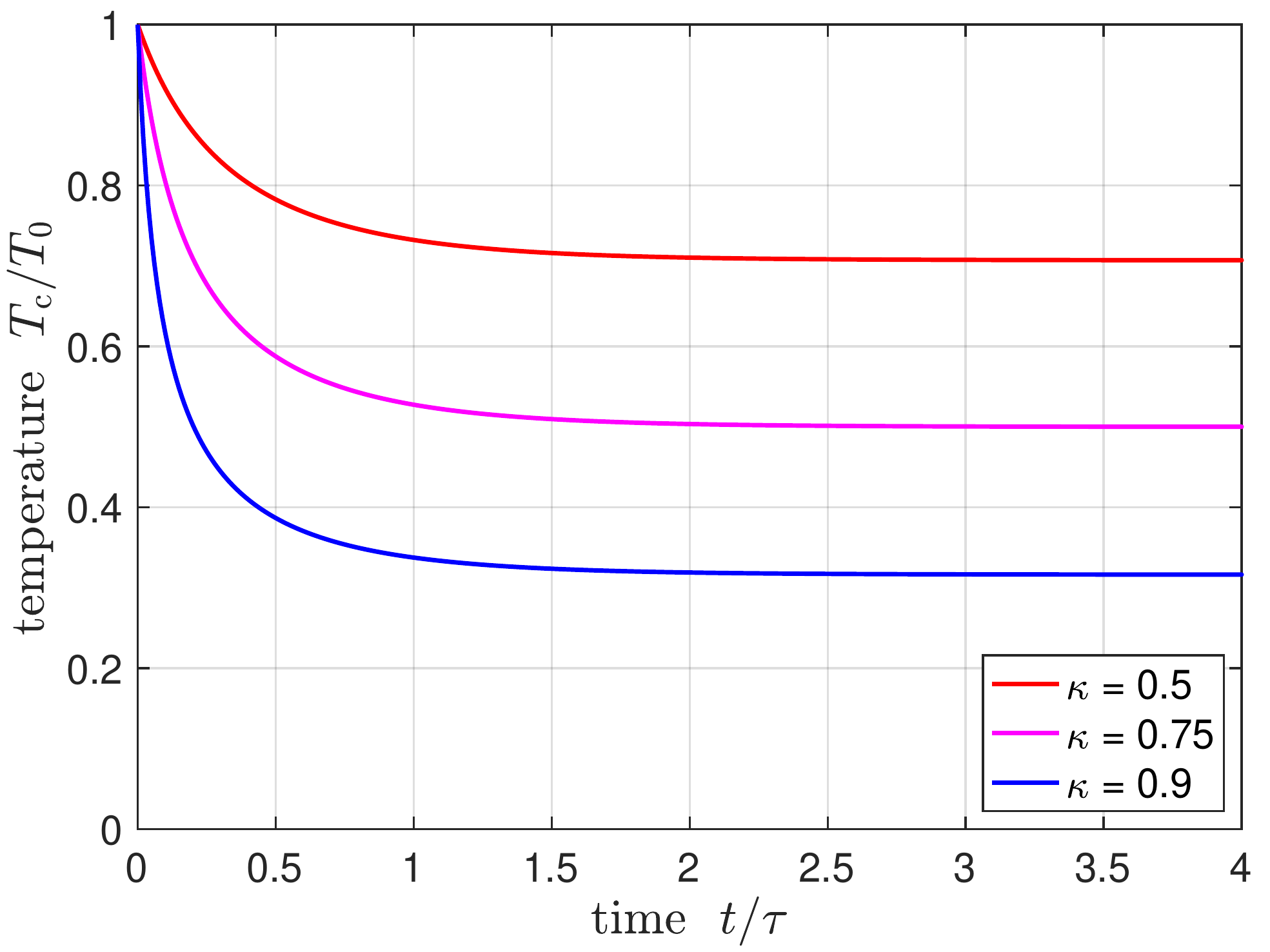}}
\put(-7.85,-.1){a.}
\put(.35,-.1){b.}
\end{picture}
\caption{Bonding thermodynamics: a.~heat influx $q_\mrm^\mrc(t)$ and b.~temperature drop $T_\mrc(t)$ during endothermic bonding}
\label{f:ex2c}
\end{center}
\end{figure}

Endothermic  bonding occurs for $K_0=0$ in the above model (which implies $U_\mrc = -\Psi_\mrc$ according to \eqref{e:Psi_bt} and \eqref{e:Ucex}). 
The mean influx, given in \eqref{e:qc_m2} and \eqref{e:qc_m2a}, now becomes
\eqb{l}
q_\mrm^\mrc = \ds\frac{K_2}{2T_0^2}\,(1-\phi)^2\,\Big(T_\mrc\,\dot T_\mrc - \frac{1}{\tau}\, T_\mrc^2\Big)\,.
\label{e:qmc_ex2b}\eqe
In order to solve \eqref{e:dphi_ex}, a temperature dependent reaction rate is considered in the form $c_\mrr = c_{\mrr 0}T_0^2/T_\mrc^2$ with $c_{\mrr 0} =$ const.
Thus \eqref{e:phi_ex} is still the solution of bonding ODE \eqref{e:dphi_ex}.
The time scale now becomes $\tau=n_1/(c_{\mrr 0}K_2)$.
From \eqref{e:Tk} now follows
\eqb{l}
\ds\frac{\dot T_\mrc}{T_\mrc} = \ds\frac{1}{\tau}\,\frac{\kappa\,e^{-2t/\tau}}{\kappa\,e^{-2t/\tau} - 1}\,,\quad $with$~~\kappa := \ds\frac{K_2}{2\,T_0\,C_k}\,. \\
\eqe
Integrating this from the initial condition $T_\mrc(0) = T_0$ gives the temperature drop
\eqb{l}
T_\mrc(t) = T_0\,\sqrt{\ds\frac{1-\kappa}{1-\kappa\,e^{-2t/\tau}}}\,,
\eqe
where the parameter $\kappa$ must be smaller than unity for the temperature to remain physical.
According to \eqref{e:qmc_ex2b}, this now leads to
\eqb{l}
q^\mrc_\mrm(t) = -\ds\frac{K_2}{2\tau}\frac{1-\kappa}{(1-\kappa\,e^{-2t/\tau})^2}\,e^{-2\,t/\tau}\,,
\eqe
and satisfies $q_\mrm^\mrc < 0$.
Fig.~\ref{f:ex2c} shows the energy outflux and temperature drop of the contact interface for the endothermic case.
%-----------------------------------------------------------------
%-----------------------------------------------------------------

\subsection{Debonding thermodynamics}\label{s:test3}

\begin{figure}[h]
\begin{center} \unitlength1cm
\begin{picture}(0,5.7)
\put(-7.9,.4){\includegraphics[height=51mm]{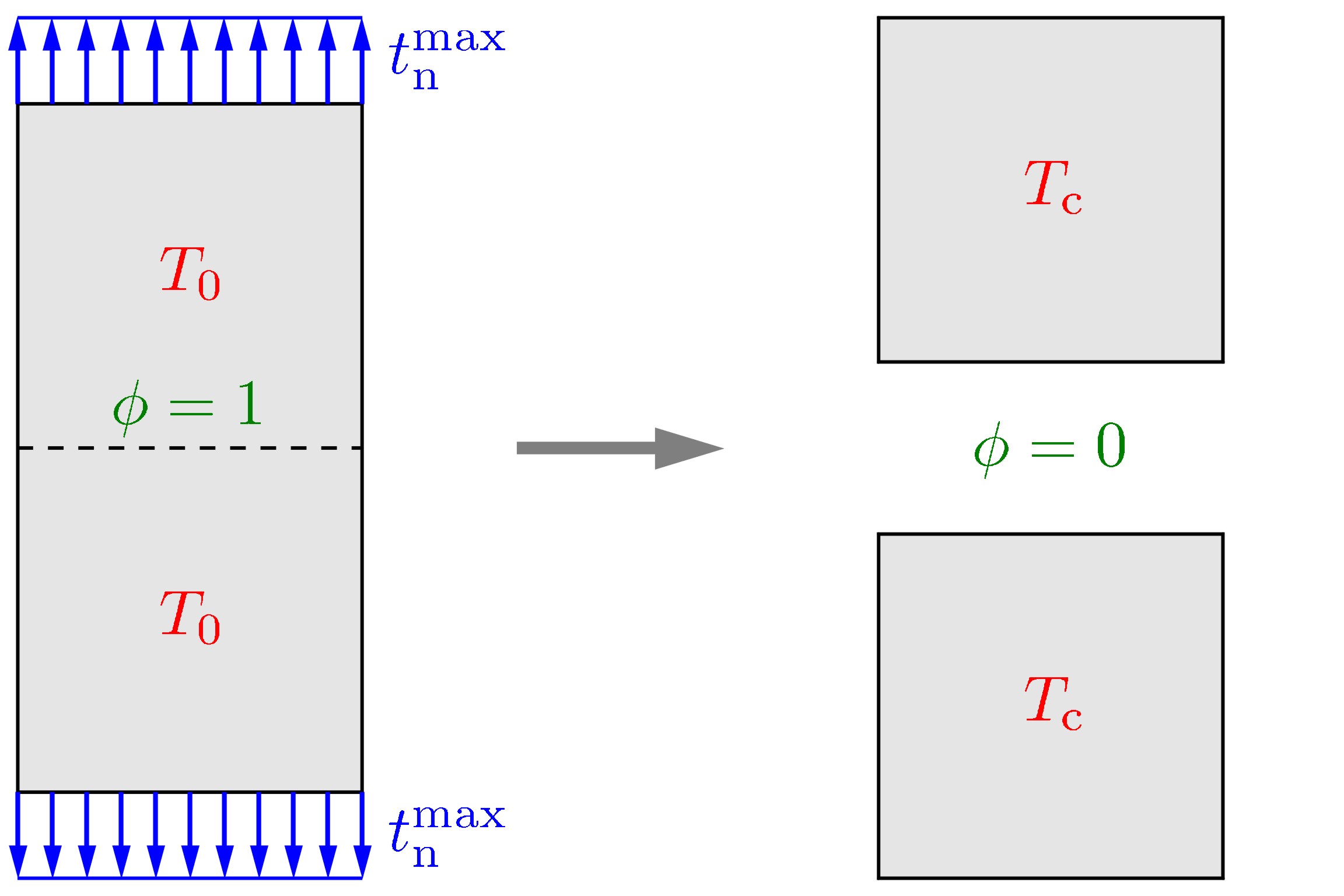}}
\put(0.15,-.1){\includegraphics[height=58mm]{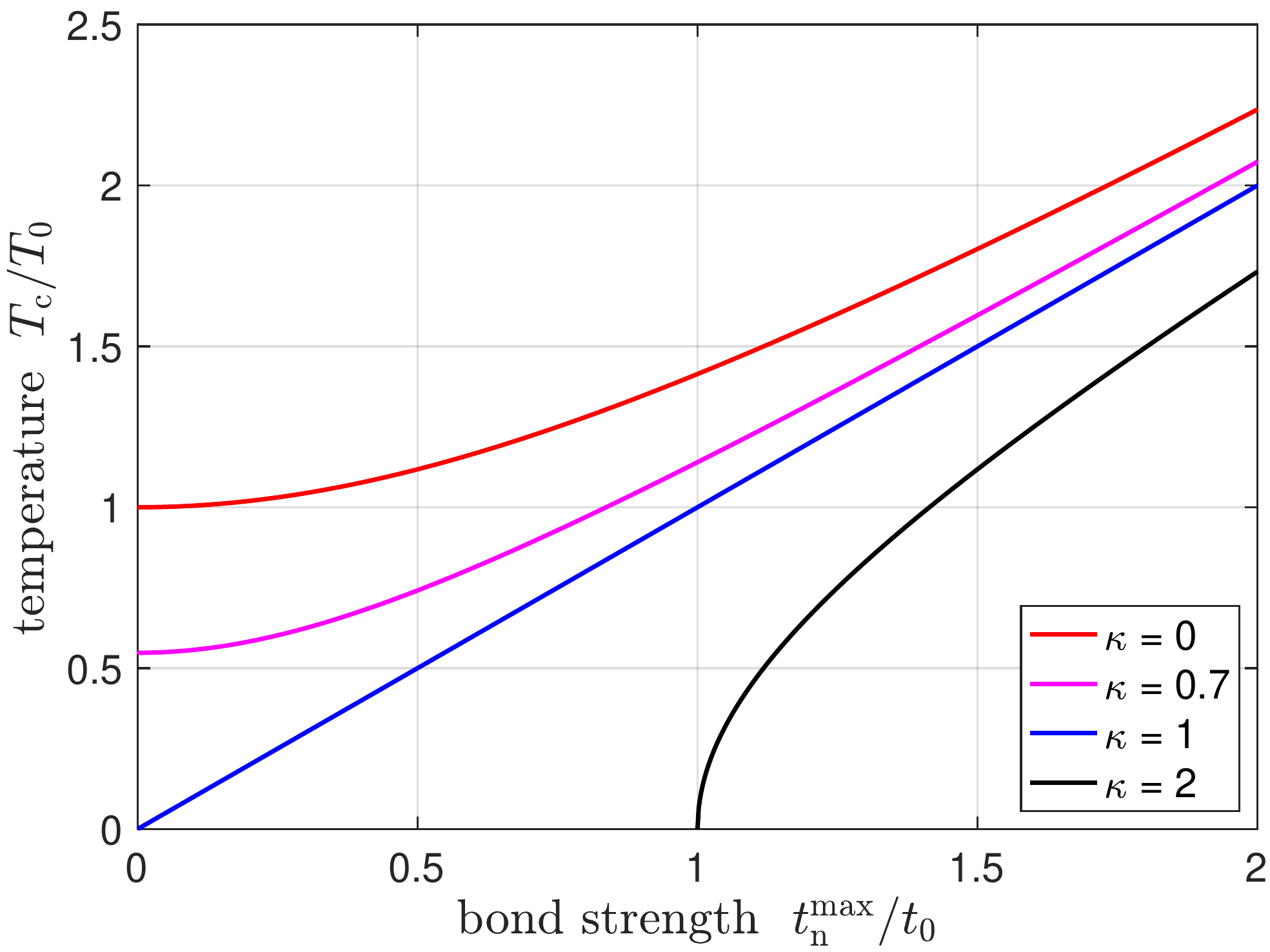}}
\put(-7.85,-.1){a.}
\put(.35,-.1){b.}
\end{picture}
\caption{Debonding thermodynamics: a.~model problem; b.~temperature change during debonding.}
\label{f:ex3}
\end{center}
\end{figure}

The last test case illustrates thermo-chemo-mechanical coupling by calculating the temperature change due to mechanical debonding.
The test case is illustrated in Fig.~\ref{f:ex3}a: Two blocks initially bonded and at temperature $T_0$ are pulled apart leading to debonding and rising temperature. 
Now the contact energy density
\eqb{l}
\Psi_\mrc = -\ds\frac{C_\mrc}{2T_0}(T_\mrc-T_0)^2 + \frac{K_0}{2} (\phi-1)^2
\label{e:Psi_t3}\eqe
is used.
When the bond breaks, elastic strain energy is converted into surface energy and kinetic energy.
The former corresponds to the stored bond energy (and is given by the second term in \eqref{e:Psi_t3}).
If viscosity is present, the latter then transforms into thermal energy.
Waiting long enough, the kinetic energy $\bbK$ completely dissipates into heat.
The temperature rise can then be calculated from the energy balance.
Setting the energy (per contact area) before and after debonding equal, gives
\eqb{l}
U_\mathrm{mech} + \ds\frac{C_\mathrm{tot}}{2T_0}\,T_0^2 = \frac{C_\mathrm{tot}}{2T_0}\,T_\mrc^2 + \frac{K_0}{2}\,,
\eqe
where $U_\mathrm{mech}$ is the mechanical energy at debonding and $C_\mathrm{tot} := C_1 + C_2 + C_\mrc$ is the total heat capacity of the system.
Considering linear elasticity gives $U_\mathrm{mech} = \frac{1}{2}(t_\mrn^\mathrm{max})^2/k_\mathrm{eff}$, 
where $k_\mathrm{eff}= (H_1/E_1+H_2/E_2)^{-1}$ is the effective stiffness of the two-body system based on height $H_k$ and Young's modulus $E_k$. \
This then leads to the temperature change
\eqb{l}
T_\mrc = T_0\,\sqrt{1 - \kappa + (t_\mrn^\mathrm{max}/t_0)^2}\,,
\eqe
with the positive constants $\kappa := K_0/(T_0\,C_\mathrm{tot})$ and $t_0^2 := T_0\,C_\mathrm{tot}\,k_\mathrm{eff}$.
It is shown in Fig.~\ref{f:ex3}b for various values of $\kappa$ and $t_\mrn^\mathrm{max}/t_0$.
%-----------------------------------------------------------------
%-----------------------------------------------------------------
Increasing the bond energy (represented by $\kappa$) leads to lower final temperatures, while higher bond strengths $t_\mrn^\mathrm{max}$ lead to higher final temperatures.
As a consequence, the final temperature can either be higher or lower than the initial temperature.
But note that $t_\mrn^\mathrm{max}/t_0 > \kappa-1$ to ensure $T_\mrc>0$.
This implies that for large $\kappa>1$ higher stresses are needed to break the bond.

\section{Conclusion}\label{s:concl}

This work has presented a unified continuum theory for coupled nonlinear thermo-chemo-mechanical contact as it follows from the fundamental balance laws and principles of irreversible thermodynamics.
It highlights the analogies between the different physical field equations, and it discusses the coupling present in the balance laws and constitutive relations.
Of particular importance is \eqref{e:qc_m2}, the equation for the mean contact heat influx.
It identifies how mechanical dissipation, chemical dissipation and interfacial entropy changes lead to interfacial heat generation.
This is illustrated by analytically solved contact test cases for steady sliding, exothermic bonding, endothermic bonding and forced bond-breaking.
\\
The proposed theory applies to all contact problems characterized by single-field or coupled thermal, chemical and mechanical contact.
There are several applications of particular interest to the present authors that are planned to be studied in future work.
One is the pressure-dependent curing thermodynamics of adhesives \citep{sain18}.
A second is the study of the bonding thermodynamics of insect and lizard adhesion based on the viscolelastic multiscale adhesion model of \citet{sauer10a}. A third is the local modeling and study of bond strength and failure of osseointegrated implants \citep{immel19}.
There are also interesting applications that require an extension of the present theory.
An example is electro-chemo-mechanical contact interactions in batteries.
Therefore the extension to electrical contact is required, e.g.~following the framework of \citet{weissenfels10}.
Another example is contact and adhesion of droplets and cells.
Therefore the extension to surface stresses, surface mobility of bonding sites, contact angles and appropriate bond reactions is needed, e.g.~following the framework of \citet{memtheo} and \citet{sahu17}.

\bigskip

{\Large{\bf Acknowledgements}}

RAS and TXD acknowledge the support of the German Research Foundation through projects SA1822/8-1 and GSC 111.
The authors are also grateful to the ACalNet for supporting RAS for a visit to Berkeley in 2018, and they thank 
Katharina Immel and Nele Lendt for their comments on the manuscript.

%\appendix

\bibliographystyle{apalike}
\bibliography{bibliography}

\end{document}